%% file: couplingpf.tex
\pdfoutput=1
\documentclass[english,11pt]{article}
\usepackage{lmodern}

\usepackage[T1]{fontenc}
\usepackage[latin9]{inputenc}
\usepackage{geometry}
\usepackage[unicode=true,bookmarks=true,bookmarksnumbered=false,bookmarksopen=false,colorlinks=true]{hyperref}

\newcommand{\blind}{1}

\usepackage{color}
\usepackage{babel}
\usepackage{float}
\usepackage{amsthm}
\usepackage{amsmath}
\usepackage{amssymb}
\usepackage{graphicx}
\usepackage{setspace}
\usepackage{dsfont}
\usepackage{subfig}
\usepackage{nicefrac}
\usepackage{bm}

\usepackage[authoryear]{natbib}

\makeatletter

\floatstyle{ruled}
\newfloat{algorithm}{tbp}{loa}
\providecommand{\algorithmname}{Algorithm}
\floatname{algorithm}{\protect\algorithmname}

 \theoremstyle{definition}
\newtheorem{thm}{Theorem}[section]
\newtheorem{lemma}{Lemma}[section]

\newtheorem{assumption}{\protect\assumptionname}
  \providecommand{\assumptionname}{Assumption}

\usepackage{fullpage}
\usepackage{dsfont}

\newcommand{\arginf}{\operatornamewithlimits{arginf}}
\newcommand{\Prb}{\mathbb{P}}
\newcommand{\E}{\mathbb{E}}
\newcommand{\setX}{\mathbb{X}}
\newcommand{\reals}{\mathbb{R}}
\newcommand{\I}{\mathds{1}}
\newcommand{\tx}{\tilde{x}}
\newcommand{\tw}{\tilde{w}}
\newcommand{\ta}{\tilde{a}}
\newcommand{\tX}{\tilde{X}}

\newcommand{\tU}{\tilde{U}}
\newcommand{\ttheta}{\tilde{\theta}}
\newcommand{\transp}{\mathsf{T}}                      
\newcommand{\bU}{\boldsymbol{U}}
\newcommand{\btU}{\tilde{\boldsymbol{U}}}
\newcommand{\btx}{\tilde{\boldsymbol{x}}}
\newcommand{\btw}{\tilde{\boldsymbol{w}}}
\newcommand{\bta}{\tilde{\boldsymbol{a}}}
\newcommand{\bx}{\boldsymbol{x}}
\newcommand{\bw}{\boldsymbol{w}}
\newcommand{\ba}{\boldsymbol{a}}
\newcommand{\bu}{\boldsymbol{u}}
\newcommand{\br}{\boldsymbol{r}}
\newcommand{\btu}{\tilde{\boldsymbol{u}}}
\newcommand{\btr}{\tilde{\boldsymbol{r}}}

\makeatother

\begin{document}
\def\spacingset#1{\renewcommand{\baselinestretch}%
{#1}\small\normalsize} \spacingset{1}


\if1\blind
{
  \title{\bf Coupling of Particle Filters\thanks{
  		This research is financially supported by 
  		the Swedish Foundation for Strategic Research (SSF) via the
  		project \emph{ASSEMBLE} and the Swedish research Council (VR)
  		via the projects 
  		\emph{Learning of complex dynamical systems} (Contract number: 637-2014-466)
  		and
  		\emph{Probabilistic modeling of dynamical
  			systems} (Contract number: 621-2013-5524).
  	}}
    \author{\textbf{Pierre E. Jacob\thanks{Corresponding author: pjacob@fas.harvard.edu. Code available at: \href{http://github.com/pierrejacob/}{github.com/pierrejacob/}.}} \hspace{.2cm}\\
    Department of Statistics, Harvard University\\[2mm]
    \textbf{Fredrik Lindsten and Thomas B. Sch\"on}\\ 
    Department of Information Technology, Uppsala University}
  \maketitle
} \fi

\if0\blind
{
  \bigskip
  \bigskip
  \bigskip
  \begin{center}
    {\LARGE\bf Coupling of Particle Filters}
\end{center}
  \medskip
} \fi
\bigskip

\spacingset{1.20}

\begin{abstract}
Particle filters provide Monte Carlo approximations of intractable
quantities such as point-wise evaluations of the likelihood in state space
models. In many scenarios, the interest lies in the comparison of
these quantities as some parameter or input varies.
%
To facilitate such comparisons, we introduce and study methods to couple two
particle filters in such a way that the correlation between the two underlying
particle systems is increased. The motivation stems from the classic variance
reduction technique of positively correlating two estimators.
%
%
The key challenge in constructing such a coupling stems from the
discontinuity of the resampling step of the particle filter. As our first contribution, we
consider coupled resampling algorithms. Within bootstrap particle filters, they improve the
precision of finite-difference estimators of the score vector and
boost the performance of particle marginal Metropolis--Hastings
algorithms for parameter inference.
%
%
The second contribution arises from the use of these coupled resampling
schemes within conditional particle filters, allowing for unbiased estimators of smoothing functionals.
The result is a new smoothing strategy that operates by averaging a number of independent and
unbiased estimators, 
which allows for 1) straightforward
parallelization and 2) the construction of accurate error estimates.
Neither of the above is possible with existing particle smoothers.
\end{abstract}

\noindent%

\section{Introduction\label{sec:introduction}}

In the context of nonlinear state space models, particle filters provide
efficient approximations of the distribution of a latent process
$(x_{t})_{t\geq 0}$, given noisy and partial observations
$(y_t)_{t\geq 1}$
\citep{doucet:defreitas:gordon:2001,cappe:ryden:2004,doucet2011tutorial}. We
assume that the latent process takes values in $\mathbb{X}\subset
\mathbb{R}^{d_x}$, 
and that the observations are in $\mathbb{Y}\subset
\mathbb{R}^{d_y}$ for some $d_x,d_y \in\mathbb{N}$.  The model specifies an
initial distribution $m_0(dx_{0}|\theta)$ and a transition kernel
$f(dx_{t}| x_{t-1},\theta)$ for the Markovian latent process.
Conditionally upon the latent process, the observations are independent
and their distribution is given by a measurement kernel $g(dy_{t}| x_{t},\theta)$. The model is
parameterized by $\theta\in\Theta\subset \mathbb{R}^{d_\theta}$, for $d_\theta\in\mathbb{N}$. Filtering 
consists in approximating the distribution $p(dx_{t}|
y_{1:t},\theta)$ for all times $t\geq 1$, whereas smoothing
consists in approximating the distribution $p(dx_{0:T}|y_{1:T},\theta)$ for
a fixed time horizon $T$, where
for $s,t\in\mathbb{N}$, we write $s:t$ for the set $\{s,\ldots,t\}$, and $v_{s:t}$
for the vector $(v_s,\ldots,v_t)$.

The bootstrap particle filter
\citep{gordon:salmon:smith:1993} generates weighted samples denoted by 
$(w_{t}^{k},x_{t}^{k})_{k=1}^{N}$, for all
$t\in\mathbb{N}$, where the particle locations $(x_{t}^{k})_{k=1}^{N}$ are samples in
$\mathbb{X}$ and the weights $(w_{t}^{k})_{k=1}^{N}$ are non-negative reals summing
to one. The number $N\in\mathbb{N}$ of particles is specified by the user---the computational cost of the algorithm is linear in $N$, while the
approximation of $p(dx_t|y_{1:t},\theta)$ by  $\sum_{k=1}^{N}w_{t}^{k}\delta_{x_{t}^{k}}(dx_{t})$
becomes more precise as $N$ increases   
\citep[e.g.][and references therein]{del2004feynman,whiteley2011stability}. 
An important by-product of the particle filter for statistical inference is the
likelihood estimator, defined as
$\hat p^N(y_{1:t}|\theta) := \prod_{s=1}^{t}N^{-1}\sum_{k=1}^{N} g(y_s|x_s^k,\theta)$. The likelihood estimator is
known to have expectation equal to the likelihood
$p(y_{1:t}|\theta)$, and its variance has been extensively
studied \citep{del2004feynman,CerDelGuy2011nonasymptotic,berard2014lognormal}. The estimator
is at the core of the particle marginal Metropolis--Hastings (MH) algorithm
\citep{andrieu:doucet:holenstein:2010,doucet2015efficient}, 
in which particle filters are run within an MH scheme,
enabling inference in a large class of state space models.

We consider methods to couple particle filters. A coupling
of particle filters, given two parameter values  $\theta$ and $\tilde\theta$,
refers to a pair of particle systems, denoted by
$(w_{t}^{k},x_{t}^{k})_{k=1}^{N}$ and $(\tw_{t}^{k},\tx_{t}^{k})_{k=1}^{N}$,
such that: 1) marginally, each system has the same distribution as if it was
generated by a particle filter, respectively given $\theta$ and $\ttheta$, and
2) the two systems are in some sense correlated. 
The same couplings can be applied to pairs of conditional particle filters \citep{andrieu:doucet:holenstein:2010}, 
which are conditioned on different reference trajectories, instead of different parameters.
In the case of particle filters, the goal is to introduce positive correlations between likelihood
estimators $\hat p ^N(y_{1:t}|\theta)$ and $\hat p
^N(y_{1:t}|\ttheta)$, which improves the performance of score estimators 
and of MH schemes \citep{deligiannidis2015correlated,dahlin2015accelerating}.
In the case of conditional particle filters, couplings lead to 
a new algorithm for smoothing, which is
trivial to parallelize, provides unbiased estimators of smoothing
functionals and accurate estimates of the associated Monte Carlo
error.

Correlating estimators is a classic Monte Carlo technique for variance
reduction, and can often be achieved by using common random numbers
\citep{kahn1953methods,asmussen2007stochastic,glasserman1992some}.
Particle filters are randomized
algorithms which can be written as a deterministic function of some random
variables and a parameter value. However, they are discontinuous
functions of their inputs, due to the resampling steps.  This discontinuity
renders theoretical guarantees supporting the use of common random numbers
such as Proposition 2.2 in \citet{glasserman1992some} inapplicable.  Despite various
attempts \citep[see][and references therein]{pitt2002smooth,lee2008towards,malik2011particle},
there are no standard ways of coupling particle filters. 
Our proposed strategy relies on common random numbers for the initialization and propagation steps, while
the resampling step is performed jointly for a pair of particle systems, using
ideas inspired by maximal couplings and optimal transport ideas.

Coupled resampling schemes and coupled particle filters are described in Section~\ref{sec:couplingparticlesystems}.
In Section~\ref{sec:methods}, they are shown to lead to various methodological developments:
in particular, they are instrumental in the construction of a new smoothing estimator, in combination with
the debiasing technique of \citet{glynn2014exact}.
In Section~\ref{sec:numerics}, numerical results illustrate the gains brought by coupled particle filters 
in a real-world prey-predator model, and Section~\ref{sec:discussion} concludes.
The appendices contain various additional descriptions, proofs and extra  numerical results.

\section{Coupled resampling \label{sec:couplingparticlesystems}}

\subsection{Common random numbers \label{sub:discontinuityresampling}}

Within particle filters, random variables are used to initialize, to resample and
to propagate the particles. We describe bootstrap particle filters in that light.
Initially, we sample $x_{0}^{k}\sim m_0(dx_0|\theta)$ for all $k\in 1:N$,
or equivalently, we compute $x_{0}^{k} = M(U_{0}^{k},\theta)$ where $M$ is a function and
$U_{0}^{1:N}$ random variables. The initial weights $w_0^k$ are set to $N^{-1}$.
Consider now step $t\geq 0$ of the algorithm.
In the resampling step,
a vector of ancestor variables $a_t^{1:N} \in \{1,\ldots,N\}^N$ is sampled.
The resampling step can be written $a_{t}^{1:N} \sim r(da^{1:N}| w_t^{1:N})$, for some distribution $r$. 
The propagation step consists in drawing $x^k_{t+1} \sim f(dx_{t+1}|x^{a_t^k}_t,\theta)$,
or equivalently, computing $x^k_{t+1} = F(x^{a^k_t}_{t}, U_{t+1}^k, \theta)$,
where $F$ is a function and $U_{t+1}^{1:N}$ random variables.  
The next weights are computed as $w_{t+1}^k\propto g(y_{t+1}|x_{t+1}^k,\theta)$,
then normalized to sum to one; and the algorithm proceeds. 
We refer to $U_{t}^{1:N}$ for all $t$ as the process-generating variables.
The resampling distribution $r$ is an algorithmic choice;
a standard condition for its validity is that, under $r$,
$\mathbb{P}(a_t^{k}=j)=w_t^{j}$;
various schemes satisfying this condition exist \citep[e.g.][]{douc2005comparison,murray2015parallel}.  

Consider a pair of particle filters given $\theta$ and $\ttheta$,
producing particle systems 
$(w_{t}^{k},x_{t}^{k})_{k=1}^{N}$ and $(\tw_{t}^{k},\tx_{t}^{k})_{k=1}^{N}$,
that we want to make as correlated as possible.
Assume that the state space is one-dimensional, that $u\mapsto M(u,\theta)$ and $u\mapsto M(u,\ttheta)$ are increasing and right-continuous, and that
$\mathbb{E}[M^2(U_0,\theta)]<\infty$ and $\mathbb{E}[M^2(U_0,\ttheta)]<\infty$.
Then Proposition 2.2 of \citet{glasserman1992some} states that 
the correlation between $M(U_0,\theta)$ and $M(V_0,\ttheta)$ is maximized,
among all choices of joint distributions for $(U_0,V_0)$ that have the same marginal distributions, 
by choosing $U_0=V_0$ almost surely. This justifies the use of
common random numbers for the initialization step.
Likewise, if the propagation function $F:(x_{t},U_{t+1},\theta)\mapsto x_{t+1}$ is continuous in
its first and third arguments, and
if the particle locations $x_{t}$ and $\tx_{t}$ are similar, then,
intuitively, $x_{t+1} = F(x_{t},U_{t+1},\theta)$ and $\tx_{t+1} = F(\tx_{t},U_{t+1},\ttheta)$ should be similar as well. 
The difficulty comes from the resampling step.
We can write $a_t^{1:N} = R(w_t^{1:N},U_{R,t})$, where $U_{R,t}$ 
are random variables, typically uniformly distributed. Since the resampling function 
$(w_t^{1:N},U_{R,t}) \mapsto a_t^{1:N} = R(w_t^{1:N}, U_{R,t})$ takes values in the discrete space $\{1,\ldots,N\}^N$, 
it cannot be a continuous function of its arguments.
In other words, even if we use the same random numbers for $U_{R,t}$, a small difference between the weight vectors
$w_t^{1:N}$ and $\tw_t^{1:N}$ might lead to sampling, for instance,
$a_t^k = i$ in the first system and $\tilde{a}_t^k = i+1$ in the second
system; and $x_t^i$ and $\tx_t^{i+1}$ have no reason to be
similar.  This leads to discontinuities in by-products of the particle system,
such as the likelihood estimator $\hat p^N(y_{1:T}|\theta)$ as a function of $\theta$,
for fixed common random numbers.
We thus separate the randomness in process-generating variables 
from the resampling step,  and consider resampling algorithms designed to correlate the particles in both systems.

\subsection{Coupled resampling and coupled particle filters\label{sub:resampling-and-couplings}}

We use bold fonts to 
denote vectors of objects indexed by $k\in 1:N$, for instance $(\bw_t, \bx_t) = (w_t^k, x_t^k)_{k=1}^N$ or
$\bU_{t} = U_t^{1:N}$,
and we drop the temporal notation whenever possible, for clarity. We consider the problem of jointly resampling
$(\bw,\bx)$ and $(\btw,\btx)$. 
A joint distribution  on $\{1,\ldots,N\}^{2}$ is characterized by a matrix $P$ with non-negative entries
$P^{ij}$, for $i,j\in\{ 1,\ldots,N\}$, that sum to one. The value $P^{ij}$ represents the
probability of sampling the pair $(i,j)$. We consider the set
$\mathcal{J}(\bw,\btw)$ of matrices $P$ such that 
$P\mathds{1}=\bw$ and $P^\transp\mathds{1}=\btw$, where $\mathds{1}$ denotes a
column vector of $N$ ones.
Pairs $(\ba,\bta)$
distributed according to $P\in\mathcal{J}(\bw,\btw)$ are such that 
$\mathbb{P}(a^{k}=j)=w^{j}$ and $\mathbb{P}(\ta^{k}=j)=\tw^{j}$ for all $k$ and $j$.
The choice $P=\bw\,\btw^\transp$ corresponds to an independent coupling of $\bw$ and
$\btw$. Sampling from this matrix $P$ is done by sampling $\ba$ with probabilities
$\bw$ and $\bta$ with probabilities $\btw$, independently.

Any choice of probability matrix $P\in\mathcal{J}(\bw,\btw)$ leads to a coupled resampling scheme,
and to a coupled bootstrap particle filter that proceeds as follows.
The initialization and propagation steps are performed as in the standard particle filter,
using common process-generating variables $\bU_{0:T}$ and the parameter values $\theta$ and $\ttheta$
respectively. 
At each step $t\geq 0$, the resampling step involves
computing a matrix $P_{t}$ in $\mathcal{J}(\bw,\btw)$,
possibly using all the variables generated thus far.
Then the pairs of ancestors $(\ba_{t},\bta_{t})$ are sampled from $P_{t}$.

Coupled resampling schemes can be applied in generic particle methods beyond
the boostrap filter. We illustrate this generality by coupling conditional
particle filters.  Given a trajectory $X=x_{0:T}$, referred to as the
reference trajectory, and process-generating variables $\bU_{0:T}$, the
conditional particle filter defines a distribution on the space of
trajectories, as follows.  At the initial step, we compute $x_0^k =
M(U_0^k,\theta)$ for all $k\in 1:N-1$, we set $x_0^N = x_0$, and $w_0^k =
N^{-1}$ for all $k$.  At each step $t$, we draw $a_{t}^{1:N-1} \sim
r(da^{1:N-1}|w_t^{1:N})$ from a multinomial distribution, and set $a_t^N = N$;
other resampling schemes can be implemented, as detailed in
\citet{ChopinS:2015}.  The propagation step computes $x_{t+1}^k=
F(x_t^{a_t^k},U_{t+1}^k,\theta)$ for $k\in1:N-1$ and sets $x_{t+1}^N =
x_{t+1}$.  The weighting step computes $w_{t+1}^k \propto
g(y_{t+1}|x_{t+1}^k,\theta)$, for all $k\in1:N$.  The procedure guarantees that
the reference trajectory $x_{0:T}$ is among the trajectories produced by the
algorithm. At the final step, we draw $b_T$ with probabilities $\bw_T$ and
retrieve the corresponding trajectory, denoted $X'$.  The coupled conditional
particle filter acts similarly, producing a pair of trajectories $(X',\tX')$
given a pair of reference trajectories $X=x_{0:T}$ and $\tX = \tx_{0:T}$. The
initialization and propagation steps follow the conditional particle filter for
each system, using common random numbers $\bU_{0:T}$.  For the resampling step,
we compute a probability matrix $P_t\in\mathcal{J}(\bw_t,\btw_t)$, based on the
variables generated thus far, and we sample pairs of ancestor variables
$(a_t^k,\ta_t^k)_{k=1}^{N-1}$. We then set $a_t^N = N$ and $\ta_t^N = N$.  At
the final step, we draw a pair of indices $(b_T,\tilde{b}_T)$ from $P_T$, a
probability matrix in $\mathcal{J}(\bw_T,\btw_T)$, and retrieve the
corresponding pair of trajectories. 
The coupled conditional particle filter leads to a new smoothing algorithm, described in Section
\ref{sec:methods:smoother}.

We now investigate particular choices of matrices $P\in\mathcal{J}(\bw_{t},\btw_{t})$ with
the aim of correlating a pair of particle systems. 

\subsection{Transport resampling\label{sub:transport-resampling}}

Intuitively, we want to choose $P\in\mathcal{J}(\bw,\btw)$ such that,
upon sampling ancestors from $P$, the resampled particles are as similar as possible
between the two systems.  Similarity between
locations can be encoded by a distance
$d:\mathbb{X}\times\mathbb{X}\to\mathbb{R}^{+}$, for instance the Euclidean
distance in $\mathbb{X}\subset\mathbb{R}^{d_x}$.   
The expected
distance between the resampled particles $\bx^{\ba}$ and $\btx^{\bta}$,
conditional upon $(\bw, \bx)$ and $(\btw, \btx)$, is given by $\sum_{i = 1}^N
\sum_{j = 1}^N P^{ij} d(x^{i},\tx^{j})$.  Denote by $D$
the distance matrix with entries $D^{ij} = d(x^{i},\tx^{j})$. The optimal
transport problem considers a matrix $P^\star$ that minimizes the expected distance
over all $P\in\mathcal{J}(\bw,\btw)$.
Computing $P^\star$, either exactly or approximately, is the topic of a rich
literature.  Exact algorithms compute $P^\star$ in order $N^{3}\log N$ operations,
while recent methods introduce regularized solutions $P^\varepsilon$,
where $\varepsilon\in(0,\infty)$ is such that $P^\varepsilon \to P^\star$ when $\varepsilon\to0$.
The regularized solution $P^\varepsilon$ is then approximated by an iterative algorithm,
yielding a matrix $\hat{P}$ in order $N^{2}$
operations \citep{cuturi2013sinkhorn,benamou2014iterative}. 
Computing the distance matrix $D$ and sampling from a generic
probability matrix $P$  already cost $N^{2}$ operations in general, thus the overall cost
is in $N^{2}$ operations.
We denote by $\hat{P}$ the matrix obtained by Cuturi's approximation \citep{cuturi2013sinkhorn}.

Unlike the exact solution $P^\star$ and its regularized approximation $P^\varepsilon$, an approximate solution $\hat{P}$
might not belong to $\mathcal{J}\left(\bw,\btw\right)$. 
Directly using such a $\hat{P}$ in a coupled particle filter would result in a bias, for instance
in the likelihood estimator.
However, we can easily construct a matrix $P\in\mathcal{J}(\bw,\btw)$ that is close 
to $\hat{P}$.  
Introduce $\boldsymbol u = \hat{P}\mathds{1}$ and $\tilde{\boldsymbol u} = \hat{P}^\transp \mathds{1}$,
the marginals of $\hat{P}$.
We compute a new matrix $P$ as
$P = \alpha \hat{P} + (1-\alpha) \br \btr^\transp$
for some $\alpha \in [0,1]$ and some probability vectors $\br$ and $\btr$.
The marginal constraints yield a system to solve for $\br$, $\btr$ and $\alpha$.
We obtain $\br = (\bw-\alpha \boldsymbol u)/(1-\alpha)$,
$\btr = (\btw-\alpha \btu)/(1-\alpha)$, 
and $0 \leq \alpha \leq \min_{i\in 1:N} \min(w^i/u^i, \tilde w^i/\tilde u^i)$.
To make the best use of the transport matrix $\hat{P}$, we select $\alpha$ to attain the upper bound.
Following \citet{cuturi2013sinkhorn}, we choose $\varepsilon$ as a small proportion of the median of the distance 
matrix $D$.
As a stopping criterion for the iterative algorithm yielding $\hat{P}$,
we can select $\alpha$ as a desired value close to one, and run the iterative algorithm until the value can be chosen,
i.e. until $\alpha \leq \min_{i\in 1:N} \min(w^i/u^i, \tilde w^i/\tilde u^i)$.

The computational cost of transport resampling is potentially
prohibitive,  but it is model-independent and linear in 
the dimension $d_x$ of the state space. Furthermore 
the active research area of numerical transport might 
provide faster algorithms in the future.  
Thus, for complex dynamical systems, the cost of transport resampling might still be
negligible compared to the cost of the propagation steps.

\subsection{Index-coupled resampling \label{sub:index-coupled-resampling}}

Next we consider a computationally cheaper alternative to transport resampling
termed \emph{index-coupled} resampling. This scheme was used by
\citet{ChopinS:2015}  in their theoretical analysis of the conditional particle filter. 
It has also been used by \citet{jasra2015multilevel} in the setting of multilevel Monte Carlo.
Its
computational cost is linear in $N$.  The idea of index-coupling is to
maximize the probability of sampling pairs $(a, \tilde{a})$ such that
$a = \tilde{a}$, by computing the matrix $P\in \mathcal{J}(\bw,\btw)$
with maximum entries on its diagonal. 
The scheme is intuitive at the initial step of the algorithm, assuming that $\theta$ and $\ttheta$ are
similar.
At step $t$, the same random number $U_t^k$ is used to compute $x^k_{t}$ and $\tx^k_{t}$
from their ancestors. Therefore, by sampling $a_t^k = \ta_t^k$, we select pairs that were
computed with common random numbers at the previous step, and give them common random numbers $U_{t+1}^k$
again. 
The scheme maximizes the number of consecutive steps where common random numbers are given to each pair.
We describe how to
implement the scheme, in the spirit of maximal couplings \citep{lindvall2002lectures},
before providing more intuition. 

First, for all $i\in 1:N$, $P$ has to satisfy $P^{ii}\leq \min(w^i,\tw^i)$,
otherwise one of the marginal constraints would be violated. We tentatively
write $P = \alpha \, \text{diag}(\boldsymbol \mu) + (1-\alpha)R$, where $\boldsymbol\nu =
\min(\bw,\btw)$ (element-wise), $\alpha = \sum_{i=1}^N \nu^i$, $\boldsymbol\mu = \boldsymbol\nu
/ \alpha$ and $R$ is a residual matrix with zeros on the diagonal. Matrices $P$
of this form have maximum trace among all matrices in $\mathcal{J}(\bw,\btw)$.  We
now look for $R$ such that $P\in \mathcal{J}(\bw,\btw)$ and such that
sampling from $P$ can be done linearly in $N$. From the marginal constraints,
the matrix $R$ needs to satisfy, for all $i\in1:N$, 
$\nu^{i} + (1-\alpha) \sum_{j=1}^N R^{ij} = w^i$
and $\nu^{i} + (1-\alpha)\sum_{j=1}^N R^{ji} = \tw^i$.
Among all the matrices $R$ that satisfy these constraints, the choice $R = \br \btr^\transp$,
where $\br = (\bw - \boldsymbol{\nu})/(1-\alpha)$ and $\btr = (\btw - \boldsymbol{\nu})/(1-\alpha)$,
is such that we can sample pairs of indices from $R$ by sampling from $\br$ and 
$\btr$ independently, for a linear cost in $N$. Thus we define the index-coupled matrix $P$ as
\begin{equation}
    P = \alpha \; \text{diag}(\boldsymbol \mu) + (1-\alpha) \; \br \btr^\transp.
    \label{eq:indexmatchingmatrix}
\end{equation}

Under model assumptions, using common random numbers to propagate a pair of particles 
will result in the pair of states getting closer. We can formulate
assumptions on the function  $(x,\theta)\mapsto \mathbb{E}[F(x,U,\theta)]$
as a function of both of its arguments, where the expectation is with respect to $U$.
We can assume for instance that it is Lipschitz in both arguments.
In an auto-regressive model where $F(x,U,\theta) = \theta X + U$,
the Lipschitz constant is $x$ as a function of $\theta$ and $\theta$ as a function of $x$. One can then find
conditions \citep[see e.g.][]{diaconis1999iterated} such 
that the distance between the two propagated particles will decrease down to a value proportional to the distance 
between $\theta$ and $\ttheta$, when common random numbers are used to propagate the pair.

We will see in Section \ref{sec:numerics}
that index-coupled resampling can perform essentially as well as transport resampling in real-world models. 

\subsection{Existing approaches\label{sub:existing-coupled-resampling}}

Various attempts have been made to modify particle filters so that
they produce correlated likelihood estimators. A detailed review is given in \citet{lee2008towards}.
We describe the method proposed in
\citet{pitt2002smooth}, which is
based on sorting the particles.  Consider first the univariate case, $d_x = 1$. We can
sort both particle systems in increasing order of $x^{1:N}$ and $\tx^{1:N}$
respectively, yielding $(w^{(k)}, x^{(k)})$ and $(\tw^{(k)},
\tx^{(k)})$ for $k\in1:N$, where the parenthesis indicate that the samples
are sorted. Then, we can draw $a^{1:N}$ and $\ta^{1:N}$ by
inverting the empirical cumulative distribution function associated with these
sorted samples, using common random numbers. 
We might sample $a^k$ and $\ta^k$ such that $a^k \neq \ta^k$,
but $a^k$ and $\ta^k$ will still be close and thus $x^{(a^k)}$ and $\tx^{(\ta^k)}$ will be similar,
thanks to the sorting.
The method can be extended to multivariate spaces using the Hilbert
space-filling curve as mentioned in \citet{deligiannidis2015correlated},
following \citet{gerber2015sequential}. That is, we use the pseudo-inverse
of the Hilbert curve to map the $d_x$-dimensional particles to the interval $[0,1]$, where they can be sorted
in increasing order.  
We refer to this approach as sorted
resampling, and use the implementation provided by the function
\texttt{hilbert\_sort} in \citet{cgal:eb-16a}.   The cost of sorted
resampling is of order $N\log N$.

\subsection{Numerical illustration \label{sub:likelihoodcurves}}

We first illustrate the effect of coupled resampling schemes in estimating likelihood curves 
for a multivariate hidden auto-regressive model.
The process starts as $x_0\sim \mathcal{N}(0, I_{d_x})$, where $I_{d_x}$ is the
identity matrix of dimension $d_x \times d_x$, the transition is defined
by $x_t \sim \mathcal{N}(A x_{t-1}, I_{d_x})$, where $A^{ij}$ is
$\theta^{|i-j|+1}$, as in \citet{guarniero2015iterated}.  Finally, the
measurement distribution is defined by $y_t \sim \mathcal{N}(x_{t}, I_{d_x})$.

\begin{figure}
    \centering
    \includegraphics[width=1.0\textwidth]{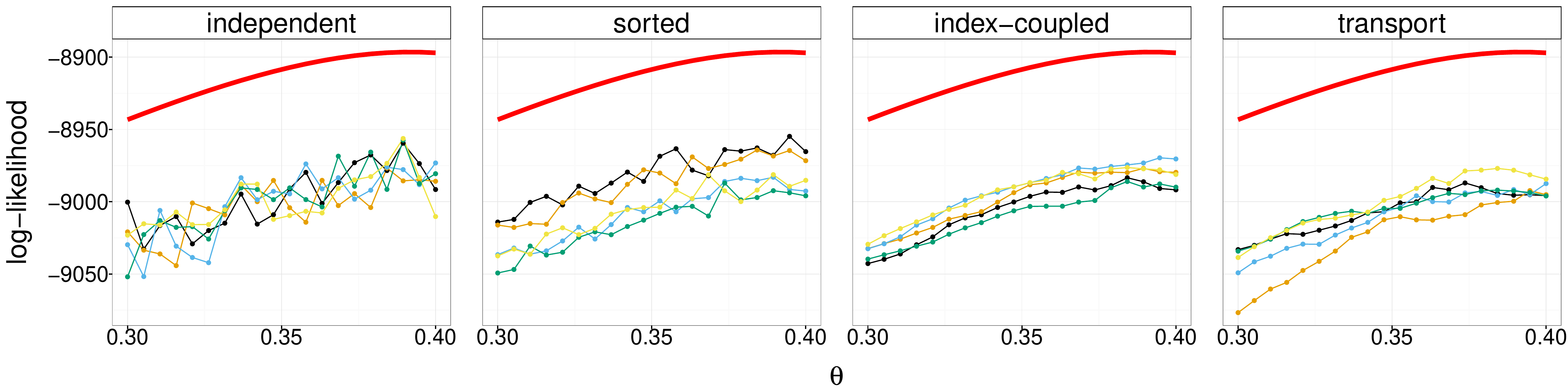}
    \caption{Log-likelihood estimators for various values of $\theta$ in a hidden auto-regressive model with $d_x = 5$, $T = 1,000$
        and $N=128$.
        From left to right: independent estimators, common random numbers with sorted resampling, with index-coupled
    resampling and with transport resampling. The thick red line
indicates the exact log-likelihood computed using Kalman filters.
    \label{fig:profiles:d5:coupled} }
\end{figure}

We generate $T = 1,000$ observations, with parameter $\theta = 0.4$ and with $d_x =5$. 
We consider a sequence of parameter values $\theta_1,\ldots,\theta_L$.  
We run a standard particle filter given $\theta_1$, and then for each $\ell\in2:L$,
we run a particle filter given $\theta_{\ell}$ conditionally upon the variables
generated by the previous particle filter given $\theta_{\ell-1}$; more details are given in Appendix \ref{sec:appendix:jointconditional}. We use $N=128$ particles,
and try various coupled resampling schemes. The transport resampling scheme uses $\varepsilon = 0.01\times\text{median}(D)$ and $\alpha = 0.99$.
The estimated log-likelihoods are shown in Figure \ref{fig:profiles:d5:coupled}
for five independent runs, and compared to the exact log-likelihood obtained by Kalman filters.

All the filters under-estimate the log-likelihood by a significant amount,
indicating that more particles would be necessary to obtain precise 
estimates for any given $\theta$.  However, we see that the shape of the log-likelihood curve is still
approximately recovered when using common random numbers and certain coupled
resampling schemes, indicating that we can compare log-likelihood values for different parameters,
even with comparably small numbers of particles.

\section{Methodological developments using coupled resampling \label{sec:methods}}

In this section, we develop the use of coupled resampling schemes for score estimation, for sampling algorithms 
targeting the posterior distribution of the parameters, and for latent state estimation. For the latter,
a new smoother is proposed, conditions for its validity are given and experiments in a toy example 
are presented, showing its potential advantages compared to standard smoothing methods. 

\subsection{Finite difference estimators of the score \label{sub:correlation-likelihood-finitedifference}}

Consider the estimation of the log-likelihood gradient, also called the score and denoted by
$\nabla_{\theta}\log p(y_{1:T}|\theta)$. We focus on univariate parameters for simplicity.
A finite difference estimator \citep{asmussen2007stochastic} of the score at the value $\theta$ is given by 
$ D^N_{h}(\theta)=(\log \hat p^N (y_{1:T}|\theta + h ) - \log \hat p^N (y_{1:T}|\theta - h ))/(2h)$,
where $h>0$ is a perturbation parameter.
If $h$ is small, the variances of the two log-likelihood estimators 
can be assumed approximately equal, and thus $\mathbb{V}(D^N_h(\theta))$
is approximately equal to 
$(2h)^{-1} \times \mathbb{V}(\log \hat p^N(y_{1:T}|\theta))\times (1 - \rho_h^N(\theta))$,
where $\rho_h^N(\theta)$ denotes the correlation between $\log \hat p^N(y_{1:T}|\theta+h)$
and $\log \hat p^N(y_{1:T}|\theta-h)$. Thus, compared to using independent estimators
with the same variance, the variance of the finite difference estimator 
can be divided by $1/(1 - \rho_h^N(\theta))$.
We refer to this number as the gain.  It corresponds to how many times more particles should be used in order to attain the same
accuracy using independent particle filters. The bias
of the gradient estimator is unchanged by the use of coupled resampling schemes,  since they 
do not change the marginal distributions of each particle filter. 

We run coupled particle filters at $\theta-h$ and $\theta+h$ for $\theta = 0.3$
and various $h$, and different coupled resampling schemes, over $1,000$
independent experiments in the hidden auto-regressive model of Section
\ref{sub:likelihoodcurves}. The correlations between the log-likelihood
estimators are shown in Figure \ref{table:finitediff}, as well as the gains. We
see that the variance can be divided by approximately $500$ for small values of
$h$, but only by approximately $10$ for larger values of $h$.  Index-coupled
resampling appears to perform better than transport resampling for small values
of $h$, perhaps due to the approximation introduced in the regularized
transport problem; a more detailed investigation of the transport
regularization is given in the next section.  Here the tuning parameters of
transport resampling were set to $\varepsilon = 5\% \times \text{median}(D)$
and $\alpha = 99\%$.
\begin{figure}
\caption{\label{table:finitediff} Correlation (rounded to two decimals) and gain (variance reduction) factors,
as a function of the perturbation parameter $h$, in a five-dimensional hidden auto-regressive model with $T = 1,000$,
$N=128$ and $R=1,000$ experiments.} 
\centering
\fbox{
\begin{tabular}{llrr}
  \hline
  h & method & correlation & gain \\ 
  \hline
  0.001 & sorted   & 0.90 & 9.6 \\ 
  0.001 & index-coupled   & 1.00 & 527.5 \\ 
  0.001 & transport   & 1.00 & 321.7 \\ 
  \hline
  0.025 & sorted   & 0.88 & 8.3 \\ 
  0.025 & index-coupled   & 0.96 & 25.2 \\ 
  0.025 & transport   & 0.97 & 33.0 \\ 
  \hline
  0.05 & sorted   & 0.84 & 6.2 \\ 
  0.05 & index-coupled   & 0.91 & 11.1 \\ 
  0.05 & transport   & 0.92 & 12.7 \\ 
   \hline
\end{tabular}}
\end{figure}

\subsection{Correlated particle marginal Metropolis--Hastings \label{sub:correlation-likelihood-pmmh}}

We now turn to the particle marginal MH algorithm (PMMH)
\citep{andrieu:doucet:holenstein:2010}, for parameter inference
in state space models.  Denoting the prior parameter distribution by $p(d\theta)$, the
algorithm generates a Markov chain $(\theta^{(i)})_{i\geq1}$ targeting the
posterior distribution $p(d\theta| y_{1:T})\propto
p(d\theta)p(y_{1:T}|\theta)$.  At iteration $i\geq1$, a parameter $\ttheta$ is
proposed from a Markov kernel $q(d\theta|\theta^{(i-1)})$, and accepted as the
next state of the chain $\theta^{(i)}$ with probability 
\begin{equation}
\min\left(1,
\frac{\hat p^N (y_{1:T}|\ttheta )}{ \hat p^N (y_{1:T}|\theta^{(i-1)})} 
\frac{p(\ttheta)}{p(\theta^{(i-1)})}
    \frac{q(\theta^{(i-1)}|\ttheta)}{q(\ttheta|\theta^{(i-1)})}\right),
    \label{eq:MHratio}
\end{equation}
where $\hat p^N (y_{1:T}|\ttheta )$ is the likelihood estimator produced 
by a filter given $\ttheta$.  In order for this
algorithm to mimic the ideal underlying MH algorithm, the ratio of likelihood estimators must be an
accurate approximation of the exact ratio of likelihoods \citep{andrieu2015convergence}.
The benefit of correlated likelihood estimators within pseudo-marginal
algorithms is the topic of recent works
\citep{deligiannidis2015correlated,dahlin2015accelerating}, following 
\citet{leecommentonpmcmc} in the discussion of
\citet{andrieu:doucet:holenstein:2010}.

A correlated particle marginal MH algorithm works on the joint space of the parameter $\theta$,
the process-generating variables $\bU_t$ for all $t\in 0:T$,
and the ancestor variables $\ba_t$, for all $t\in 0:T-1$.
Denote by $\varphi$ the distribution of $\bU_{0:T}$, assumed to be standard multivariate 
normal for simplicity, and let $\phi$ be a  Markov kernel leaving $\varphi$ invariant.
Consider any iteration $i\geq 1$ of the algorithm; the current state of the
Markov chain contains $\theta^{(i-1)} = \theta$, $\bU_{0:T}^{(i-1)} =
\bU_{0:T}$, $\ba_{0:T-1}^{(i-1)} = \ba_{0:T-1}$, and the associated
likelihood estimator is $\hat p^N (y_{1:T}|\theta)$. The particles
$(\bw_t, \bx_t)$, for all $t\in 0:T$, are deterministic given $\theta$, $\bU_{0:T}$ and
$\ba_{0:T-1}$.  The algorithm proceeds in the following way.
\begin{enumerate}
    \item A parameter value is proposed: $\ttheta \sim q(d\ttheta|\theta)$, as
        well as new process-generating variables: $\btU_{0:T} \sim \phi(d\btU_{0:T}|\bU_{0:T})$. 
    \item A particle filter is run given $\ttheta$, using $\btU_{0:T}$ and 
        conditionally upon the current particle filter. That is, at each resampling step,
        a matrix $P_t$ is computed using $(\bw_t, \bx_t)$ and $(\btw_t, \btx_t)$,
        and the new ancestors $\bta_t$ are sampled conditional upon $\ba_{t}$.
        The algorithm produces ancestor variables $\bta_{0:T-1}$ and a likelihood estimator $\hat p^N(y_{1:T}|\ttheta)$.
    \item With the probability given by Eq.~\eqref{eq:MHratio}, the chain moves
        to the state with parameter $\ttheta$, variables  $\btU_{0:T}$,
        ancestors $\bta_{0:T-1}$ and likelihood estimator $\hat p^N
        (y_{1:T}|\ttheta)$. Otherwise, the current state of the chain is
        unchanged.
\end{enumerate}

Appendix \ref{sec:appendix:jointconditional} contains further details on the
conditional sampling of a particle filter as required by step (b) above.
Appendix \ref{sec:appendix:validitity:PMMH} contains conditions on the coupled
resampling scheme for the algorithm to be exact, which are verified for sorted
and index-coupled schemes, as well as for a slightly modified transport scheme.

In the hidden auto-regressive model, we specify a standard normal
prior on $\theta$. The distribution $\varphi$ of the
process-generating variables is a multivariate normal
distribution, and we choose the kernel $\phi$ to be auto-regressive: $\tU =
\rho U + \sqrt{1-\rho^2} \mathcal{N}(0,I)$, with $\rho = 0.999$. 
We use a normal random walk with a standard deviation of $0.01$ for the proposal on $\theta$.  
We run each algorithm $100$ times for $M=20,000$ iterations, starting the chain
from a uniform variable in $[0.37,0.41]$, taken to be in the bulk of the posterior distribution.
Figure \ref{table:cpmmh} shows the obtained average acceptance rates and effective sample sizes,
defined as $M$ divided by the integrated autocorrelation time and obtained with the function \texttt{effectiveSize} of the \texttt{coda} package. With 
index-coupled resampling, the effective sample size can reach acceptable levels with fewer particles,
compared to standard PMMH or compared to sorted resampling (as used by \citet{deligiannidis2015correlated}).

Transport resampling is considerably more expensive for a given choice of $N$. For $N=128$, we show the acceptance
rates and effective sample sizes obtained over $20$ independent experiments with various levels of approximation to
the optimal transport problem. When $\epsilon$ is close to zero and $\alpha$ is close to one,
we can achieve greater effective sample sizes with transport resampling than with the other schemes, for a fixed $N$.

\begin{figure}
  \caption{Average acceptance rates (AR) and effective sample sizes (ESS) of standard and correlated PMMH, obtained for various numbers of particles (left), and for various regularization parameters $\varepsilon$
    and stopping criteria $\alpha$
    for the transport resampling scheme with $N=128$ (right), in a hidden auto-regressive model
    with $d_x = 5$, $T = 1,000$ and $M=20,000$ iterations.  \label{table:cpmmh}} 
  \begin{minipage}{.5\linewidth}
    \fbox{%
      \begin{tabular}{llrr}
        \hline
        N & method & AR (\%) & ESS \\ 
        \hline
        64 & indep. & 0.06 (0.02) & 29 (95) \\ 
  64 & sorted   & 0.12 (0.06) & 25 (50) \\ 
  64 & index-c. & 0.87 (0.23) & 41 (16) \\ 
        \hline
  128 & indep. & 0.06 (0.02) & 23 (41) \\ 
  128 & sorted   & 0.18 (0.10) & 26 (28) \\ 
  128 & index-c. & 2.00 (0.41) & 98 (30) \\ 
        \hline
  256 & indep. & 0.07 (0.03) & 16 (24) \\ 
  256 & sorted   & 0.42 (0.23) & 39 (25) \\ 
  256 & index-c. & 4.67 (0.49) & 244 (58) \\ 
        \hline
      \end{tabular}}
      \caption*{Standard PMMH (indep.), 
      and with sorted and index-coupled resampling (index-c.),
  over $100$ experiments.}
  \end{minipage}%
  \hspace*{0.3cm}
  \begin{minipage}{.5\linewidth}
    \fbox{%
      \begin{tabular}{llrr}
        \hline
        $\varepsilon$ & $\alpha$ & AR (\%) & ESS\\ 
        \hline
        0.10 & 0.95 & 1.06 (0.29) & 83 (34) \\ 
  0.10 & 0.99 & 1.12 (0.28) & 85 (23) \\ 
        \hline
  0.05 & 0.95 & 2.13 (0.47) & 133 (44) \\ 
  0.05 & 0.99 & 3.86 (0.49) & 220 (54) \\
        \hline
      \end{tabular}}
      \caption*{With $N = 128$ and transport \\ resampling,
      over $20$ experiments.}
  \end{minipage} 
\end{figure}

\subsection{A new smoothing method\label{sec:methods:smoother}}

Next, we turn to an application of \emph{coupled conditional particle filters} for the task of smoothing.
The parameter $\theta$ is fixed and removed from the notation.
Denote by $h$ a generic test function on $\mathbb{X}^{T+1}$, of
which we want to compute the expectation with respect to the smoothing distribution $\pi(dx_{0:T})=p(dx_{0:T}|y_{1:T})$; we write 
$\pi(h)$ for $\int_{\mathbb{X}^{T+1}} h(x_{0:T}) \pi(dx_{0:T})$.

\subsubsection{Algorithm}

We build upon the debiasing technique of \citet{glynn2014exact}, which follows
a series of unbiased estimation techniques
\citep[see][and references therein]{Rhee:Glynn:2012,vihola2015unbiased}.
The Rhee--Glynn estimator introduced in 
\citet{glynn2014exact} uses the kernel of a Markov chain with invariant
distribution $\pi$, in order to produce unbiased estimators of $\pi(h)$.
In the setting of smoothing, the conditional particle filter defines a
Markov kernel leaving the smoothing distribution invariant 
\cite{andrieu:doucet:holenstein:2010}; extensions include 
backward sampling \citep{whiteleycommentonpmcmc} and ancestor
sampling \citep{LindstenJS:2014}.
The conditional particle filter kernel has
been extensively studied in
\citet{ChopinS:2015,andrieuvihola2013uniform,LindstenDM:2015}.
The use of conditional particle filters within the Rhee--Glynn estimator 
naturally leads to the problem of coupling two conditional
particle filters. 
%

The Rhee--Glynn construction adapted to our context goes as follows.
We draw two trajectories $X^{(0)}$ and $\tX^{(0)}$ from two independent particle filters,
which we denote by 
$X^{(0)} \sim \text{PF}(\bU^{(0)})$ and  $\tX^{(0)} \sim \text{PF}(\btU^{(0)})$, with $\bU^{(0)}\sim \varphi$ and $\btU^{(0)}\sim \varphi$
denoting the process-generating variables.
Note that even for fixed process-generating variables the sampled trajectories are random, due to the randomness of the resampling steps.
We apply one step of the conditional particle filter to the first trajectory: we sample process-generating variables $\bU^{(1)}\sim \varphi$ and 
write $X^{(1)}\sim \text{CPF}(X^{(0)},\bU^{(1)})$. Then, for all $n \geq 2$,
we apply the coupled conditional particle filter (CCPF)
to the pair of trajectories, which is written 
$(X^{(n)},\tilde{X}^{(n-1)}) \sim \text{CCPF}(X^{(n-1)},\tilde{X}^{(n-2)}, \bU^{(n)})$,
where $\bU^{(n)}\sim \varphi$. The resulting chains are such that
\begin{enumerate}
    \item  marginally, $(X^{(n)})_{n\geq 0}$ and $(\tX^{(n)})_{n\geq 0}$ have the same distributions as if they were generated
        by conditional particle filters, and thus converge under mild assumptions to the smoothing distribution;
    \item for each $n\geq 0$, $X^{(n)}$ has the same distribution as $\tX^{(n)}$, since the variables $(\bU^{(n)})_{n\geq 0}$ are independent
        and identically distributed;
    \item under mild conditions stated below, at each iteration $n\geq 2$, there is a non-zero
        probability that $X^{(n)} = \tX^{(n-1)}$. 
        We refer to this event as a meeting, and introduce the meeting time $\tau$, defined as 
        $\tau = \inf\{n\geq 2: X^{(n)} = \tX^{(n-1)}\}$.
\end{enumerate}
We then define the Rhee--Glynn smoothing estimator as 
\begin{equation}
    H= h(X^{(0)}) + \sum_{n=1}^{\tau} h(X^{(n)}) - h(\tX^{(n-1)}). \label{eq:RGestimator}
\end{equation}
This is an unbiased estimator of $\pi(h)$ with finite variance and finite computational cost, under conditions given below.
The full procedure is described in Algorithm \ref{alg:rheeglynnsmoother}. To estimate the smoothing functional $\pi(h)$,
one can sample $R$ estimators, $H^{(r)}$ for $r\in 1:R$, and take the empirical average $\bar{H}=R^{-1}\sum_{r=1}^R H^{(r)}$; it is unbiased and converges to $\pi(h)$
at the standard Monte Carlo rate as $R\to\infty$. 

\begin{algorithm}
    \begin{itemize}
        \item \textsf{Draw $\bU^{(0)}\sim \varphi$ and $X^{(0)}\sim\text{PF}(\bU^{(0)})$, draw $\bU^{(1)}\sim \varphi$,  and draw $X^{(1)} \sim \text{CPF}(X^{(0)}, \bU^{(1)})$.}
        \item \textsf{Draw $\btU^{(0)}\sim \varphi$ and $\tX^{(0)}\sim\text{PF}(\btU^{(0)})$.}
        \item \textsf{Compute $\Delta^{(0)} =  h(X^{(0)})$ and $\Delta^{(1)} = h(X^{(1)}) - h(\tilde{X}^{(0)})$, set $H = \Delta^{(0)} + \Delta^{(1)}$.}
        \item \textsf{For $n = 2, 3,\ldots $,}
            \begin{itemize}
                \item \textsf{Draw $\bU^{(n)}\sim \varphi$ and $(X^{(n)},\tilde{X}^{(n-1)}) \sim \text{CCPF}(X^{(n-1)},\tilde{X}^{(n-2)}, \bU^{(n)})$.}
                \item \textsf{Compute $\Delta^{(n)} = h(X^{(n)}) - h(\tX^{(n-1)})$, set $H \leftarrow H + \Delta^{(n)}$.}
                \item \textsf{If $X^{(n)} = \tX^{(n-1)}$, then $n$ is the meeting time $\tau$: exit the loop.}
            \end{itemize}
        \item \textsf{Return $H$.}
    \end{itemize}
\protect\caption{Rhee--Glynn smoothing estimator. \label{alg:rheeglynnsmoother}}
\end{algorithm}

Popular smoothing techniques include the fixed-lag smoother and the forward
filtering backward smoother \citep[see][for recent reviews]{doucet2011tutorial,LindstenS:2013,kantas2015particle}.
The Rhee--Glynn smoothing estimator sets itself apart in the following
way, due to its form as an average of independent unbiased estimators. 
\begin{enumerate}
    \item Complete parallelization of the computation of the terms $H^{(r)}$ is possible.
        On the contrary, particle-based methods are not entirely parallelizable
        due to the resampling step \citep{murray2015parallel,lee2015forest}. 
    \item 
        Error estimators can be constructed based on the central limit theorem,
        allowing for an empirical assessment of the performance of the estimator.
        Error estimators for particle smoothers
        have not yet been proposed, although see \citet{lee2015variance}.
\end{enumerate}

\subsubsection{Theoretical properties\label{sec:newsmoother:theory}} 

We give three sufficient conditions for the validity of Rhee--Glynn smoothing estimators.
\begin{assumption}
\label{assumption:upperbound}  
    The measurement density of the model is bounded from above:
    \[
        \exists \bar{g} < \infty, \quad \forall y\in \mathbb{Y}, \quad \forall x\in\mathbb{X},  \quad g(y | x, \theta) \leq \bar{g}.  
    \]
\end{assumption}
That bound limits the influence of the reference trajectory in the conditional particle filter.
\begin{assumption}
    \label{assumption:couplingmatrix}
    The resampling probability matrix $P$, constructed from the weight vectors $\bw$ and $\btw$, is such that 
    \[
        \forall i\in \{1,\ldots,N\}, \quad P^{ii} \geq w^i \; \tw^i.
    \]
    Furthermore, if $\bw = \btw$, then $P$ is a diagonal matrix with entries given by $\bw$.
\end{assumption}
One can check that the condition holds for independent and index-coupled resampling schemes. 
The second part of Assumption~\ref{assumption:couplingmatrix}
ensures that if two reference trajectories are equal, an application of the coupled conditional particle filter
returns two identical trajectories.

\begin{assumption}
\label{assumption:mixing}  
Let $(X^{(n)})_{n \geq 0}$ be a Markov chain generated by the conditional particle filter. 
The test function $h$ is such that 
\[\mathbb{E}\left[h(X^{(n)})\right] \xrightarrow[n\to \infty]{} \pi(h).\]
Furthermore, there exists $\delta > 0$, $n_0 < \infty$ and $C<\infty$ such that 
\begin{align*}
    \forall n\geq n_{0},\quad & \mathbb{E}\left[h(X^{(n)})^{2+\delta}\right]\leq C.
\end{align*}
\end{assumption}
This assumption relates to the validity of the conditional particle filter to estimate $\pi(h)$,
addressed under general assumptions in
\citet{ChopinS:2015,andrieuvihola2013uniform,LindstenDM:2015}. Up to the term
$\delta>0$ which can be arbitrarily small, the assumption is a requirement if we want to estimate $\pi(h)$
using conditional particle filters while ensuring a finite variance.

Our main result states that the proposed estimator is unbiased and has a finite variance.
Similar results can be found in Theorem 1 in \citet{rhee:phd}, Theorem 2.1 in \citet{McLeish:2012},
Theorem 7 in \citet{vihola2015unbiased} and in \citet{glynn2014exact}.

\begin{thm}
    Under Assumptions \ref{assumption:upperbound}-\ref{assumption:couplingmatrix}-\ref{assumption:mixing},
    the Rhee--Glynn smoothing estimator $H$, given in Eq.~\eqref{eq:RGestimator},
    is an unbiased estimator of $\pi(h)$ with 
    \begin{align*}
        \mathbb{E}[H^2] = 
        \sum_{n = 0}^{\infty} \mathbb{E}\left[(\Delta^{(n)})^2\right] + 2\sum_{n = 0}^{\infty} \sum_{\ell = n + 1}^{\infty} \mathbb{E}\left[\Delta^{(n)} \Delta^{(\ell)} \right] < \infty,
    \end{align*}
    where $\Delta^{(0)} = h(X^{(0)})$ and for $n\geq 1$, $\Delta^{(n)} = h(X^{(n)}) - h(\tX^{(n-1)})$.
    \label{thm:finitevariance}
\end{thm}

The proof is given in Appendix~\ref{sec:appendix:validitity:RG}.  The theorem uses
univariate notation for $H$ and $\Delta_n$, but the Rhee--Glynn
smoother can be applied to estimate multivariate smoothing functionals,
for which the theorem can be interpreted component-wise.

\subsubsection{Practical considerations \label{sec:newsmoother:practical}} 

For a fixed computational budget, the only tuning parameter is the number of particles $N$, 
which implicitly sets the number of independent estimators $R$ that can be obtained within the budget.
The computational cost of
producing an unbiased estimator $H$ is of order $NT\times
\mathbb{E}[\tau]$, and the expectation of $\tau$ is seen empirically to
decrease with $N$, so that the choice of $N$ is not obvious; in practice we recommend choosing a value of $N$ large
enough so that the meeting time occurs within a few steps, but other considerations
such as memory cost could be taken into account.  The memory cost for each estimator is of
order $T + N\log N$ in average \citep{jacob2015path}.  This memory
cost holds also when using ancestor sampling \citep{LindstenJS:2014}, whereas backward sampling
\citep{whiteleycommentonpmcmc} results in a memory cost of $N T$. 
As in \citet{glynn2014exact}, we can appeal to
\citet{glynn1992asymptotic} to obtain a central limit theorem parameterized by
the computational budget instead of the number of samples.  

The performance of the proposed estimator is tied to the meeting time. As in \citet{ChopinS:2015}, the
coupling inequality \citep{lindvall2002lectures} can be used to relate  the
meeting time with the mixing of the underlying conditional particle filter
kernel. Thus, the proposed estimator is expected to work in the same situations
where the conditional particle filter works.  It can be seen as a framework to
parallelize conditional particle filters and to obtain reliable confidence intervals.
Furthermore, any improvement in the conditional particle filter directly translates into a 
more efficient Rhee--Glynn estimator. 

The variance of the proposed estimator can first be reduced by a Rao--Blackwellization argument.  
In the $n$-th term of the sum in Eq.~\eqref{eq:RGestimator}, 
the random variable $h(X^{(n)})$ is obtained by applying the test function $h$ to 
a trajectory drawn among $N$ trajectories, say $x_{0:T}^{1:N}$, with probabilities $w_T^{1:N}$.
Thus the random variable $\sum_{k=1}^N w_T^{k}h(x_{0:T}^{k})$ is a conditional expectation
of $h(X^{(n)})$ given $x_{0:T}^{1:N}$ and $w_T^{1:N}$, which
has the same expectation as $h(X^{(n)})$. Any term $h(X^{(n)})$ or
$h(\tX^{(n)})$ in $H$ can be replaced by similar conditional expectations.
This enables the use of all the particles generated by the conditional particle filters.
A further variance reduction technique is discussed in Appendix \ref{sec:appendix:furtherVR}.

\subsubsection{A hidden auto-regressive model with an unlikely observation \label{sec:methods:smoother:numerics}}

We consider the first example of \citet{ruiz2016particle}. The latent process
is defined as $x_{0}\sim\mathcal{N}\left(0,\tau_{0}^{2}\right)$ and $x_{t}=\eta
x_{t-1}+\mathcal{N}\left(0,\tau^{2}\right)$; we take $\tau_{0}=0.1$, $\eta=0.9$
and $\tau=0.1$ and consider $T=10$ time steps. The process is observed only at
time $T$, where $y_{T}=1$ and we assume
$y_{T}\sim\mathcal{N}\left(x_{T},\sigma^{2}\right)$, with $\sigma=0.1$. The
observation $y_{T}$ is unlikely under the latent process distribution.
Therefore the filtering distributions and the smoothing distributions have
little overlap, particular for times $t$ close to $T$. 

We consider the problem of estimating the smoothing means, and run $R=10,000$ independent Rhee--Glynn estimators,
with various numbers of particles, with ancestor sampling \citep{LindstenJS:2014} and without variance reduction. For comparison,
we also run a bootstrap particle filter $R$ times, with larger numbers
of particles. This compensates for the fact that the Rhee--Glynn estimator
requires a certain number of iterations, each involving a coupled
particle filter. The average meeting times  for each value of $N$ are:
$10.6$ $(25.1)$  for $N = 128$, 
$8.9$ $(17.0)$   for $N = 256$,
$7.3$ $(10.8)$ for $N = 512$,
$6.1$ $(7.3)$ for $N = 1024$.

\begin{figure}
\caption{Confidence intervals on the smoothing means, obtained with $R=10,000$
Rhee--Glynn smoothers (left), and bootstrap 
particle filters (right). The true smoothing means are shown using black dots.
(Note that the estimators for different times are dependent since they are obtained from the same trajectories.)
\label{fig:unlikelyobs:confidenceintervals}}
  \centering
\begin{minipage}{.45\textwidth}
    \centering
  \includegraphics[width=0.95\textwidth]{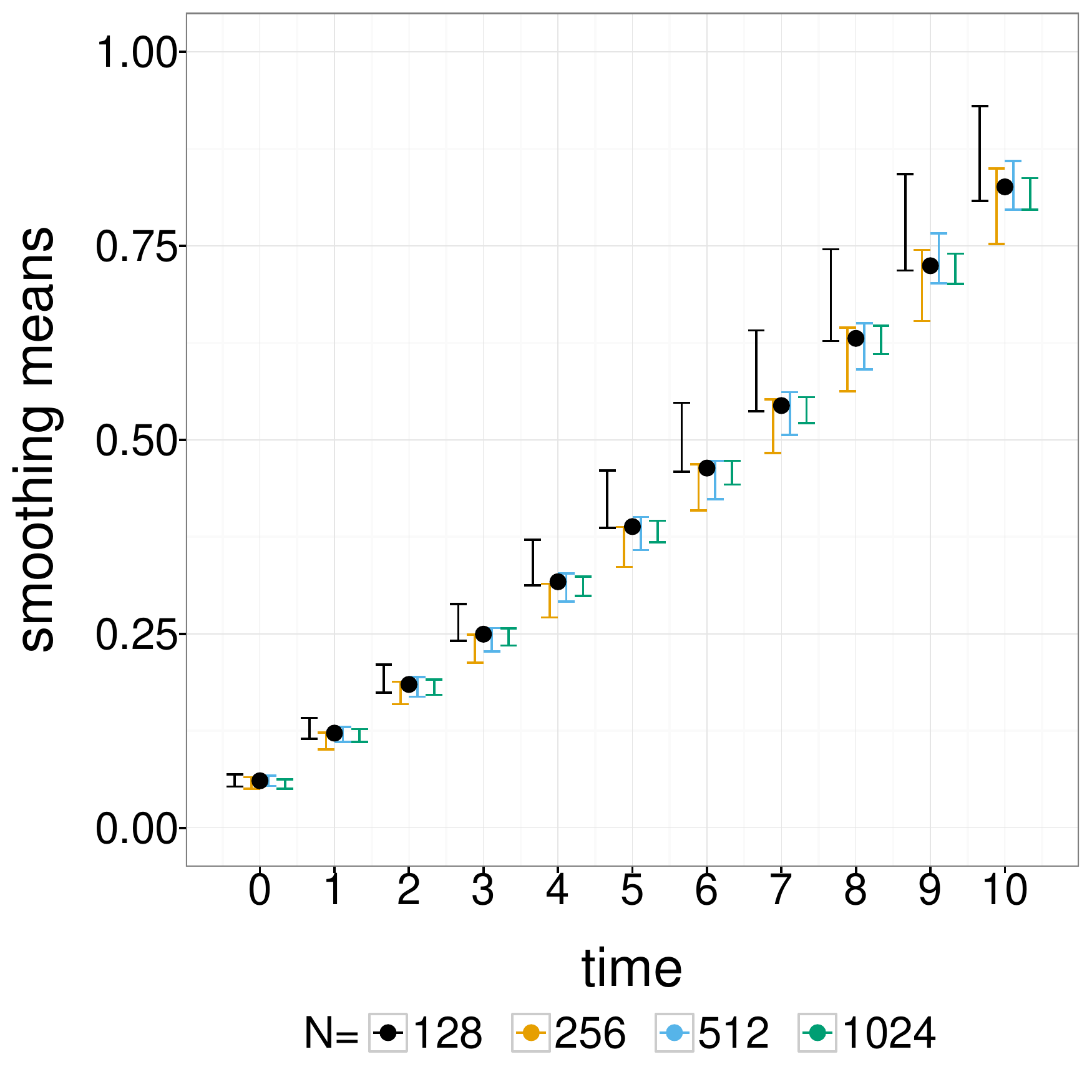}
  \caption*{Rhee--Glynn estimators.}
  \end{minipage}
  \hspace*{0.2cm}
\begin{minipage}{.45\textwidth}
  \includegraphics[width=0.95\textwidth]{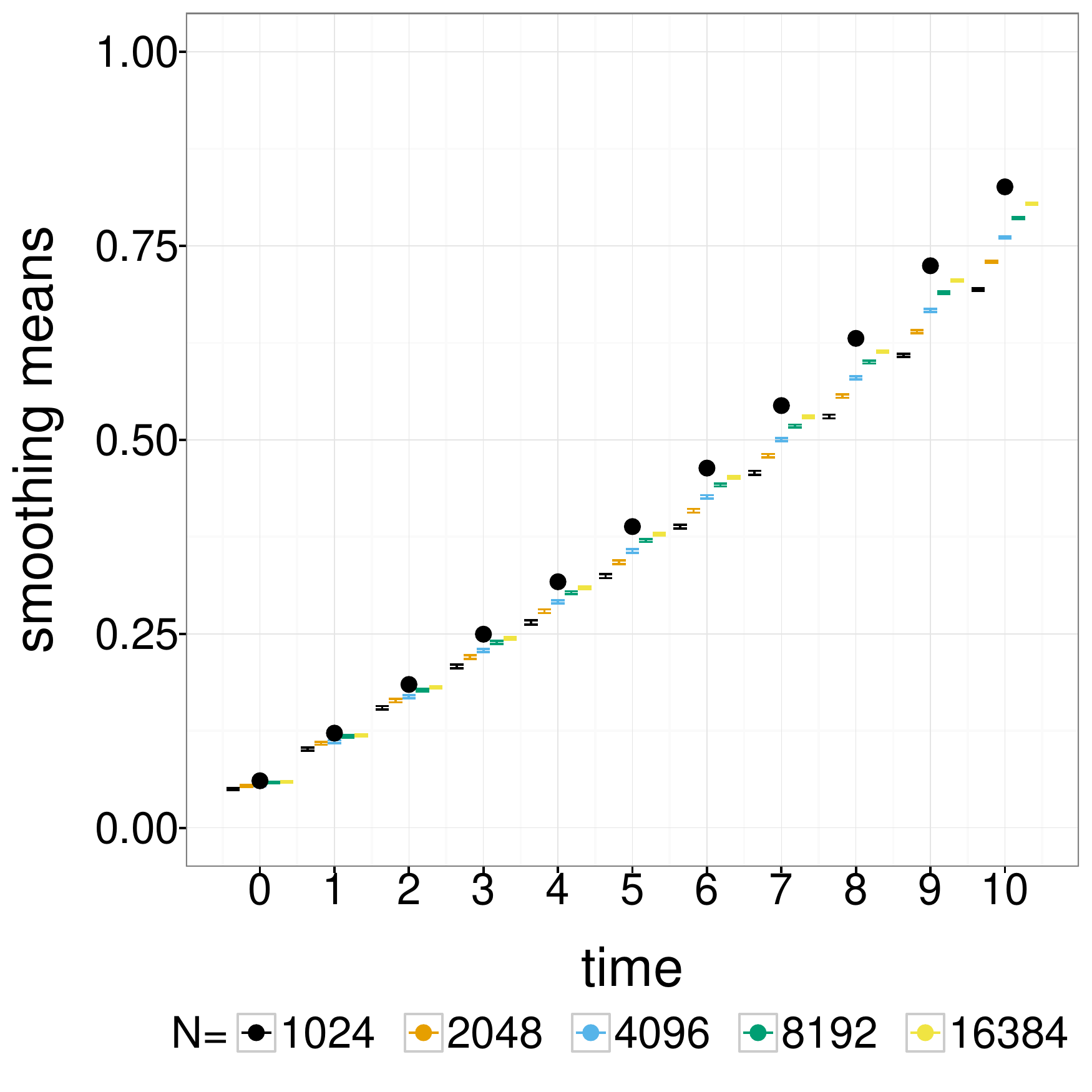}
  \caption*{Estimators obtained by particle filters.}
  \end{minipage}
\end{figure}

%
For each method, we compute a confidence interval as $[\hat{x}_{t}-2\hat{\sigma}_{t}/\sqrt{R},\hat{x}_{t}+2\hat{\sigma}_{t}/\sqrt{R}]$
at each time $t$, where $\hat{x}_{t}$ is the mean of the $R$ estimators
and $\hat{\sigma}_{t}$ is the standard deviation. The results are
shown in Figure \ref{fig:unlikelyobs:confidenceintervals}. The exact
smoothing means are obtained analytically and shown by black dots.
The Rhee--Glynn estimators lead to reliable confidence intervals. Increasing $N$ reduces
the width of the interval and the average meeting time. On the other
hand, standard particle smoothers with larger numbers of
particles still yield unreliable confidence
intervals. The poor performance
of standard particle smoothers 
is to be expected in the setting of highly-informative observations
\citep{ruiz2016particle,del2015sequential}. 

\section{Numerical experiments in a prey-predator model\label{sec:numerics}}

We investigate the performance of the correlated particle marginal Metropolis--Hastings algorithm and
of the Rhee--Glynn smoother for a nonlinear non-Gaussian model.
We consider the Plankton--Zooplankton model of \cite{jones2010bayesian}, which is an example
of an implicit model: the transition density is intractable \citep{breto2009time,jacob2015sequential}.
The hidden state $x_t = (p_t, z_t)$ represents the population size of phytoplankton and zooplankton, and the transition from time $t$ to $t+1$
is given by a Lotka--Volterra equation,
\[\frac{dp_t}{dt} = \alpha p_t - c p_t z_t , \quad \text{and}\quad \frac{dz_t}{dt} = e c p_t z_t -m_l z_t -m_q z_t^2,\]
where the stochastic daily growth rate $\alpha$ is drawn from $\mathcal{N}(\mu_\alpha,\sigma_\alpha^2)$ at every integer time~$t$.
The propagation of each particle involves solving numerically the above equation, here using a Runge-Kutta method in  
the \texttt{odeint} library \citep{ahnert2011odeint}. 
The initial distribution is given by
$\log p_0  \sim \mathcal{N}(\log 2 , 1)$ and $\log z_0  \sim \mathcal{N}(\log 2, 1)$.
The parameters $c$ and $e$ represent the clearance rate of the prey and the
growth efficiency of the predator. Both $m_l$ and $m_q$ parameterize the  
mortality rate of the predator.  
The observations $y_t$ are noisy measurements of the phytoplankton $p_t$, $\log y_t \sim \mathcal{N}(\log
p_t, 0.2^2)$; $z_t$ is not observed.
We generate $T = 365$ observations using $\mu_\alpha = 0.7, \sigma_\alpha = 0.5$, 
$c = 0.25$, $e = 0.3$, $m_l = 0.1$, $m_q = 0.1$. 

\subsection{Correlated particle marginal Metropolis--Hastings \label{sub:pz:correlation-likelihood-pmmh}}

\begin{figure}
  \caption{Density plots obtained with standard PMMH with $N=1,024$ (left), with $N=128$ (middle left),
      and with correlated PMMH with $N=128$ and index-coupled resampling (middle right) and transport resampling (right), for the parameter $\mu_\alpha$
      of the phytoplankton--zooplankton model with $T=365$ observations. The results from $10$ independent
      experiments with $M=100,000$ iterations are overlaid. \label{figure:pz:densityplots}} 
  \centering
  \begin{minipage}{.24\linewidth}
      \centering
      \includegraphics[width=1.0\textwidth]{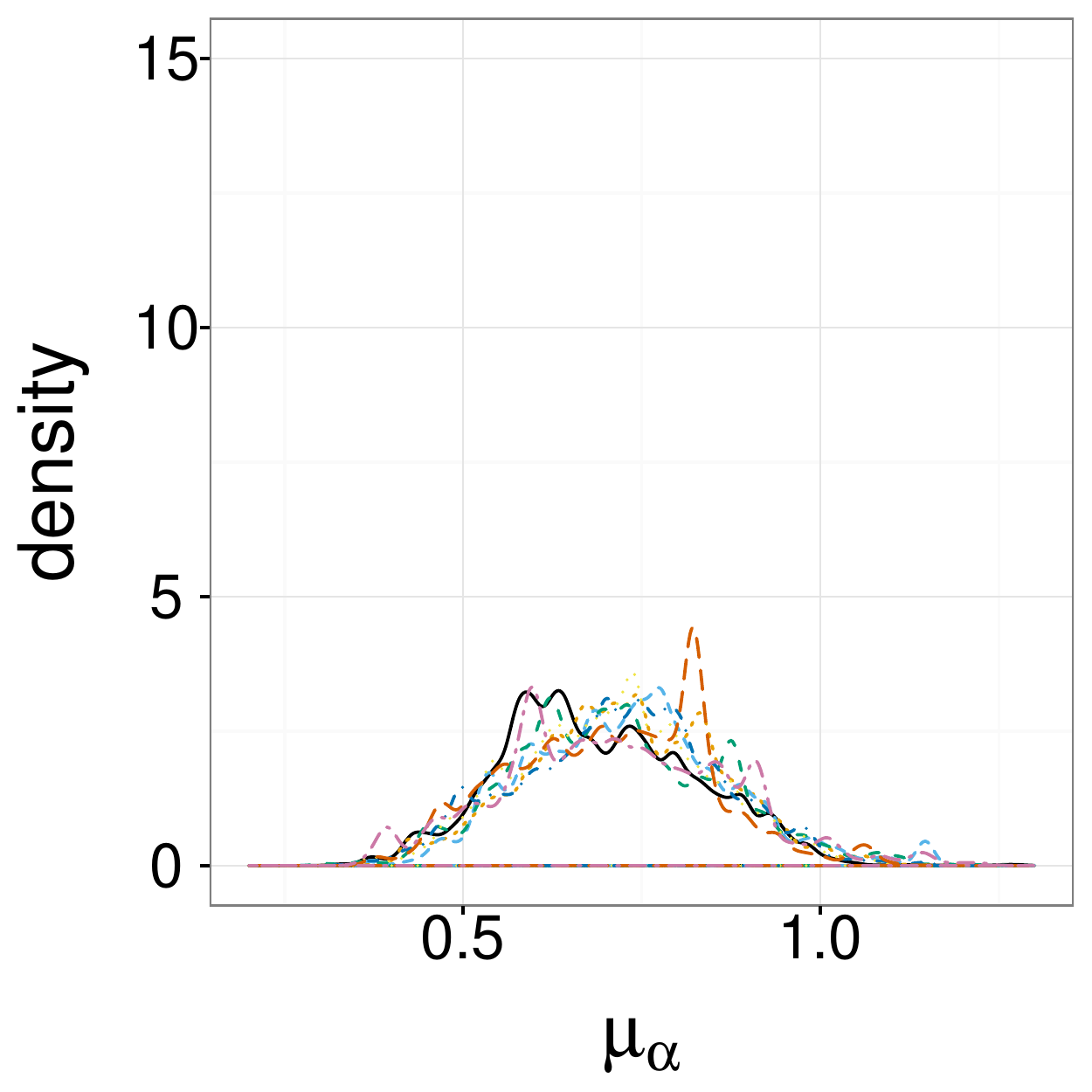}
      \caption*{PMMH, $N=512$.}
  \end{minipage}%
  \begin{minipage}{.24\linewidth}
      \centering
      \includegraphics[width=1.0\textwidth]{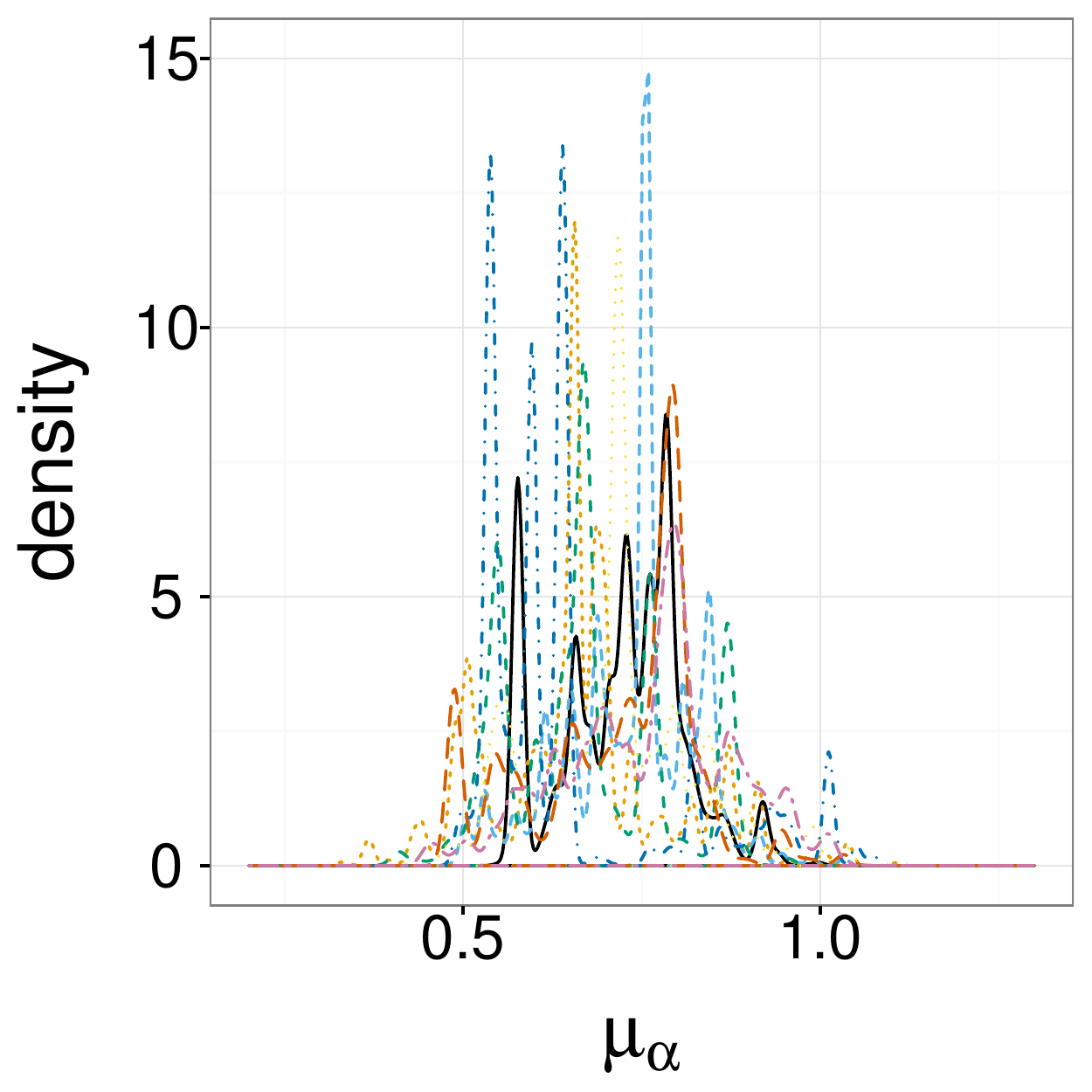}
      \caption*{PMMH, $N=128$.}
  \end{minipage} 
  \begin{minipage}{.24\linewidth}
      \centering
      \includegraphics[width=1.0\textwidth]{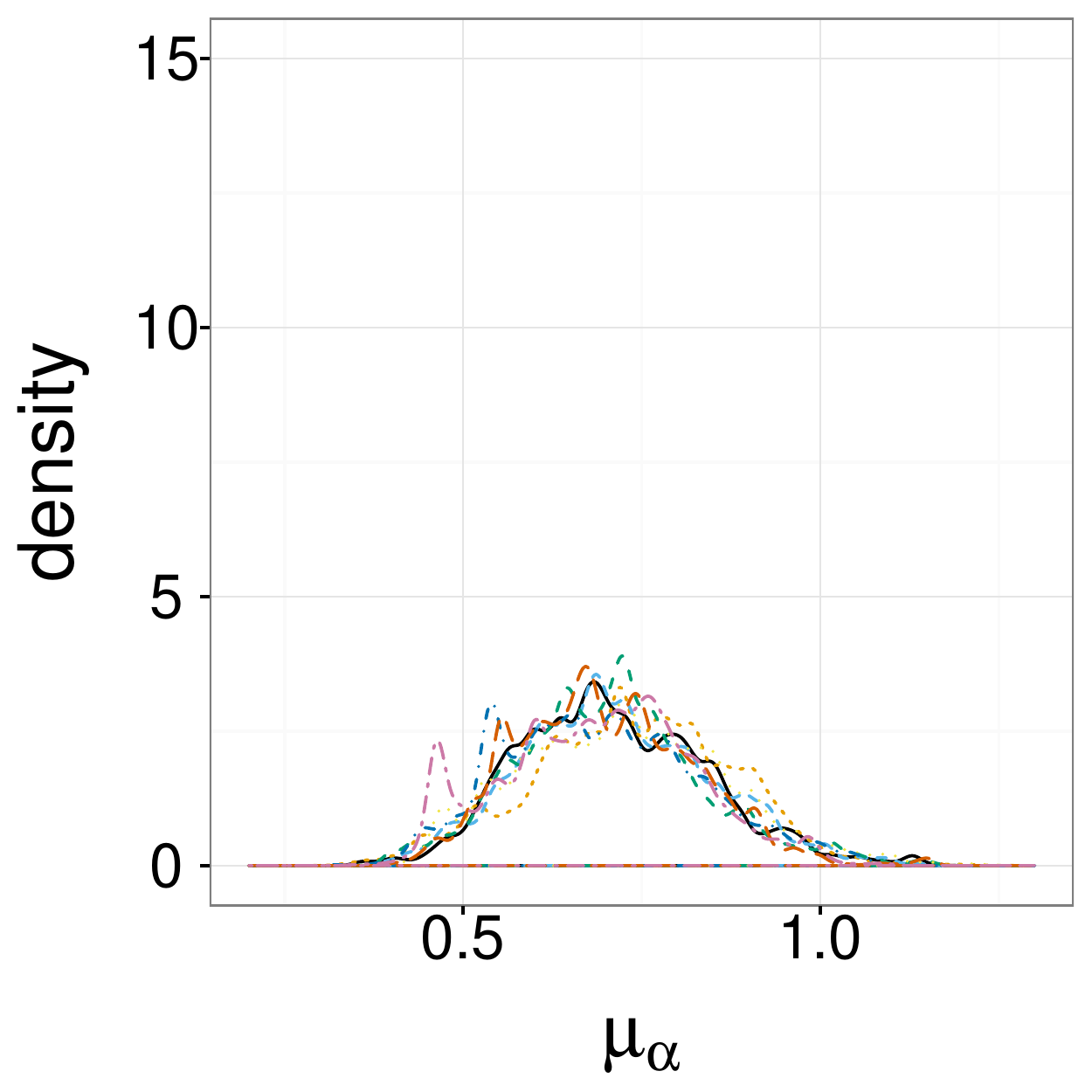}
      \caption*{Correlated PMMH, $N=128$, index-c.}
  \end{minipage} 
  \hspace*{0.1cm}
  \begin{minipage}{.24\linewidth}
      \centering
      \includegraphics[width=1.0\textwidth]{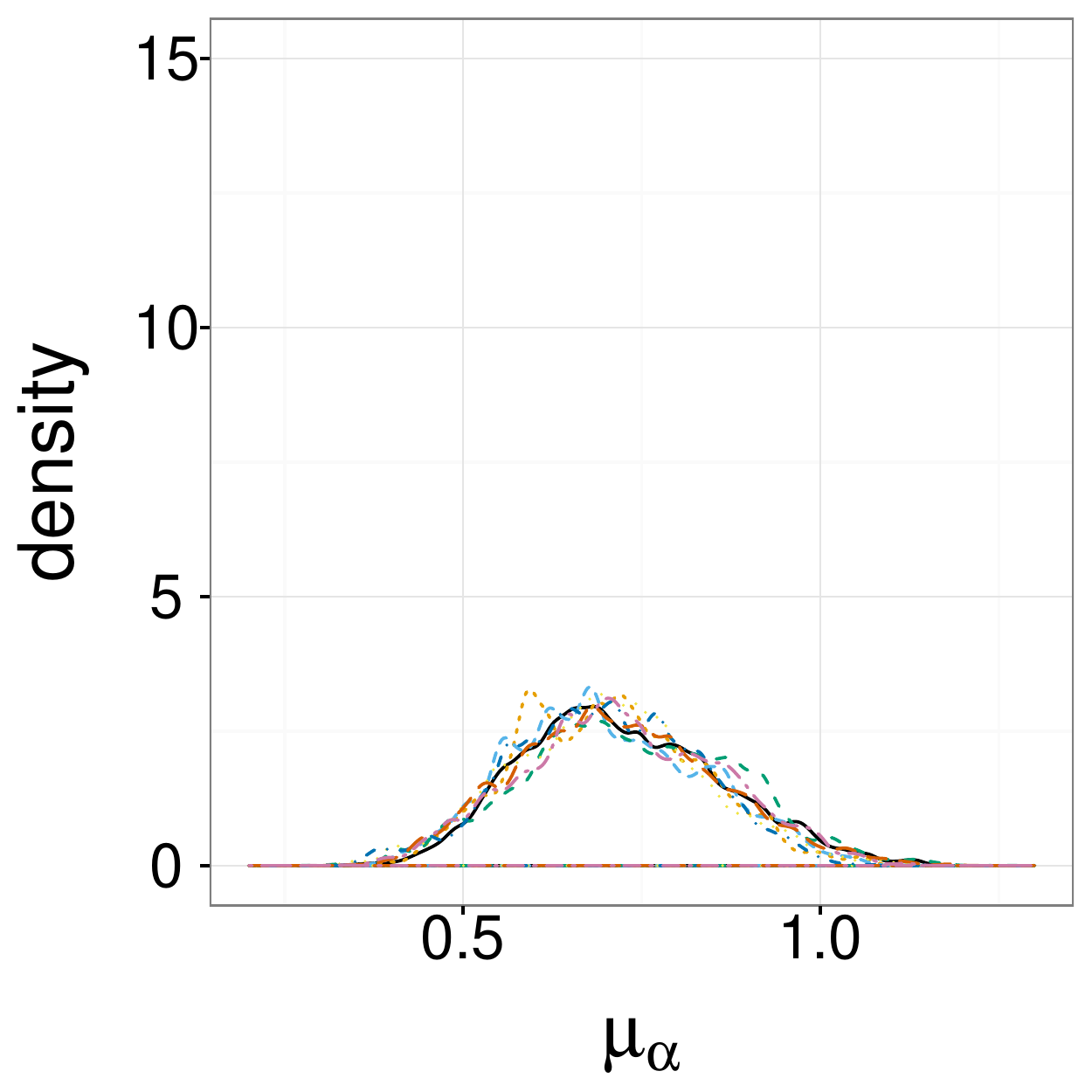}
      \caption*{Correlated PMMH, $N=128$, transport.}
  \end{minipage} 
\end{figure}

The parameter is $\theta = (\mu_\alpha, \sigma_\alpha, c, e, m_l, m_q)$. 
We specify a centered normal prior on $\mu_\alpha$ with variance $100$,
an exponential prior on $\sigma_\alpha$ with unit rate, and uniform priors in
$[0,1]$ for the four other parameters.  With logarithm and logistic transforms, we map
$\theta$ to $\mathbb{R}^6$. For the Metropolis--Hastings proposal distribution,
we use a normal random walk with a covariance matrix chosen as one sixth of
the covariance of the posterior, obtained from long pilot runs. We start the
Markov chains at the (transformed) data-generating parameter.  We then run the
particle marginal Metropolis--Hastings with $N=512$ particles and $M=100,000$
iterations, $10$ times independently, and obtain a mean acceptance rate of
$4.5\%$, with standard deviation of $0.3\%$, and an effective sample size averaged over
the parameters (ESS) of $106$ $(54)$.  With only $N=128$ particles, we 
obtain a mean acceptance of $0.4\%$ ($0.1\%$) and an ESS of $19$ $(10)$. Density estimators of the posterior
of $\mu_\alpha$ with these two samplers are shown on the left-most plots of Figure \ref{figure:pz:densityplots}.

We investigate whether we can obtain better posterior approximations using
the correlated Metropolis--Hastings algorithm, still with $N=128$  particles and $M=100,000$ iterations.
We set the correlation coefficient 
for the propagation of the process-generating variables
to $\rho = 0.99$.  We consider the use
of index-coupled, sorted and transport resampling. For the latter we choose
$\varepsilon = 0.1\times \text{median}(D)$ and $\alpha = 0.95$. 
For index-coupled resampling, we obtain an acceptance of $4.2\%$ ($0.5\%$),
with sorted resampling $5.0\%$ ($0.5\%$) and with transport resampling $5.4\%$
($0.4\%$). For the ESS, we obtain $108$ $(45)$ for index-coupled resampling,
$113$ $(61)$ for sorted resampling and $117$ $(53)$ for transport resampling.
We display the density estimators of the posterior of $\mu_\alpha$ in 
the right-most panels of Figure \ref{figure:pz:densityplots}, for index-coupled and transport
resampling.
Similar results are obtained with sorted resampling (not shown). However,
results in Section~\ref{sub:correlation-likelihood-pmmh} indicate
that sorted resampling would be less efficient in higher dimension. 
We conclude that the correlated algorithm with $N=128$ particles give posterior
approximations that are comparable to those obtained with a standard PMMH
algorithm that uses four times more particles; thus important computational
savings can be made. With the provided \texttt{R} implementation, the algorithm with $N=128$ and transport
resampling takes around $1000$ minutes per run, compared to $200$ minutes for
the other schemes, and around $700$ minutes for the standard algorithm with
$N=512$.

\subsection{Rhee--Glynn smoother \label{sub:pz:smoother}}

Next, we consider the problem of estimating the mean population of zooplankton at each time $t\in0:T$,
given a fixed parameter taken to be the data-generating one.
The intractability of the transition density precludes the use 
of ancestor or backward sampling, or the use of forward filtering backward sampling.

We draw $R = 1,000$ independent Rhee--Glynn smoothing estimators, using
$N=4,096$ particles.  The observed meeting times have a median of $4$, a mean of $4.7$ and a
maximum of $19$.  
The estimator $\hat{z}_{t}$ of the smoothing mean of $z_t$ at each time~$t$
is obtained by averaging $R = 1,000$ independent estimators.
We compute the Monte Carlo variance $\hat{v}_{t}$ of $\hat{z}_{t}$ at each time, and define 
the relative variance as $\hat{v}_{t}/(\hat{z}_t^2)$. 

We combine the Rhee--Glynn estimator (denoted by ``unbiased''
below) with the variance reduction technique of Section \ref{sec:newsmoother:practical}
(denoted by ``unbiased+RB''). Furthemore, we use the variance reduction of Appendix \ref{sec:appendix:furtherVR},
denoted by ``unbiased+RB+m'', with $m$ chosen to be the median of the meeting time, i.e. $m=4$. 
The latter increases the average meeting time from $4.7$ to $5.1$. We compare
the resulting estimators with a fixed-lag smoother \citep{doucet2011tutorial}
with a lag parameter $L=10$, and with a standard particle filter storing the complete trajectories.

\begin{figure}
    \centering
    \includegraphics[width=1.0\textwidth]{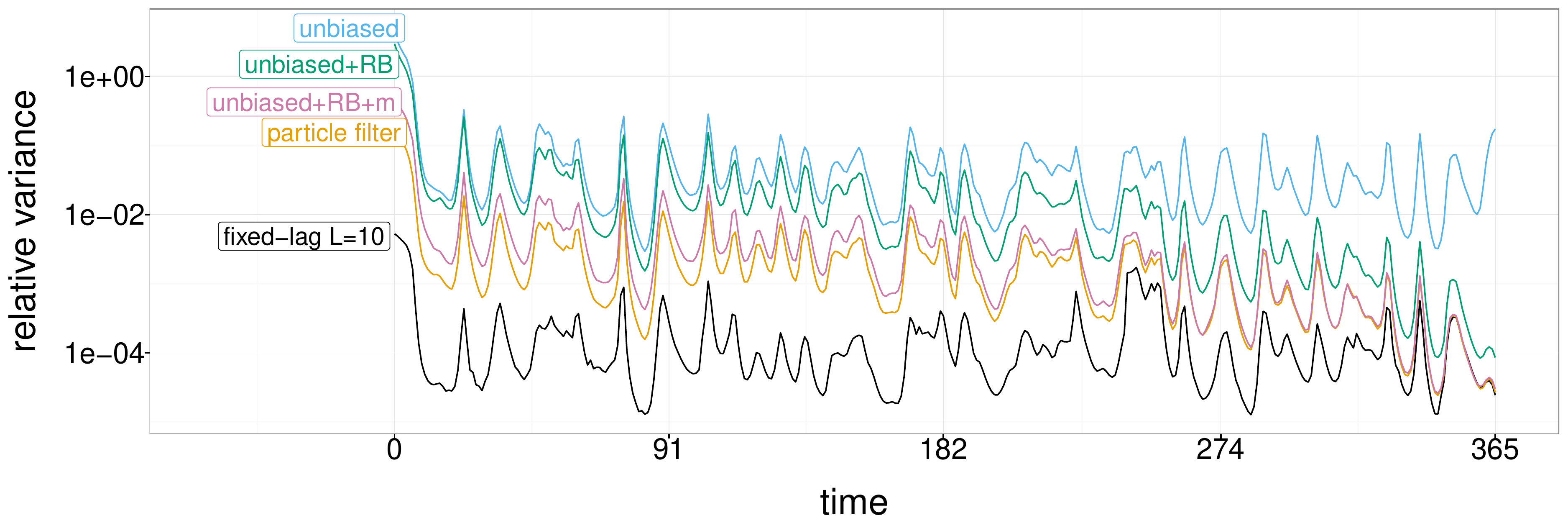}
    \caption{Comparison of the relative variance of the standard particle filter, 
        a fixed-lag smoother with lag $L=10$, and the proposed unbiased method,
        with Rao--Blackwellization (RB) and variance reduction (RB+m),
        for the estimation of the mean of the zooplankton population $z_t$,
        for the phytoplankton--zooplankton model with $T = 365$ observations. }
    \label{fig:pz:fixedlag}
\end{figure}

We use the same number of particles $N = 4,096$ and compute $R = 1,000$
estimators for each method.  The relative variance  is shown in Figure
\ref{fig:pz:fixedlag}.  First we see that the variance reduction techniques
have a significant effect, particularly for $t$ close to $T$ but also for small~$t$. In particular, the estimator $H_{m,\infty}$ with Rao--Blackwellization
(``unbiased+RB+m'') achieves nearly the same relative variance as the particle
filter.  The cost of these estimators can be computed as the number of
iterations $\max(m,\tau)$, times twice the cost of a particle filter for each
coupled particle filter.  In the present setting where the average number of
iterations is around five, we conclude that removing the bias from the standard
particle filter can be done for an approximate ten-fold increase in
computational cost. As expected the fixed-lag smoother leads to a significant
decrease in variance.  For this model, the incurred bias is negligible for
$L=10$ (not shown), which, however, would be hard to tell if we did not have
access to either unbiased methods or long runs of asymptotically exact methods. 

In this model, standard particle filters and fixed-lag approximations perform
well, leading to smaller mean squared error than the proposed estimators, for a
given computational cost. However, the proposed estimators are competitive, the
tuning of the algorithm is minimal, and unbiasedness prevents the
possibility of over-confident error bars as in Section~\ref{sec:methods:smoother:numerics}.  Therefore the proposed method trades an
extra cost for convenience and reliability. 

\section{Discussion\label{sec:discussion}}

Coupled particle filters can be beneficial in multiple settings.
Coupled bootstrap particle filters can be helpful in score estimation and parameter inference,
while coupled conditional particle filters lead to a new smoothing algorithm.
The attractive features of the Rhee--Glynn smoother include simple parallelization and accurate error bars;
these traits would be shared by perfect samplers, which aim at the more 
ambitious task of sampling exactly from the smoothing distribution \citep{leedoucetperfectsimulation}.

We have shown the validity of the Rhee--Glynn estimator under mild conditions, 
and its behaviour as a function of the
time horizon and the number of particles deserves further analysis. 
Numerical experiments in Appendix \ref{appendix:smoother:numerics:investigation}
investigate the effect of the time horizon and of the number of particles, among other effects.  
Furthermore, together with Fisher's
identity \citep{poyiadjis2011particle}, the proposed smoother produces unbiased estimators of the
score, for models where the transition density is tractable. This could in turn help maximizing the 
likelihood via stochastic gradients.

Another topic of future research might be 
the development of coupling ideas outside the context
of state space models,  following the growing popularity of particle methods 
in varied settings
\citep[see e.g.][]{DelDouJas:2006,BouchardCSJ:2012,NaessethLS:2014}. 

\noindent \textbf{Appendices}

Appendix \ref{sec:appendix:jointconditional} describes the sampling of a second particle filter given a first one.
Appendix~\ref{sec:appendix:validitity:PMMH} provides conditions for the validity of the correlated particle marginal
Metropolis--Hastings, and Appendix~\ref{sec:appendix:validitity:RG} for the validity of the Rhee--Glynn smoother.
Appendix~\ref{sec:appendix:furtherVR} describes an additional variance reduction technique for the Rhee--Glynn smoother.
Appendix \ref{appendix:approximate-transport} provides a description of the approximation of the transport problem.
Appendix \ref{appendix:smoother:numerics:investigation} provides extensive
numerical experiments for the proposed smoother and Appendix \ref{appendix:pseudocode:pf} gives pseudo-code descriptions. 
\texttt{R} functions to reproduce the figures of the article 
are provided at \href{http://github.com/pierrejacob/}{github.com/pierrejacob/} in \texttt{CoupledPF} and \texttt{CoupledCPF}.

\noindent \textbf{Acknowledgements}

Similar ideas and other applications of coupled resampling schemes have been independently proposed in \citet{sen2016coupling}.
The first author thanks Mathieu Gerber and Marco Cuturi for
helpful discussions. This work was initiated during the workshop on
\emph{Advanced Monte Carlo methods for complex inference problems} at
the Isaac Newton Institute for Mathematical Sciences, Cambridge, UK
held in April 2014. We would like to thank the organizers for a great
event which led to this work.

\bibliographystyle{rss}
\bibliography{Biblio}

\appendix

\section{Joint or conditional sampling of particle filters \label{sec:appendix:jointconditional}}

If we are interested in estimating two likelihoods $p(y_{1:T}|\theta)$ and
$p(y_{1:T}|\ttheta)$, for known values of $\theta$ and $\ttheta$, we can run a coupled particle filter 
in one forward pass, as described in Section~\ref{sub:resampling-and-couplings}. Likewise,
the coupled conditional particle filter given two reference trajectories can be run in one forward pass.
However,  we
might need to correlate $\hat p^N(y_{1:T}|\theta)$ with $\hat p^N(y_{1:T}|\ttheta)$ for
various values of $\ttheta$ which are not known in advance, 
as in the setting of Metropolis--Hastings schemes in Section~\ref{sub:correlation-likelihood-pmmh}. 

To address this situation, we can run a first particle filter given $\theta$, and store
all the generated particles $(\bw_t, \bx_t)$, ancestors $\ba_t$ and
random numbers $\bU_{t}$ for all $t$.  We can later run a second particle filter, given $\ttheta$,
conditionally on the variables generated by the first filter. 
At each resampling step, a probability matrix $P_t$ is 
computed given the variables generated thus far.
The ancestry vector $\bta_t$ is then sampled according to $P_t$,
conditionally upon the ancestors $\ba_t$ from the first filter.  Conditional sampling from $P_t$
can be done in order $N$ operations for index-coupled resampling, in order
$N\log N$ for sorted resampling and in order $N^2$ for generic coupled resampling schemes such as transport resampling.

Storing all the generated variables incurs a memory cost of order $N\times T$.  
By carefully storing and resetting the state of the random number
generator, one can in principle re-compute the first particle filter 
during the run of the second one, and thus the memory cost can be reduced to $N$, in exchange of a
two-fold increase in computational cost and a more sophisticated implementation
\citep[see e.g.][for a similar discussion]{jun2012entangled}.

\section{Validity of correlated particle marginal MH \label{sec:appendix:validitity:PMMH}}

We give a sufficient condition on the coupled resampling scheme for the correlated
particle marginal MH algorithm to target the same
distribution as the standard particle marginal MH algorithm.

Let $r(d\ba_t|\bw_{t})$ denote the probability distribution of the ancestors $\ba_t$ at step~$t$.
Since the weights $\bw_{t}$ are deterministic functions of $\ba_{0:t-1}$, $\bU_{0:t}$, and $\theta$,
we can also write $r(d\ba_t|\ba_{0:t-1},\bU_{0:t},\theta)$.
In the proposed algorithm,
$(\bw_t,\bx_t)$ and $(\btw_t,\btx_t)$
are used to compute a probability matrix $P_t$ and then 
$\bta_t$ are sampled conditionally on $\ba_t$ as described in Appendix \ref{sec:appendix:jointconditional}.
We denote that conditional distribution by 
$c(\bta_t |  \bta_{0:t-1}, \btU_{0:t}, \ttheta, \ba_{0:t}, \bU_{0:t},  \theta)$.

\begin{lemma}
    Assume that the marginal and conditional resampling distribution,
    respectively $r$ and $c$, associated with the coupled resampling scheme, are such that
\begin{align}
    & r(\ba_t | \ba_{0:t-1}, \bU_{0:t}, \theta) c(\bta_t |  \bta_{0:t-1}, \btU_{0:t}, \ttheta, \ba_{0:t}, \bU_{0:t},  \theta)\nonumber\\
    = \; & r(\bta_t | \bta_{0:t-1}, \btU_{0:t}, \ttheta) c(\ba_t | \ba_{0:t-1}, \bU_{0:t}, \theta, \bta_{0:t}, \btU_{0:t}, \ttheta).\label{eq:conditionvaliditycoupledresamplingpmmh}
\end{align}
Furthermore, assume that under the marginal resampling distribution $r$, $\mathbb{P}(a_t^k = j) = w_t^j$ 
for all $k$ and $j$ in $1:N$.
    Then the Markov kernel defined on $\theta$, $\bU_{0:T}$ and $\ba_{0:T-1}$ by the 
correlated particle marginal MH algorithm
has the same invariant distribution as the standard particle marginal MH.
    \label{lemma:validity:correlatedpmmh}
\end{lemma}

We first provide the proof of Lemma \ref{lemma:validity:correlatedpmmh}, 
and then we show that the condition of Eq.~\eqref{eq:conditionvaliditycoupledresamplingpmmh} is satisfied for sorted, index-coupled and
transport resampling schemes.
\begin{proof}[Lemma \ref{lemma:validity:correlatedpmmh}]
The extended target distribution of the particle marginal MH 
algorithm has density
\begin{align}
    \bar\pi(\theta,\bU_{0:T}, \ba_{0:T-1}) 
    = \frac{p(\theta|y_{1:T}) \varphi(\bU_{0:T-1}) \prod_{t=0}^{T-1} r(\ba_t|\ba_{0:t-1},\bU_{0:t},\theta) \widehat p^N(y_{1:T}|\theta)}{
        p(y_{1:T}|\theta)},
    \label{eq:pmmh:extendedtarget}
\end{align}
which is just a change of notation compared to \citet{andrieu:doucet:holenstein:2010}.
The condition $\mathbb{P}(a_t^k = j) = w_t^j$ for all $k$ and $j$ in $1:N$ ensures
that the marginal distribution on $\theta$ is indeed the posterior distribtuion $p(d\theta|y_{1:T})$.

We denote by $\xi$ all the auxiliary variables generated by the particle filter:
$\bU_{t}$ for all $t\in 0:T$, and $\ba_t$ for all $t\in 0:T-1$.
The extended target distribution of Eq.~\eqref{eq:pmmh:extendedtarget} can 
be rewritten 
    $\bar\pi(\theta,\xi)  = p(\theta|y_{1:T}) m_\theta(\xi) \widehat p^N(y_{1:T}|\theta)/p(y_{1:T}|\theta)$,
where $m_\theta(\xi)$ is the distribution of $\xi$ defined by a run of the particle filter. 

Rewriting the procedure described in Section
\ref{sub:correlation-likelihood-pmmh}, from the state $(\theta, \xi)$, we
sample $\ttheta \sim q(d\ttheta|\theta)$ and  $\tilde{\xi} \sim
K_{\theta,\ttheta}(d\tilde{\xi}|\xi)$ from a Markov kernel on the space of
$\xi$, which may depend on $\theta$ and $\ttheta$.  The particle marginal MH algorithm 
uses $K_{\theta,\ttheta}(\tilde{\xi}|\xi) = m_{\ttheta}(\tilde{\xi})$. Other kernels leaving
$\bar\pi(d\theta,d\xi)$ invariant can be constructed, a sufficient condition
being the standard detailed balance:
\begin{align}
  \label{eq:pmmh:validity:detailed-balance}
  m_\theta(\xi) K_{\theta,\ttheta}(\tilde{\xi}|\xi) &= m_{\ttheta}(\tilde{\xi}) K_{\ttheta,\theta}(\xi|\tilde{\xi}),
  &
  \forall& \theta,\ttheta,\xi,\tilde{\xi}.
\end{align}
We consider kernels $K_{\theta,\ttheta}$ of the form  
\begin{align*}
    K_{\theta,\ttheta}(\tilde{\xi}|\xi) = \phi(\btU_{0:T}|\bU_{0:T})
    \prod_{t=0}^{T-1} c(\bta_t | \bta_{0:t-1}, \btU_{0:t}, \ttheta, \ba_{0:t},  \bU_{0:t}, \theta).
\end{align*}
In this expression, $\phi$ is a Markov kernel in detailed balance with respect to $\varphi$,
the distribution of the process-generating variables.
The condition of Eq.~\eqref{eq:conditionvaliditycoupledresamplingpmmh} implies
Eq.~\eqref{eq:pmmh:validity:detailed-balance}. $\blacksquare$
\end{proof}

For independent and sorted resampling, the conditional sampling of $\bta_t$ does 
not require any variable from the first particle system, so that we have 
\begin{align*}
    c(\bta_t |  \bta_{0:t-1}, \btU_{0:t}, \ttheta, \ba_{0:t}, \bU_{0:t},  \theta) &=  r(\bta_t | \bta_{0:t-1}, \btU_{0:t}, \ttheta),
\end{align*}
and the condition of Eq.~\eqref{eq:conditionvaliditycoupledresamplingpmmh} is satisfied.

For general coupled resampling schemes, under conditional 
sampling,
for each $k\in 1:N$, $\ta_t^{k}$ is distributed according to 
the $a_t^k$-th row of  $P_t$ defined by a coupled resampling scheme,
e.g. Eq.~\eqref{eq:indexmatchingmatrix} for index-coupled resampling.
The conditional probability $c(\bta_t |  \bta_{0:t-1}, \btU_{0:t}, \ttheta, \ba_{0:t}, \bU_{0:t},  \theta)$
takes the form $\prod_{k=1}^N P^{a_t^k \ta_t^k}_t / w^{a_t^k}_t$.
For the index-coupled probability matrix of Eq.~\eqref{eq:indexmatchingmatrix}, 
Eq.~\eqref{eq:conditionvaliditycoupledresamplingpmmh} is satisfied,
with respect to $r(\ba_t|\ba_{0:t-1},\bU_{0:t},\theta)=\prod_{k=1}^N w_t^{a_t^k}$,
because we obtain the transpose of $P_t$ if 
we swap $\bw$ and $\btw$ in its construction.

For transport resampling, the distance matrix $D$ in Section
\ref{sub:transport-resampling} is such that we obtain its transpose if $\bx$ and
$\btx$ are swapped in its construction. Thus, the optimal transport probability matrix $P_t$
is such that we obtain its transpose if $(\bw,\bx)$ and $(\btw, \btx)$ are swapped in its definition. Therefore,
if $\ta_t^{k}$ is distributed according to $P_t^{a_t^k\cdot}$ for each $k\in 1:N$,
then the condition of Eq.~\eqref{eq:conditionvaliditycoupledresamplingpmmh} will be satisfied.

However, if we use an approximate solution $\hat{P}$ to the transport problem, 
then the condition might not hold. This can be circumvented by symmetrizing the 
coupled resampling matrix, by computing $\hat{P}$ 
using $((\bw,\bx),(\btw,\btx))$, and $\tilde{P}$ using $((\btw,\btx),(\bw,\bx))$.
Then one can use the matrix $\hat{P}_t=(\hat P + \tilde{P}^\transp)/2$,
and sample $\ta_t^{k}$ distributed according to $\hat{P}_t^{a_t^k\cdot}$ for each $k\in 1:N$.
This ensures that the detailed balance condition holds with respect to the multinomial resampling distribution.

\section{Validity of Rhee--Glynn smoothing estimators\label{sec:appendix:validitity:RG}}

We first state a result on the probability of meeting in one step of
the coupled conditional particle filter. 
\begin{lemma}
    Under Assumptions \ref{assumption:upperbound} and \ref{assumption:couplingmatrix},  there exists $\varepsilon>0$ such that 
\[
    \forall X  \in \mathbb{X}^{T+1}, \quad \forall \tX \in \mathbb{X}^{T+1}, \quad \mathbb{P}(X' = \tX' | X, \tX) \geq \varepsilon,
\]
where $(X',\tX') \sim \text{CCPF}(X,\tX, \bU)$ and $\bU\sim \varphi$.
Furthermore, if $X = \tX$, then $X' = \tX'$ almost surely.
    \label{lemma:meetingprobability}
\end{lemma}
The constant $\varepsilon$
depends on $N$ and $T$, and on the coupled resampling scheme being used.  Lemma
\ref{lemma:meetingprobability} can be used, together with the coupling
inequality \citep{lindvall2002lectures}, to prove the ergodicity of the
conditional particle filter kernel, which is akin to the approach of
\citet{ChopinS:2015}.  The coupling inequality states that the total variation
distance between $X^{(n)}$ and  $\tX^{(n-1)}$ is less than
$2\mathbb{P}(\tau > n)$, where $\tau$ is the meeting time.  By assuming $\tX^{(0)}\sim\pi$,
$\tX^{(n)}$ follows $\pi$ at each step $n$, and we obtain a bound for the total
variation distance between $X^{(n)}$ and $\pi$.  Using Lemma
\ref{lemma:meetingprobability}, we can bound the probability
$\mathbb{P}(\tau > n)$ from above by $(1-\varepsilon)^n$, as in the proof of Theorem \ref{thm:finitevariance} below.
This implies that the computational cost of the proposed estimator
has a finite expectation for all $N\geq 2$ and~$T$.  

\subsection{Proof of Lemma \ref{lemma:meetingprobability} \label{sec:proof:meetingprobability}}

Dropping the parameter from the notation, we use $f(dx_t |x_{t-1})$ for the transition,
$m_0(dx_0)$ for the initial distribution and $g_t(x_t) =
g(y_t|x_t)$ for the measurement.  Let $\mathcal{F}_t$ denote the filtrations generated by the coupled
conditional particle filter at time $t$.  We denote by $x_{0:t}^k$, for
$k\in1:N$, the surviving trajectories at time~$t$.

Let $I_t \subseteq 1:N-1$ be the set of common particles at time $t$ defined by
    $I_t = \{j \in 1:N-1 : x_{0:t}^j  = \tilde x_{0:t}^j \}$.
The meeting probability, implicitly conditioned upon the reference trajectories $x_{0:T}$ and $\tilde{x}_{0:T}$, can be bounded by:
\begin{multline}
    \Prb(x_{0:T}^\prime = \tilde x_{0:T}^\prime) = \E\left[\I\!\left(x_{0:T}^{b_T} = \tilde x_{0:T}^{\tilde{b}_T} \right)\right]
	\geq \sum_{k=1}^{N-1} \E[\I\!\left(k \in I_T\right) P_T^{kk}] \\
	= (N-1)\E[\I\!\left(1\in I_T \right) P_T^{11}]
	\geq \frac{N-1}{ (N\bar{g})^2} \E[\I\!\left(1\in I_T \right) g_T(x_T^1) g_T(\tilde x_T^1)],
\end{multline}
where we have used Assumptions \ref{assumption:upperbound} and \ref{assumption:couplingmatrix}.
Now, let $\psi_t : \setX^t \mapsto \reals_+$ and consider
\begin{align}
	\label{eq:crude:h}
	\E[\I\!\left( 1\in I_t \right) \psi_t(x_{0:t}^1) \psi_t(\tilde x_{0:t}^1)] =
	\E[\I\!\left( 1\in I_t \right) \psi_t(x_{0:t}^1)^2],
\end{align}
since the two trajectories agree on $\{1\in I_t\}$.
We have
\begin{align}
	\I\!\left( 1\in I_t \right) \geq \sum_{k=1}^{N-1} \I\!\left(k\in I_{t-1} \right) \I\!\left(a_{t-1}^1 = \tilde a_{t-1}^1 = k \right),
\end{align}
and thus
\begin{multline}
	\label{eq:crude:h2}
	\E[\I\!\left( 1\in I_t \right) \psi_t(x_{0:t}^1)^2] \\
	\geq \E[\sum_{k=1}^{N-1} \I\!\left(k\in I_{t-1} \right) \E[ \I\!\left(a_{t-1}^1 = \tilde a_{t-1}^1 = k \right) \psi_t(x_{0:t}^1)^2 \mid \mathcal{F}_{t-1} ]] \\
	= (N-1)\E[\I\!\left(1\in I_{t-1} \right) \E[ \I\!\left(a_{t-1}^1 = \tilde a_{t-1}^1 = 1 \right) \psi_t(x_{0:t}^1)^2 \mid \mathcal{F}_{t-1}  ]].
\end{multline}
The inner conditional expectation can be computed as
\begin{multline}
	\label{eq:cruce:h2-inner}
	\E[ \I\!\left(a_{t-1}^1 = \tilde a_{t-1}^1 = 1 \right) \psi_t(x_{0:t}^1)^2 \mid \mathcal{F}_{t-1} ] \\
	=\sum_{k,\ell=1}^N P_{t-1}^{k\ell} \I\!\left(k=\ell=1\right) \int \psi_t((x_{0:t-1}^k, x_t ))^2 f(dx_t|x_{t-1}^k) \\
	= P_{t-1}^{11} \int \psi_t((x_{0:t-1}^1, x_t))^2 f(dx_t|x_{t-1}^1) \\
    \geq \frac{g_{t-1}(x_{t-1}^1) g_{t-1}(\tilde x_{t-1}^1) }{(N\bar{g})^2} \left( \int \psi_t((x_{0:t-1}^1, x_t )) f(dx_t|x_{t-1}^1) \right)^2,
\end{multline}
where we have again used Assumptions \ref{assumption:upperbound} and \ref{assumption:couplingmatrix}.
Furthermore, on $\{1\in I_{t-1}\}$ it holds that $x_{0:t-1}^1 = \tilde x_{0:t-1}^1$ and therefore, combining Eqs.~\eqref{eq:crude:h}--\eqref{eq:cruce:h2-inner} we get
\begin{multline}
	\E[\I\!\left( 1\in I_t \right) \psi_t(x_{0:t}^1) \psi_t(\tilde x_{0:t}^1)] \\
	\geq \frac{(N-1)}{(N\bar{g})^2}\E\Big[\I\!\left(1\in I_{t-1} \right) g_{t-1}(x_{t-1}^1) \int \psi_t((x_{0:t-1}^1, x_t )) f(dx_t|x_{t-1}^1) \\ \times
	g_{t-1}(\tilde x_{t-1}^1) \int \psi_t((\tilde x_{0:t-1}^1, x_t )) f(dx_t|\tilde x_{t-1}^1)
	\Big].
\end{multline}
Thus, if we define
for $t=1,\ldots,T-1$, 
$\psi_t(x_{0:t}) = g_t(x_t) \int \psi_{t+1}(x_{0:t+1}) f(dx_{t+1}|x_t)$,  
and
$\psi_T(x_{0:T}) = g_T(x_T)$, 
it follows that
\begin{align*}
	\Prb(x_{0:T}^\prime= \tilde x_{0:T}^\prime) &\geq \frac{(N-1)^\transp}{(N\bar{g})^{2T}} \E[\I\!\left(1\in I_1 \right) \psi_1(x_1^1)\psi_1(\tilde x_1^1)] \\
	&= \frac{(N-1)^\transp}{(N\bar{g})^{2T}} \E[\psi_1(x_1^1)^2] \geq \frac{(N-1)^\transp}{(N\bar{g})^{2T}} Z^2 > 0,
\end{align*}
where $Z > 0$ is the normalizing constant of the model, defined as
$\mathbb{E}[\prod_{t=1}^\transp g_t(x_t)]$ where the expectation is with respect to
the distribution $m_0(dx_0) \prod_{t=1}^\transp f(dx_t|x_{t-1})$ of the latent
process $x_{0:T}$.

We note that, for any fixed $T$, the bound goes to zero when $N\to \infty$.
The proof fails to capture accurately the behaviour of $\varepsilon$
in Lemma \ref{lemma:meetingprobability} as a function of $N$ and $T$.

\subsection{Proof of Theorem \ref{thm:finitevariance} \label{sec:proof:unbiased}}

We present a proof for a generalization of the estimator given in Section \ref{sec:methods:smoother}.
Introduce a truncation variable $G$, with support on the integers $\{0,1,2,\ldots\}$.
Define the estimator as
\begin{equation}
H=\sum_{n=0}^{G}\frac{\Delta^{(n)}}{\mathbb{P}\left(G\geq n\right)},
    \label{eq:generalized:RGestimator}
\end{equation}
where $\Delta^{(0)}=h(X^{(0)})$ and $\Delta^{(n)}=h(X^{(n)})-h(\tX^{(n-1)})$, for $n\geq1$.
We consider the following assumption on the truncation variable.

\begin{assumption}
\label{assumption:truncation} The truncation variable
$G$ is Geometric, with probability mass function $\mathbb{P}(G = n) = (1-p)^n p$, with
support on $\{0,1,2,\ldots\}$ and parameter $p\in[0,1)$, chosen such that
$p < 1-\left(1-\varepsilon\right)^{\delta/(2+\delta)}$,
where $\varepsilon$ is as in Lemma \ref{lemma:meetingprobability}
and $\delta$ as in Assumption \ref{assumption:mixing}. Furthermore, $G$ is independent of all the other
variables used in Eq.~\eqref{eq:generalized:RGestimator}.
\end{assumption}

This assumption precludes the use of a range of values of $p$ near
one, which could have been a tempting choice for computational reasons.
On the other hand, it does
not prevent the use of values of $p$ near $0$, so that we retrieve the estimator of Eq.~\eqref{eq:RGestimator}
by setting $p=0$, ensuring that Assumption
\ref{assumption:truncation} is satisfied for all values of $\varepsilon$
and $\delta$. 

We can first upper-bound $\mathbb{P}\left(\tau>n\right)$, for all $n\geq2$,
using Lemma \ref{lemma:meetingprobability} (see e.g. \citet{williams1991probability},
exercise E.10.5). We obtain for all $n\geq2$,
\begin{equation}
\mathbb{P}\left(\tau>n\right)\leq\left(1-\varepsilon\right)^{n-1}.\label{eq:meetingtime:survival2}
\end{equation}
This ensures that $\mathbb{E}[\tau]$ is finite; and that $\tau$ is almost surely finite.
We then introduce the random variables
\begin{equation}
    \forall m \geq 1 \quad Z_{m}=\sum_{n=0}^{m} \frac{\Delta^{(n)}\mathds{1}(n\leq G)}{\mathbb{P}\left(n \leq G \right)}.
    \label{eq:Z}
\end{equation}
Since $\tau$ is almost surely finite, and since $\Delta^{(n)} = 0$ for all $n \geq \tau$,
then $Z_m\to Z_\tau = H$ almost surely when $m\to\infty$. We prove that $(Z_m)_{m\geq 1}$ is a Cauchy sequence in 
$L_2$, i.e.
    $\sup_{m'\geq m} \mathbb{E}\left[ (Z_{m'} - Z_m)^2 \right]$
    goes to $0$ as $m\to\infty$. 
We write 
\begin{align*}
    (Z_{m'} - Z_m)^2 &= \sum_{n = m + 1}^{m'} \frac{(\Delta^{(n)})^2 \mathds{1}(n\leq G)}{\mathbb{P}\left(n \leq G \right)^2} 
    + 2 \sum_{n = m + 1}^{m'} \sum_{\ell = n + 1}^{m'} \frac{\Delta^{(n)} \Delta^{(\ell)} \mathds{1}(\ell \leq G)}{\mathbb{P}\left(n \leq G \right)\mathbb{P}\left(\ell \leq G \right)} 
\end{align*}
and thus, using the independence between $G$ and $(\Delta^{(n)})_{n \geq 0}$,
\begin{align*}
    \mathbb{E}\left[(Z_{m'} - Z_m)^2\right] &= \sum_{n = m + 1}^{m'} \frac{\mathbb{E}\left[(\Delta^{(n)})^2\right]}{\mathbb{P}\left(n \leq G \right)} 
    + 2 \sum_{n = m + 1}^{m'} \sum_{\ell = n + 1}^{m'} \frac{\mathbb{E}\left[\Delta^{(n)} \Delta^{(\ell)} \right]}{\mathbb{P}\left(n \leq G \right)}.
\end{align*}
To control $\mathbb{E}[(\Delta^{(n)})^{2}] = \mathbb{E}[(\Delta^{(n)})^{2}\mathds{1}\left(\tau>n\right)]$,
we use H\"older's inequality, with $p=1+\delta/2$, and $q=(2+\delta)/\delta$,
where $\delta$ is as in Assumption \ref{assumption:mixing}, 
\begin{align*}
\mathbb{E}\left[(\Delta^{(n)})^{2}\right]  & \leq\mathbb{E}\left[(\Delta^{(n)})^{2+\delta}\right]^{1/(1+\delta/2)}\left(\left(1-\varepsilon\right)^{\delta/(2+\delta)}\right)^{n-1}.
\end{align*}
Furthermore, using Assumption \ref{assumption:mixing},
there exists $C_{1}<\infty$ such that, for $n_0\in\mathbb{N}$
large enough, 
\begin{align}
    \label{eq:boundlimit} \forall n\geq n_0\quad & \mathbb{E}\left[(\Delta^{(n)})^{2+\delta}\right]^{1/(1+\delta/2)}\leq C_{1}.
\end{align}
We write $\eta = \left(1-\varepsilon\right)^{\delta/(2+\delta)}$, and take $m$ such that $m \geq n_0$.
Using Cauchy--Schwarz, we have for all $n,\ell \geq m$,
\[\mathbb{E}\left[\Delta^{(n)} \Delta^{(\ell)} \right] \leq \left(\mathbb{E}\left[(\Delta^{(n)})^2 \right] \mathbb{E}\left[(\Delta^{(\ell)})^2 \right]\right)^{1/2}\leq C_1 \eta^{(n-1)/2} \eta^{(\ell-1)/2}.\]
We can now write 
\begin{align*}
    \mathbb{E}\left[(Z_{m'} - Z_m)^2\right] &\leq C_1 \sum_{n = m + 1}^{m'} \frac{\eta^{n-1}}{\mathbb{P}\left(n \leq G \right)} 
    + 2 \sum_{n = m + 1}^{m'} \sum_{\ell = n + 1}^{m'} \frac{C_1 \eta^{(n-1)/2}\eta^{(\ell-1)/2}}{\mathbb{P}\left(n \leq G \right)}\\
    &\leq C_1 \sum_{n = m + 1}^{m'} \frac{\eta^{n-1}}{\mathbb{P}\left(n \leq G \right)} 
    + 2 C_1 \sum_{n = m + 1}^{m'} \frac{\eta^{n-1}}{\mathbb{P}\left(n \leq G \right)} \sqrt{\eta}\frac{1-\left(\sqrt{\eta}\right)^{m'}}{1-\left(\sqrt{\eta}\right)}.
\end{align*}
Under Assumption \ref{assumption:truncation}, we have $\mathbb{P}\left(n \leq G \right) = (1-p)^{n+1}$. For the above series to go to zero when
$m\to \infty$ and $m'\geq m$, it is enough that $\eta / (1-p)<1$. By definition of $\eta$, this holds if 
$\left(1-\varepsilon\right)^{\delta/(2+\delta)} < 1 - p$, 
which is part of Assumption \ref{assumption:truncation}. Thus $(Z_m)_{m \geq 1}$ is a Cauchy sequence in $L_2$.

By uniqueness of the limit, since $(Z_m)_{m \geq 1}$ goes almost surely to
$H$, $(Z_m)_{m \geq 1}$ goes to $H$ in $L_2$. This shows that $H$ has finite first two moments. We can retrieve the
expectation of $H$ by
\[ \mathbb{E}Z_{m}=\sum_{n=0}^{m}\mathbb{E}[\Delta^{(n)}]=\mathbb{E}\left[h(X^{(m)})\right] \xrightarrow[m\to \infty]{} \pi(h),
\]
according to Assumption \ref{assumption:mixing}.
We can retrieve the second moment of $H$ by  
\begin{align*}
    \mathbb{E}[Z_{m}^2] &= \sum_{n = 0}^{m} \frac{\mathbb{E}\left[(\Delta^{(n)})^2\right]}{\mathbb{P}\left(n \leq G \right)} 
    + 2 \sum_{n = 0}^{m} \sum_{\ell = n + 1}^{m} \frac{\mathbb{E}\left[\Delta^{(n)} \Delta^{(\ell)} \right]}{\mathbb{P}\left(n \leq G \right)}\\
    &\xrightarrow[m\to \infty]{} 
\sum_{n = 0}^{\infty} \frac{\mathbb{E}\left[(\Delta^{(n)})^2\right] + 2 \sum_{\ell = n + 1}^{\infty} \mathbb{E}\left[\Delta^{(n)} \Delta^{(\ell)} \right]}{\mathbb{P}\left(n \leq G \right)}.
\end{align*}

\section{Further variance reduction for the Rhee--Glynn estimator \label{sec:appendix:furtherVR}}

A  variance reduction can be achieved in the following way.
Let $M,m$ be two integers such that $M > m\geq 0$. Define
\begin{align}
    H_{m,M}&= h(X^{(m)}) + \sum_{n=m+1}^{M} h(X^{(n)}) - h(\tX^{(n-1)}) \label{eq:form1}\\
 &= h(X^{(M)}) + \sum_{n=m}^{M-1} h(X^{(n)}) - h(\tX^{(n)}),\label{eq:form2}
\end{align}
We have $\E[H_{m,M}]=\E[h(X^{(M)})]$ by Eq.~\eqref{eq:form2} and using the fact
that $X^{(n)}$ and $\tX^{(n)}$ have the same distribution.  Furthermore,
$\E[h(X^{(M)})]$ goes to $\pi(h)$ as $M\to\infty$ under Assumption
\ref{assumption:mixing}.  We consider the estimator $H_{m,\infty}$, which can
be computed in a finite time as follows.

We run Algorithm \ref{alg:rheeglynnsmoother} until step
$\max(\tau,m)$.  If $\tau\leq m+1$,  from Eq.~\eqref{eq:form1},  $H_{m,\infty}
= h(X^{(m)})$ almost surely, since $X^{(n)} = \tX^{(n-1)}$ for all $n\geq m+1$.
If $\tau > m+1$,  $H_{m,\infty}= h(X^{(m)}) +
\sum_{n=m+1}^{\tau-1} h(X^{(n)}) - h(\tX^{(n-1)})$, again using Eq.~\eqref{eq:form1}.
The estimator $H_{m,\infty}$ is thus made of a single term with large
probability if $m$ is large enough; the computational cost is of
$\max(\tau,m)$ instead of $\tau$ for the original estimator. The
intuition is that the fewer terms there are in $H_{m,\infty}$, the
smaller the variance. 

Another question is whether we can average over various choices of $m$.
We can compute $\bar H_m = \sum_{n=0}^m \alpha_n H_{n,\infty}$ where $\sum_{n=0}^m \alpha_n = 1$; this estimator is still unbiased.
It follows (after some calculations) that
\begin{align*}
  \bar H_m = \sum_{n=0}^m \alpha_n h(X^{(n)}) + \sum_{n=1}^{\tau-1} \beta_n (h(X^{(n)})-h(\tX^{(n-1)})),
\end{align*} 
where $\beta_n = \sum_{j=0}^{n-1 \wedge m} \alpha_j$; the choice of coefficients $\alpha_{0:m}$ is left for future work.

\section{Approximate transport \label{appendix:approximate-transport}}

In this section, we briefly describe the approximation to the transport problem
introduced in \citet{cuturi2013sinkhorn,CuturiDoucet}, which is explained along
with various other methods in \citet{benamou2014iterative}. 
The idea is to regularize the original transport
program, with the modified objective function 
\[
\left\langle P,D\right\rangle -\varepsilon h\left(P\right),
\]
where $h\left(P\right)=-\sum_{i,j} P^{ij}\log P^{ij}$ is the entropy of $P$,
$\left\langle P,D\right\rangle$ is the sum of the terms $P^{ij}D^{ij}$, and
$\varepsilon\in\mathbb{R}_{+}$. When $\varepsilon\to0$, minimizing the above
objective over $\mathcal{J}\left(\bw,\btw\right)$ corresponds to the original optimal
transport problem. We can write 
\begin{align*}
    \left\langle P,D\right\rangle -\varepsilon h\left(P\right) & =\varepsilon \text{KL}\left(P||\exp\left(-D/\varepsilon\right)\right),
\end{align*}
where $\text{KL}(S||Q)=\sum_{i,j}S^{ij}\log\left(S^{ij}/Q^{ij}\right)$ and
$\exp\left(S\right)$ is the element-wise exponential of $S$. 
Minimizing the regularized transport objective is equivalent to finding
the matrix $P$ with minimal KL projection on $K=\exp\left(-D/\varepsilon\right)$,
leading to the optimization problem
\begin{equation}
    P^\varepsilon = \arginf_{P\in\mathcal{J}\left(\bw,\btw\right)}\text{KL}\left(P||K\right).\label{eq:regularizedtransport}
\end{equation}
Compared to the original transport problem, this program is computationally simpler. 
By noting that $\mathcal{J}\left(\bw,\btw\right)=\mathcal{J}_{\bw}\cap\mathcal{J}_{\tilde{\bw}}$
where $\mathcal{J}(\bw)=\left\{ P:P\mathds{1}=\bw\right\} $ and $\mathcal{J}(\tilde{\bw})=\left\{ P:P^\transp\mathds{1}=\tilde{\bw}\right\} $,
we can find the solution of Eq. (\ref{eq:regularizedtransport}) by
starting from $P^{(0)}=K$, and by performing iterative KL projections 
on $\mathcal{J}(\bw)$ and $\mathcal{J}(\tilde{\bw})$, thus constructing a sequence of
matrices $P^{(i)}$ for $i\in \{1,\ldots,n\}$. When $n$ goes to infinity, the matrix $P^{(n)}$
converges to $P^\varepsilon$.
This is Algorithm 1 in \citet{cuturi2013sinkhorn},
which we state for completeness in Algorithm \ref{alg:sinkhorn} below.

The algorithm is iterative, and requires matrix-vector
multiplications which cost $\mathcal{O}(N^2)$ at every iteration. The number of steps $n$ 
to achieve a certain precision can be
taken independently of the number of particles $N$,
thanks to the convexity of the objective function \citep{cuturi2013sinkhorn}.
The overall cost is thus in $\mathcal{O}(N^2)$. Recent and future algorithms 
might reduce this cost, see for instance the algorithm of \citet{aude2016stochastic}.

\noindent \begin{center}
\textsf{}
\begin{algorithm}[H]
Input: $\bw,\tilde{\bw}$, two $N$-vectors of normalized weights, 
$x,\tilde{x}$, two sets of locations in $\mathbb{X}$; a distance $d$
on $\mathbb{X}$.

Parameters: $\varepsilon>0$ for the regularization, $n$ for the
number of iterations. 

The element-wise division between two vectors is denoted by $\slash$, $\mathds{1}$ is a column vector of ones.
\begin{enumerate}
\item \textsf{Compute the pairwise distances $D=(D^{ij})$ where $D^{ij}=d(x^{i},\tilde{x}^{j})$.}
\item \textsf{Compute $K=\exp\left(-D/\varepsilon\right)$ (element-wise) and set $v^{(0)}=\mathds{1}$.}
\item \textsf{For $i\in\left\{ 1,\ldots,n\right\} $,}
\begin{enumerate}
\item \textsf{$u^{(i)}\leftarrow \bw\slash\left(Kv^{(i-1)}\right)$,}
\item \textsf{$v^{(i)}\leftarrow\tilde{\bw}\slash\left(K^\transp u^{(i)}\right)$.}
\end{enumerate}
\item \textsf{Compute $\hat{P}$ as $\text{diag}(u^{(n)})\,K\,\text{diag}(v^{(n)})$.}
\end{enumerate}
\protect\caption{Cuturi's approximation to the optimal transport problem. \label{alg:sinkhorn}}
\end{algorithm}
\par\end{center}

There are two tuning parameters: the regularization parameter
$\varepsilon$ and the number of iterations $n$.
For $\varepsilon$, we follow \citet{cuturi2013sinkhorn} and set a small proportion
of the median of the distance matrix $D$. For instance, we can set $\varepsilon = 10\% \times \text{median}(D)$.
For the choice of $n$, we use the following adaptive criterion.

As described in Section \ref{sub:transport-resampling}, once the approximate
solution $\hat{P}$ is obtained by Algorithm \ref{alg:sinkhorn}, we need to
correct its marginals.  We compute the approximate marginals $\bu =
\hat{P}\mathds{1}$ and $\tilde{\bu} =  \hat{P}^\transp \mathds{1}$, and set 
\[\alpha =
\min_{i\in 1:N } \left( \min\left( \frac{w_i}{u_i}, \frac{\tilde w_i}{\tilde
u_i} \right) \right), \quad  \boldsymbol r = \frac{\bw-\alpha \bu}{1-\alpha}, \quad \tilde{\boldsymbol r} =
\frac{\tilde \bw-\alpha \tilde \bu}{1-\alpha}.\] The final transport probability
matrix is given by $P = \alpha \hat{P} + (1-\alpha) \boldsymbol{r} \tilde{\boldsymbol{r}}^\transp$.  When $n$
increases, $\hat{P}=P^{(n)}$ gets nearer to the regularized solution $P^\varepsilon$ which is in
$\mathcal{J}(\bw,\btw)$.  Accordingly, when $n$ increases we can take $\alpha$
close to one. This gives a heuristic approach to choose $n$: we stop
Algorithm \ref{alg:sinkhorn} when the current solution $P^{(n)} =
\text{diag}(u^{(n)})\,K\,\text{diag}(v^{(n)})$ is such that $\alpha$ computed
above is at least a certain value, for instance $90\%$. This ensures that the
transport probability matrix $\hat{P}$ is a close approximation to the regularized
transport problem.

\section{Experiments with the Rhee--Glynn smoother \label{appendix:smoother:numerics:investigation}}

We explore the sensitivity of the proposed smoother to various inputs, section by section.
The experiments are based on the hidden auto-regressive model,
with $d_x = 1$, and the data are generated with $\theta = 0.95$; except in Section \ref{sec:smoother:multimodality} 
where we use a nonlinear model.
Each experiment is replicated $R = 1,000$ times. We do not use any variance reduction technique
in this section.

\subsection{Effect of the resampling scheme \label{sec:smoother:numerics:resampling}}

First we investigate the role of the resampling scheme. 
A naive scheme is systematic resampling performed on each system
with a common uniform variable.  A second scheme is
index-coupled resampling as described in Section \ref{sub:index-coupled-resampling}.
In both cases, at the final step of the coupled conditional particle filter, we sample
two trajectory indices $(b_T,\tilde{b}_T)$ using systematic resampling with 
a common uniform variable. 

We consider a time series of length $T =
20$.  In Table \ref{table:effectresampling}, we give the average meeting time 
as a function of $N$, for both resampling schemes, with the standard deviation
between parenthesis.  First we see that the meeting time is orders of magnitude
smaller when using index-coupled resampling. Secondly, we see that the meeting time 
tends to decrease with $N$, when using index-coupled resampling, whereas it increases when
using systematic resampling.  For longer time series, we find that the
index-coupled resampling is the only viable option, and thus focus on this
scheme for the Rhee--Glynn smoother. 

\input{ar1effectresampling}

\subsection{Effect of the number of particles\label{sec:smoother:numerics:nparticles}}

We consider the effect of the number of particles $N$, on the meeting time and
on the variance of the resulting estimator.  We use a time series of length $T
= 500$, generated from the model.  As seen in the previous section, when using
index-coupled resampling, we expect the meeting time $\tau$ to occur sooner if
$N$ is larger.  On the other hand, the cost of the coupled conditional particle
filter is linear in $N$, so that the overall cost of obtaining each estimator
$H$ has expectation of order $\mathbb{E}[\tau]\times N$.  We give estimators
of this cost as a function of $N$ in Table \ref{table:effectnparticles}, as
well as the average meeting time. We see that the cost per estimator decreases
when $N$ increases, and then increases again. There seems to be an optimal
value of $N$ yielding the minimum cost.

\input{ar1effectnparticles}

\begin{figure}
  \centering
  \subfloat[Variance as a function of $t$ for different $N$.]{\includegraphics[width=0.5\textwidth]{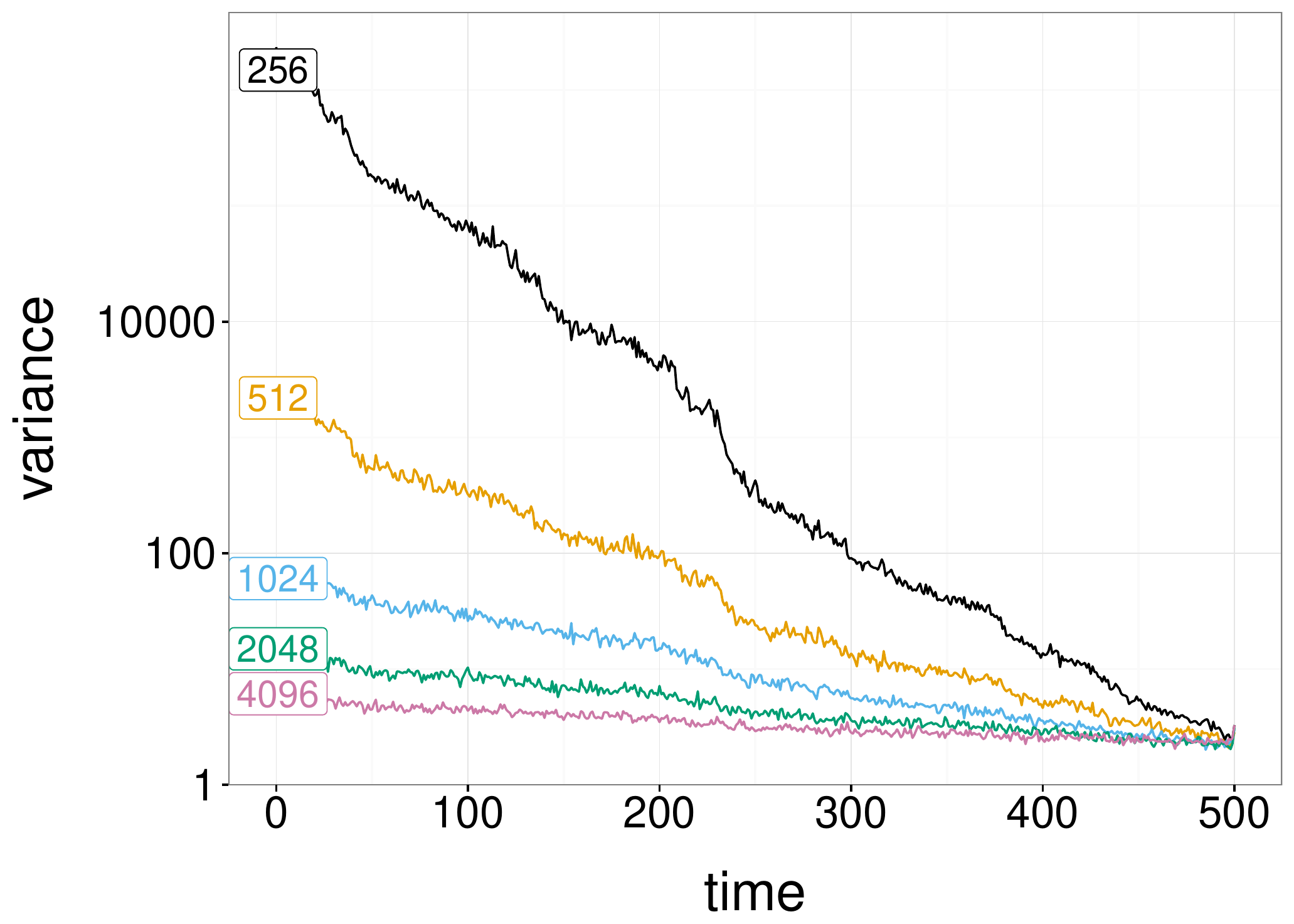}
  \label{fig:ar1effectnparticle:variance}}
  \subfloat[Efficiency as a function of $t$ for different $N$.]{\includegraphics[width=0.5\textwidth]{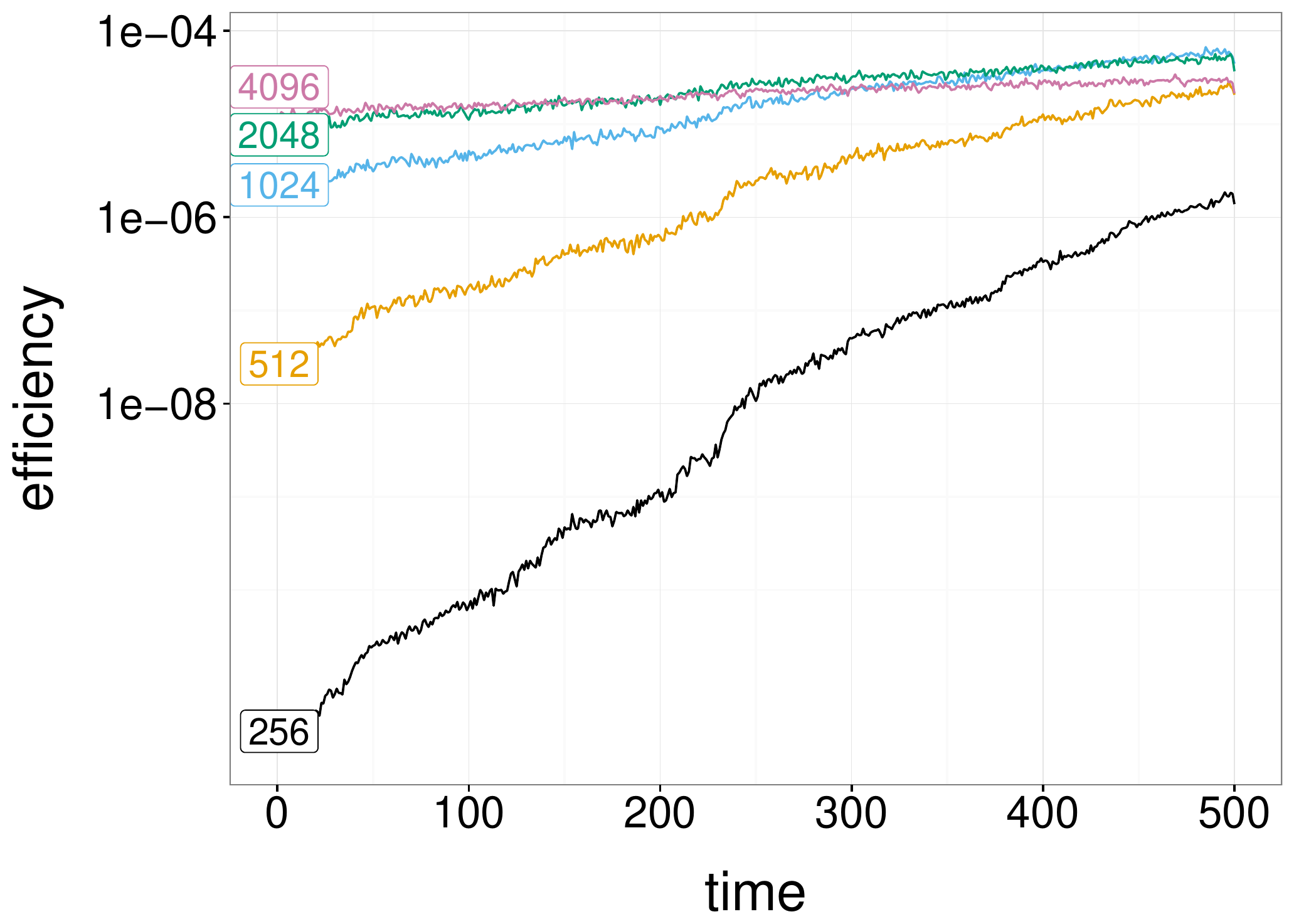}
  \label{fig:ar1effectnparticle:efficiency}}
  \caption{ Variance (left) and efficiency (right) of the estimator of the
      smoothing mean $\mathbb{E}[x_t|y_{1:T}]$ for $T = 500$, in the hidden
      auto-regressive model.  The efficiency takes into account the
      computational cost of the estimator. The y-axis is on the logarithmic scale.
      \label{fig:ar1effectnparticle:woas}}
\end{figure}

We now consider the estimators $H_{t}$ of each smoothing mean
$\mathbb{E}[x_t|y_{1:T}]$, for $t\in 0:T$, i.e. we take $h$ to be the identity
function. 
We compute the empirical variance of $H_{t}$, for each $t$, over the $R$ experiments.
To take into account both variance and computational cost, we define the efficiency as
$1/(\mathbb{V}[H_{t}]\times \mathbb{E}[\tau] \times N)$ and approximate this value for each $t$,
using the $R$ estimators.   The results are shown in Figure
\ref{fig:ar1effectnparticle:woas}.

We see that the variance explodes exponentially when $T - t$ increases (for fixed $T$ and increasing $t$;
see Section \ref{sec:smoother:numerics:horizon} for the behaviour with $T$).
From Figure \ref{fig:ar1effectnparticle:variance}, the variance is reduced when larger values of $N$ are used.
Secondly, the variance is most reduced for the estimators of the first smoothing means, 
i.e.  $\mathbb{E}[x_t|y_{1:T}]$ for small $t$. As such, the efficiency is maximized 
for the largest values of $N$ only when $t$ is small, as can be seen from Figure \ref{fig:ar1effectnparticle:efficiency}.
For values of $t$ closer to $T$, the efficiency is higher for $N = 1,024$ and $N = 2,048$ than it is for $N = 4,096$.

\subsection{Effect of the truncation variable \label{sec:smoother:numerics:truncation}}

We now consider the use of Geometric truncation variables,
as introduced in the Appendix \ref{sec:appendix:validitity:RG}.
We set $T = 500$ and $N = 512$. We try a few values of the Geometric
probability $p$, in an attempt to reduce the computation cost per estimator.
The value $p=0$ corresponds to the estimator proposed in Section \ref{sec:methods:smoother}.
The average meeting times are shown in Table \ref{table:effecttruncation}.

\input{ar1effecttruncation}

\begin{figure}
  \centering
  \subfloat[Variance as a function of $t$ for various $p$.]{\includegraphics[width=0.5\textwidth]{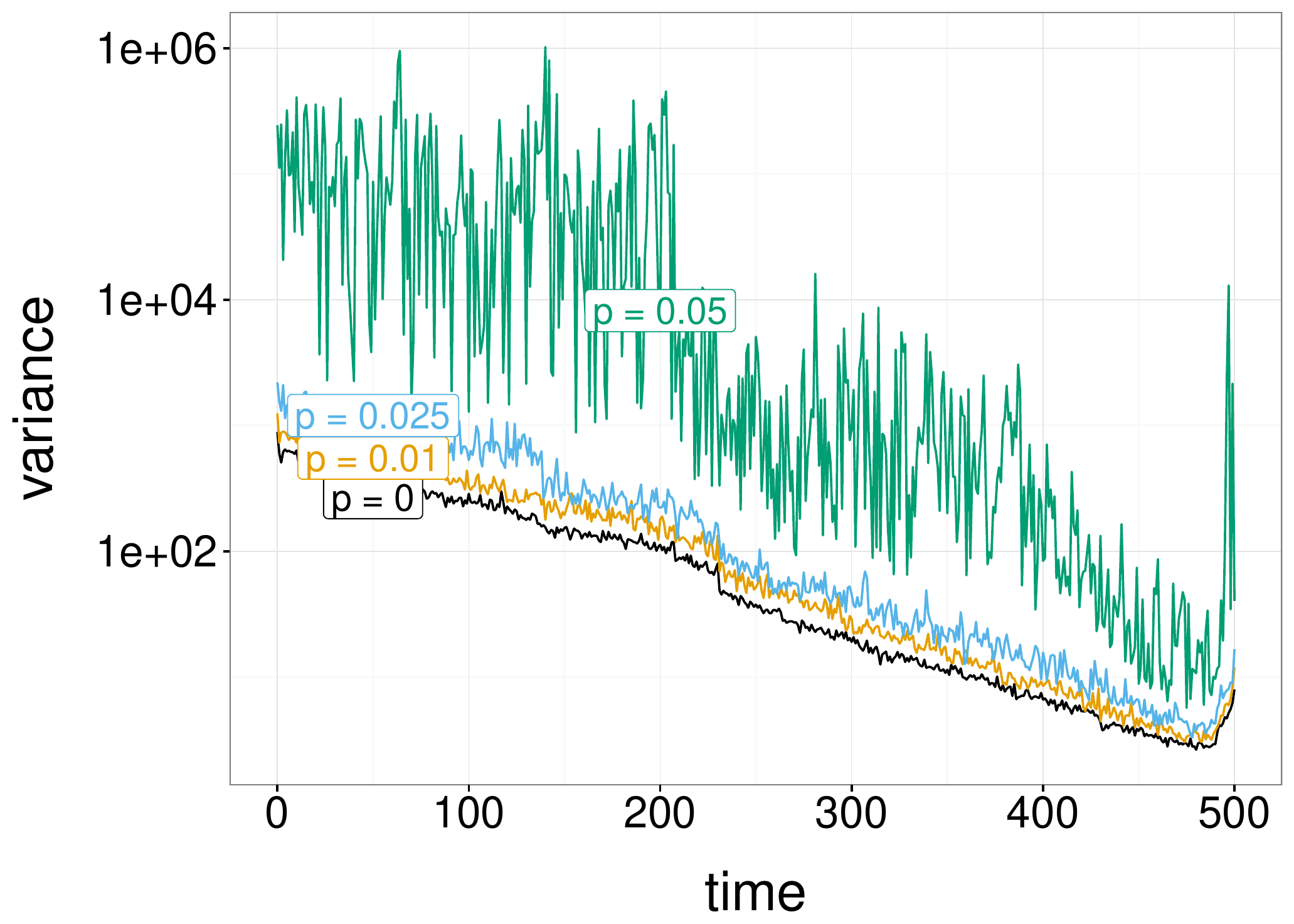}
  \label{fig:ar1effecttruncation:variance}}
  \subfloat[Efficiency as a function of $t$ for various $p$.]{\includegraphics[width=0.5\textwidth]{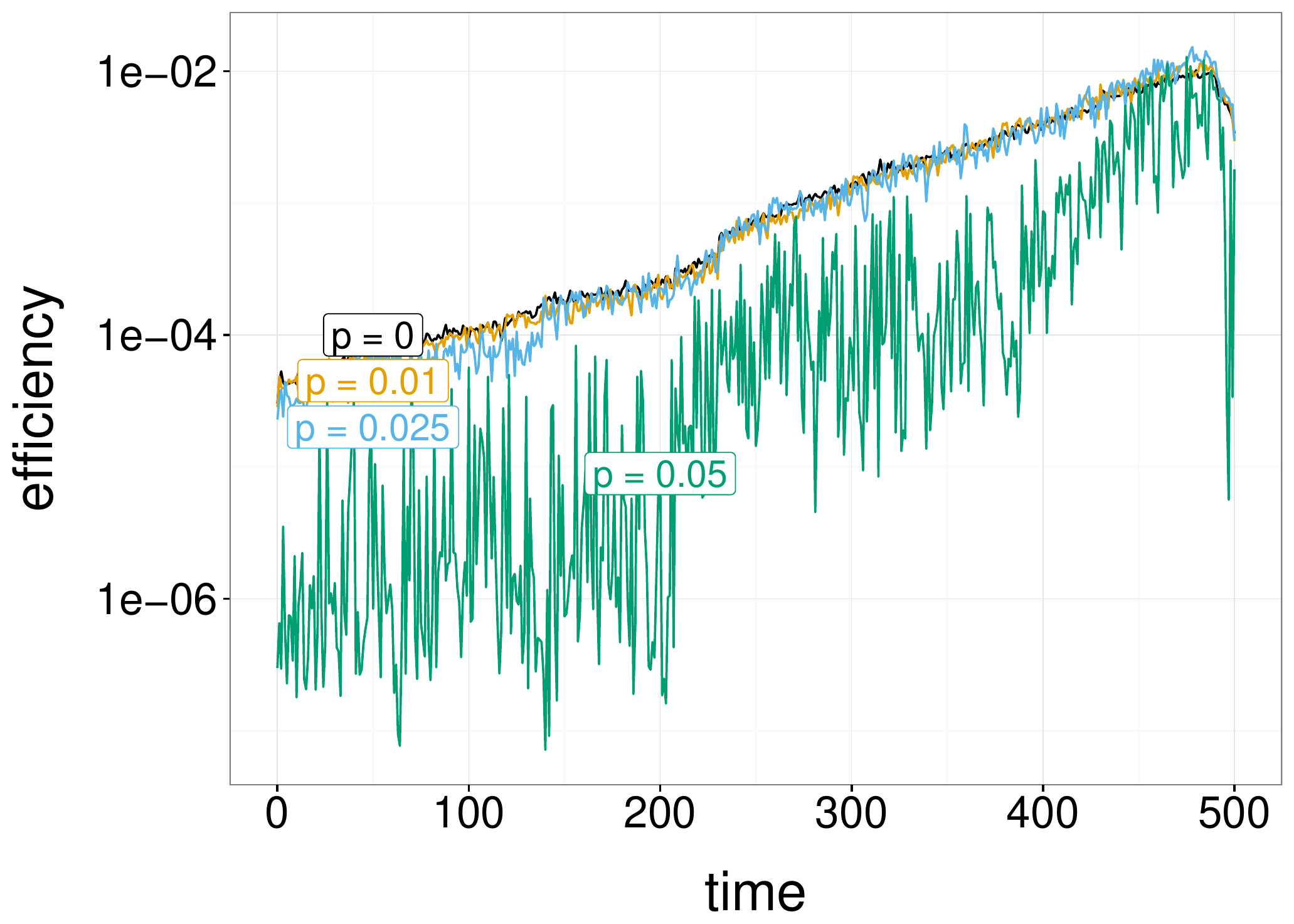}
  \label{fig:ar1effecttruncation:efficiency}}
  \caption{ Variance (left) and efficiency (right) of the smoothing estimator using a Geometric
      truncation variable with parameter $p$, for $T = 500$, and $N=512$, in the hidden auto-regressive model.
  The efficiency takes into account the computational cost of the estimator.  The y-axis is on the logarithmic scale. \label{fig:ar1effecttruncation}}
\end{figure}

We plot the variance of the estimator of the smoothing mean
$\mathbb{E}[x_t|y_{1:T}]$ against $t$ on Figure \ref{fig:ar1effecttruncation:variance}, for each $p$.  We plot the
efficiency $\mathbb{E}[\min(G,\tau)] \times \mathbb{V}[H_{t}]$ against
$t$ on Figure \ref{fig:ar1effecttruncation:efficiency}, for each $p$.
First, from Figure \ref{fig:ar1effecttruncation:variance}, we see that 
increasing $p$ leads to a higher variance. In
particular, the value $p = 0.05$ leads to a much larger variance than the other values. 
This seems to be in agreement with Assumption \ref{assumption:truncation},
which states that $p$ has to be below a certain threshold related to the meeting probability.

On Figure \ref{fig:ar1effecttruncation:efficiency}, we see that 
the increase of variance is compensated by a reduction of the
computation cost, for the smaller values of $p$. Therefore,
the three smaller values lead to the same overall efficiency.
On the other hand, the largest value $p=0.05$ leads to a significantly lower efficiency.  
Thus, there does not seem to be much benefit in using $p\neq 0$ in this example.

\subsection{Effect of ancestor sampling \label{sec:smoother:numerics:ancestorsampling}}

We consider the use of ancestor sampling, which requires being able to evaluate the transition density,
$f(x_t|x_{t-1},\theta)$, for all $x_{t-1},x_t$ and all $\theta$.
We set $T = 500$ as before, and consider different values of $N$. The average meeting times 
are displayed in Table \ref{table:effectancestor}.
We see that the meeting times are significantly reduced by using ancestor sampling, especially for smaller numbers of particles.

\input{ar1effectancestor}

\begin{figure}
  \centering
  \subfloat[Variance as a function of $t$ for different $N$.]{\includegraphics[width=0.5\textwidth]{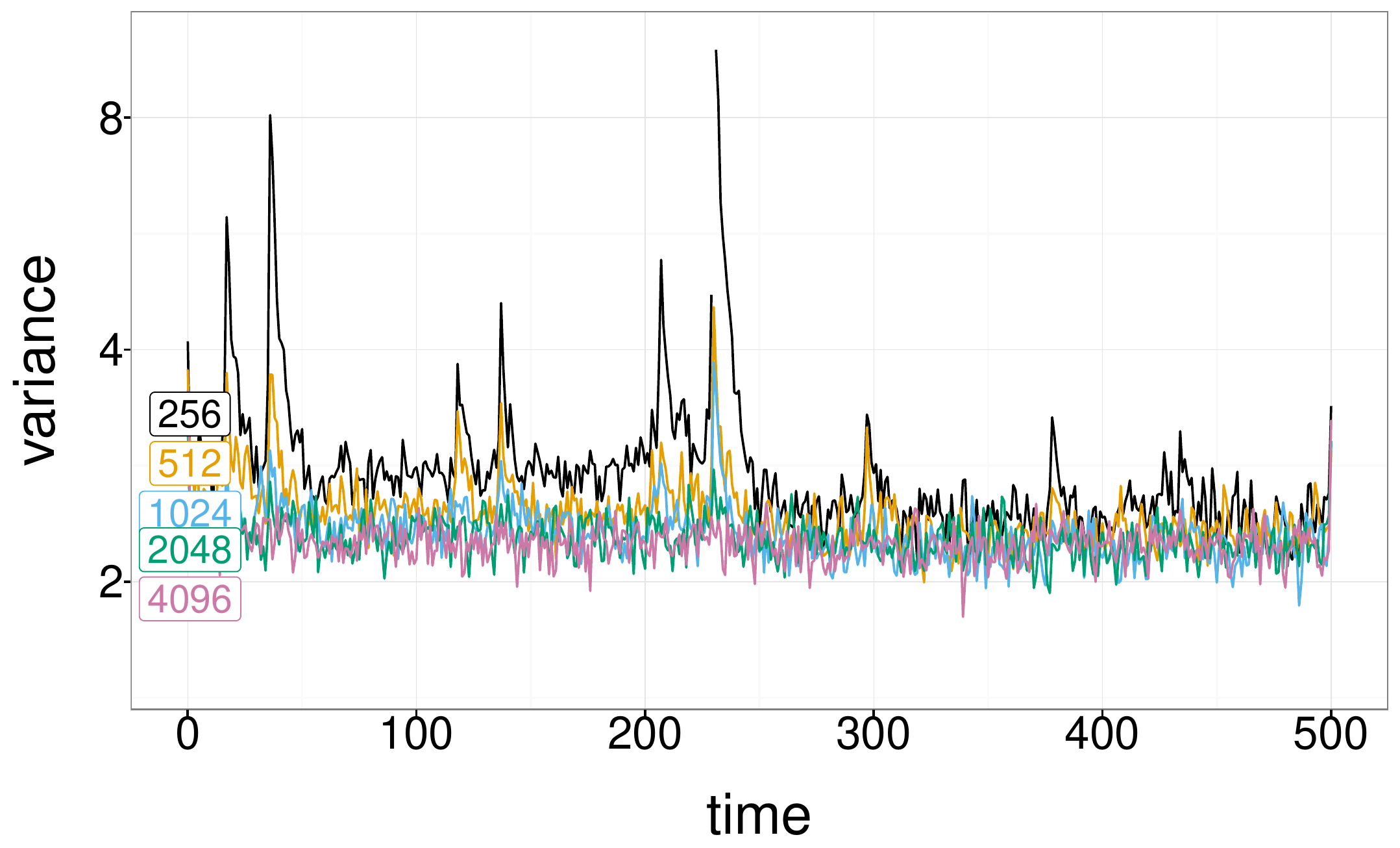}
  \label{fig:ar1effectnparticle:variance:was}}
  \subfloat[Efficiency as a function of $t$ for different $N$.]{\includegraphics[width=0.5\textwidth]{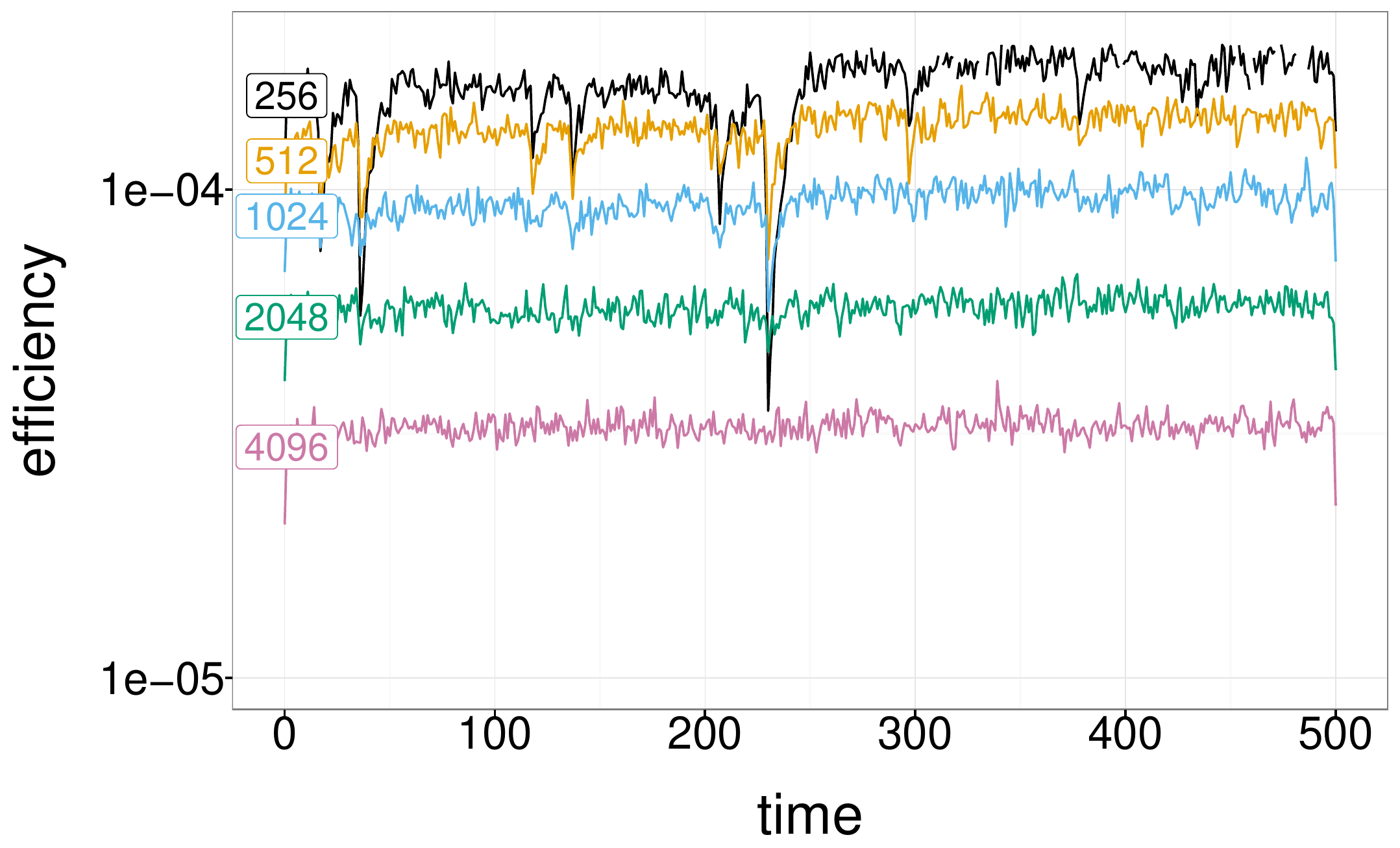}
  \label{fig:ar1effectnparticle:efficiency:was}}
  \caption{ Variance (left) and efficiency (right) of the estimator of the
      smoothing mean $\mathbb{E}[x_t|y_{1:T}]$ for $T = 500$, in the hidden
      auto-regressive model, when using ancestor sampling.  The efficiency takes into account the
      computational cost of the estimator. The y-axis is on the logarithmic scale. 
      \label{fig:ar1effectnparticle:was}}
\end{figure}

We consider variance and efficiency, here defined as $1/(\mathbb{V}[H_{t}]\times \mathbb{E}[\tau] \times N)$.
The results are shown in Figure \ref{fig:ar1effectnparticle:was}.
This is to be compared with Figure \ref{fig:ar1effectnparticle:woas} obtained without ancestor sampling.

First we see that the variance is significantly reduced by ancestor sampling. 
The variance seems to increase only slowly as $T - t$ increases, for each value of $N$. 
From Figure \ref{fig:ar1effectnparticle:efficiency:was}, we see that the 
smallest value of $N$ now leads to the most efficient algorithm. In other words, 
for a fixed computational budget, it is more efficient to produce 
more estimators with $N=256$ than to increase the number of particles
and to average over fewer estimators.

\subsection{Effect of the time horizon \label{sec:smoother:numerics:horizon}}

We investigate the effect of the time horizon $T$, that is, the total length of
the time series, on the performance of the smoother. We expect the conditional
particle filter kernel to perform less and less well when $T$ increases. To compensate for
this loss of efficiency, we increase the number of particles $N$ linearly with $N$: for $T = 64$ we use $N
= 128$, for $T = 128$ we use $N = 256$, and so forth up to $T = 1,024$ and $N = 2,048$. With that scaling, the
computational cost of each run of coupled conditional particle filter is
quadratic in $T$.  A first question is whether the meeting time is then stable
with $T$.  Table \ref{table:effecthorizon} reports the average meeting times
obtained when scaling $N$ linearly with $T$. We see that the meeting times
occur in roughly the same number of steps, implying
that the linear scaling of $N$ with $T$ is enough.

\input{tablear1horizon}

\begin{figure}
  \centering
  \subfloat[Without ancestor sampling.]{\includegraphics[width=0.5\textwidth]{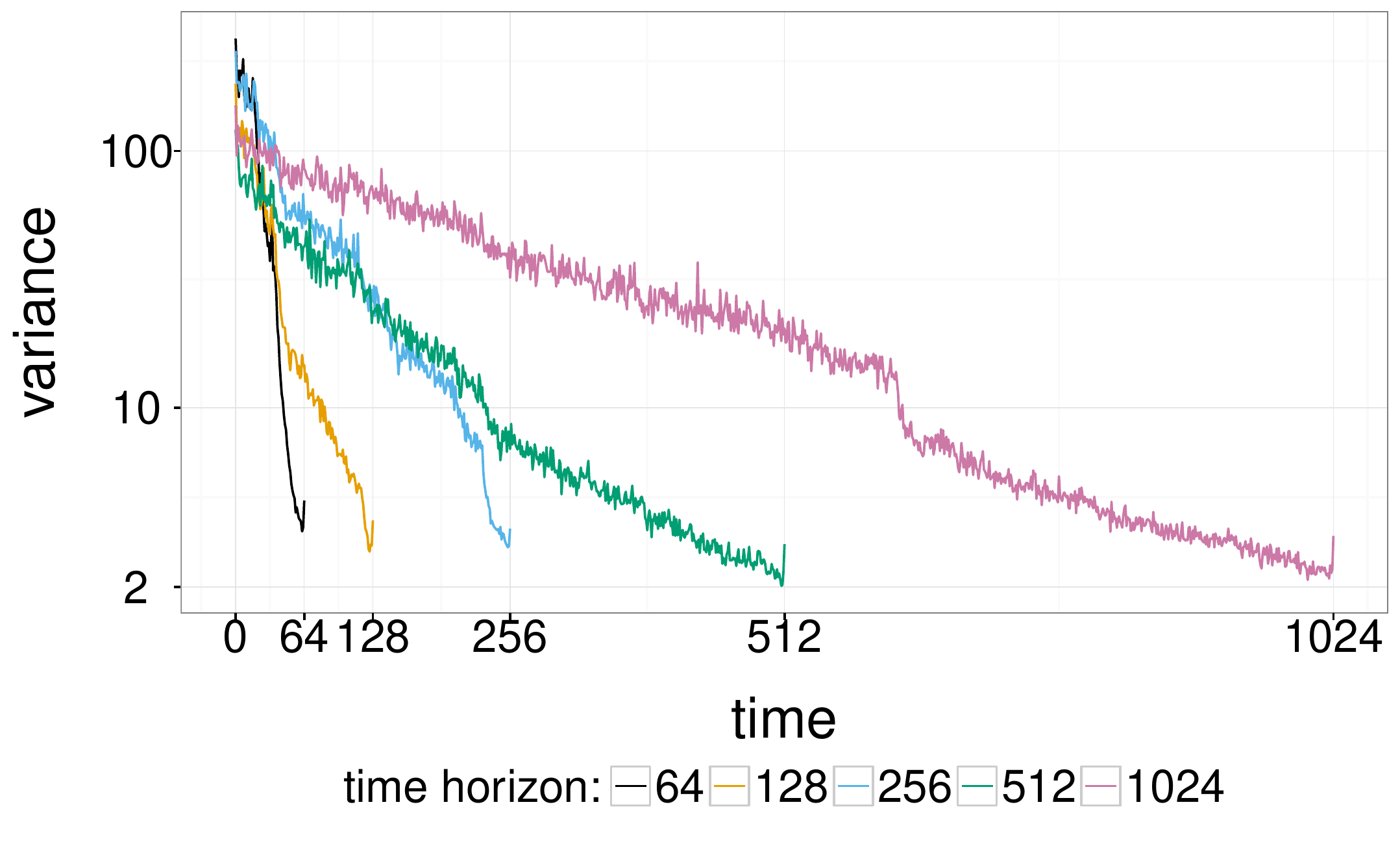}
  \label{fig:ar1effecthorizon:variance:woas}}
  \subfloat[With ancestor sampling.]{
      \includegraphics[width=0.5\textwidth]{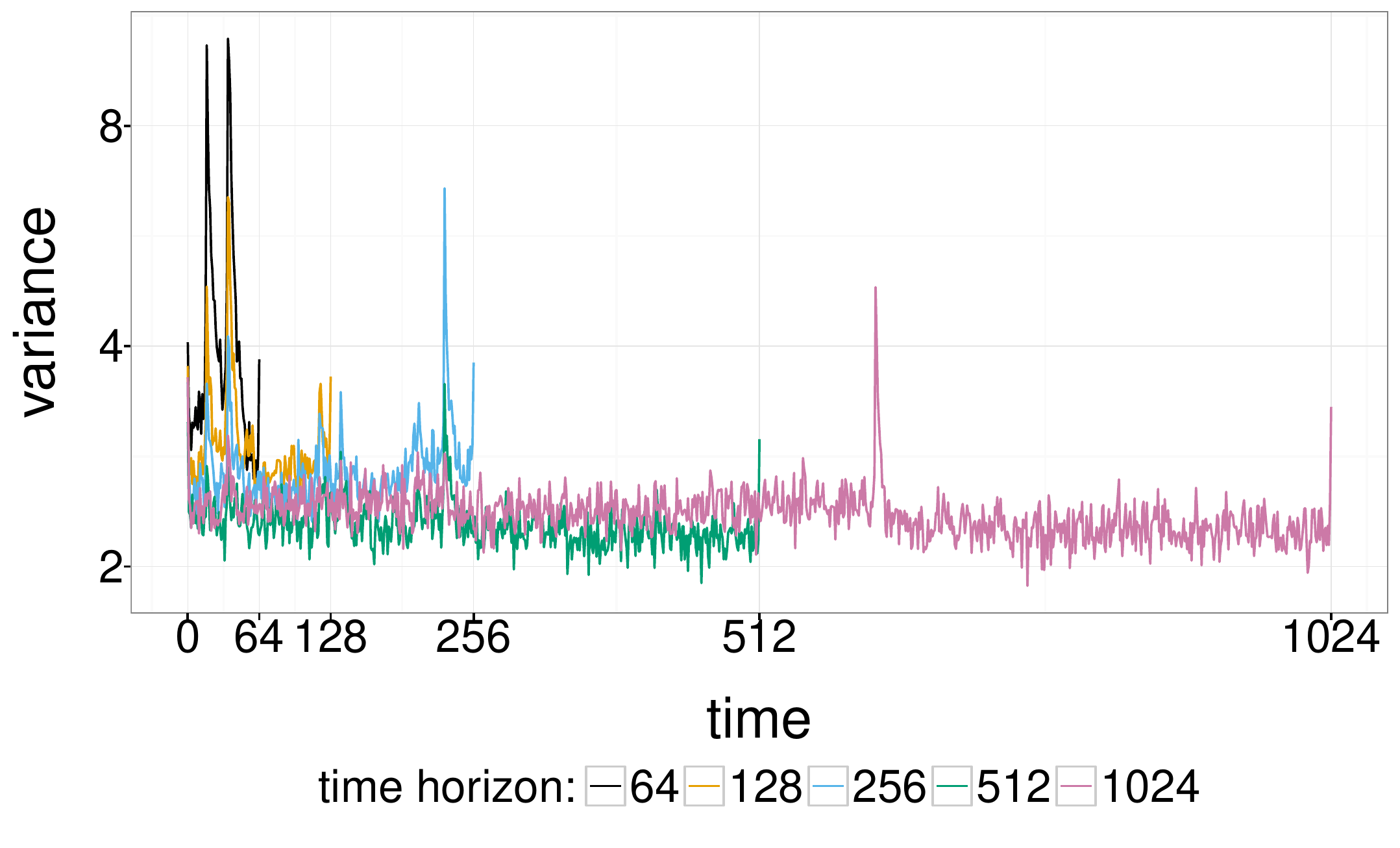}
  \label{fig:ar1effecthorizon:variance:was}}
  \caption{ Variance of the estimator of the
      smoothing mean $\mathbb{E}[x_t|y_{1:T}]$ for various time horizons,
      without (left) and with (right) ancestor sampling, in the hidden
      auto-regressive model.  The y-axis is on the logarithmic scale. 
      \label{fig:ar1effecthorizon}}
\end{figure}

A second question is whether scaling $N$ linearly with $T$ is enough to ensure
that the variance of the resulting estimator is stable.  Results are shown in
Figure \ref{fig:ar1effecthorizon}, obtained without (Figure
\ref{fig:ar1effecthorizon:variance:woas}) and with ancestor sampling (Figure
\ref{fig:ar1effecthorizon:variance:was}).  The plots show the variance of the
estimator of the smoothing means $\mathbb{E}[x_t|y_{1:T}]$ for all $t\leq T$
and various $T$.  We see that, for the values of $t$ that are less than all the time horizons,
the variance of the estimators  of
$\mathbb{E}[x_t|y_{1:T}]$ seems stable with $T$.
The experiments thus indicate that, to estimate
$\mathbb{E}[x_t|y_{1:T}]$ for all $t$, one can scale $N$ linearly in $T$
and expect the meeting time and the variance of the Rhee--Glynn estimators to be stable. 
Overall, the computational cost is then quadratic in $T$.

\subsection{Effect of multimodality in the smoothing distribution \label{sec:smoother:multimodality}}

\begin{figure}
  \centering
  \subfloat[Approximation of the smoothing distribution,
  using a conditional particle filter, at $t = 36$.]{\includegraphics[width=0.45\textwidth]{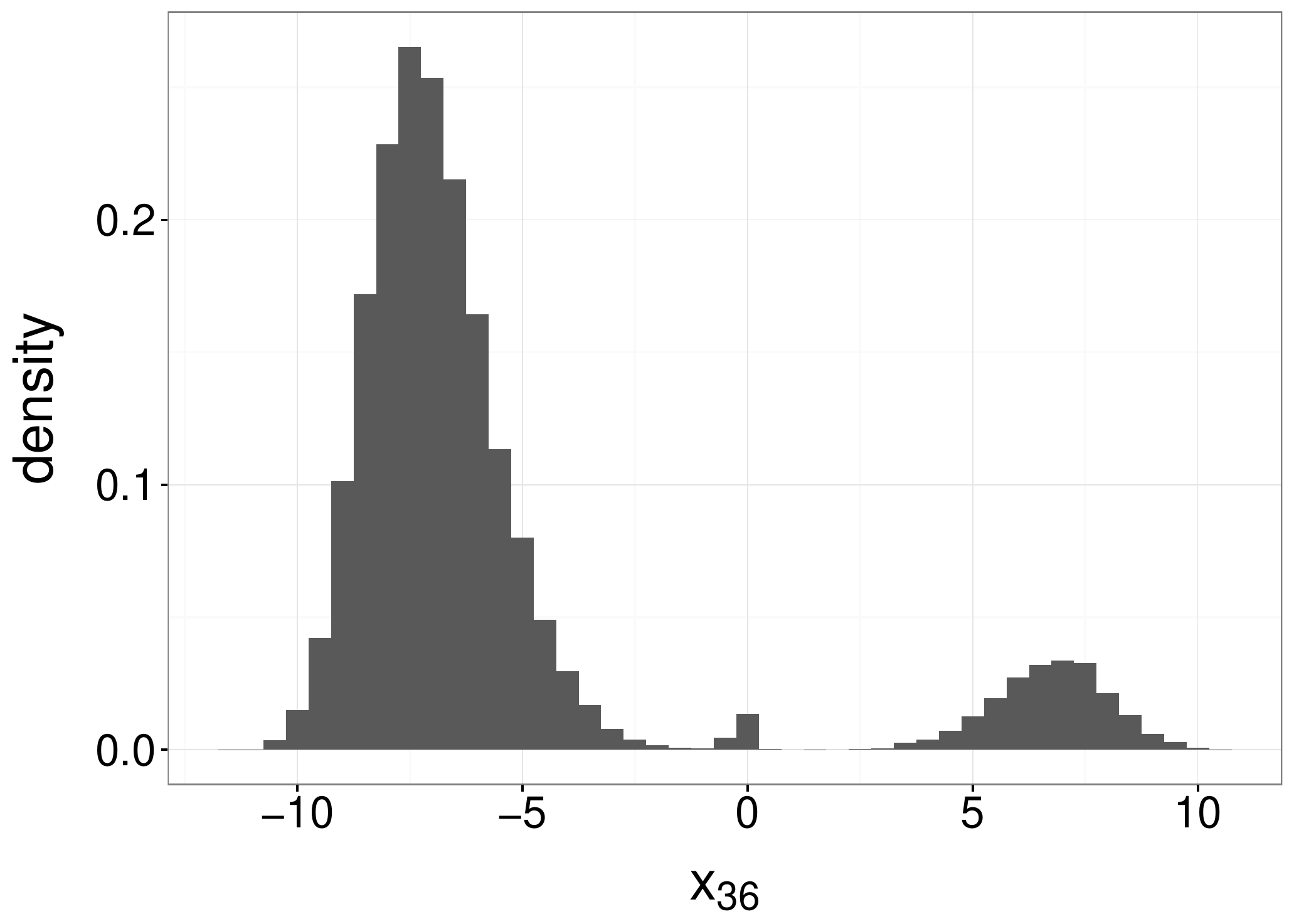}
  \label{fig:gss:pgas}}
  \subfloat[Rhee--Glynn estimators of the smoothing mean,
  and true mean in vertical dashed (red) line, at $t = 36$.]{\includegraphics[width=0.45\textwidth]{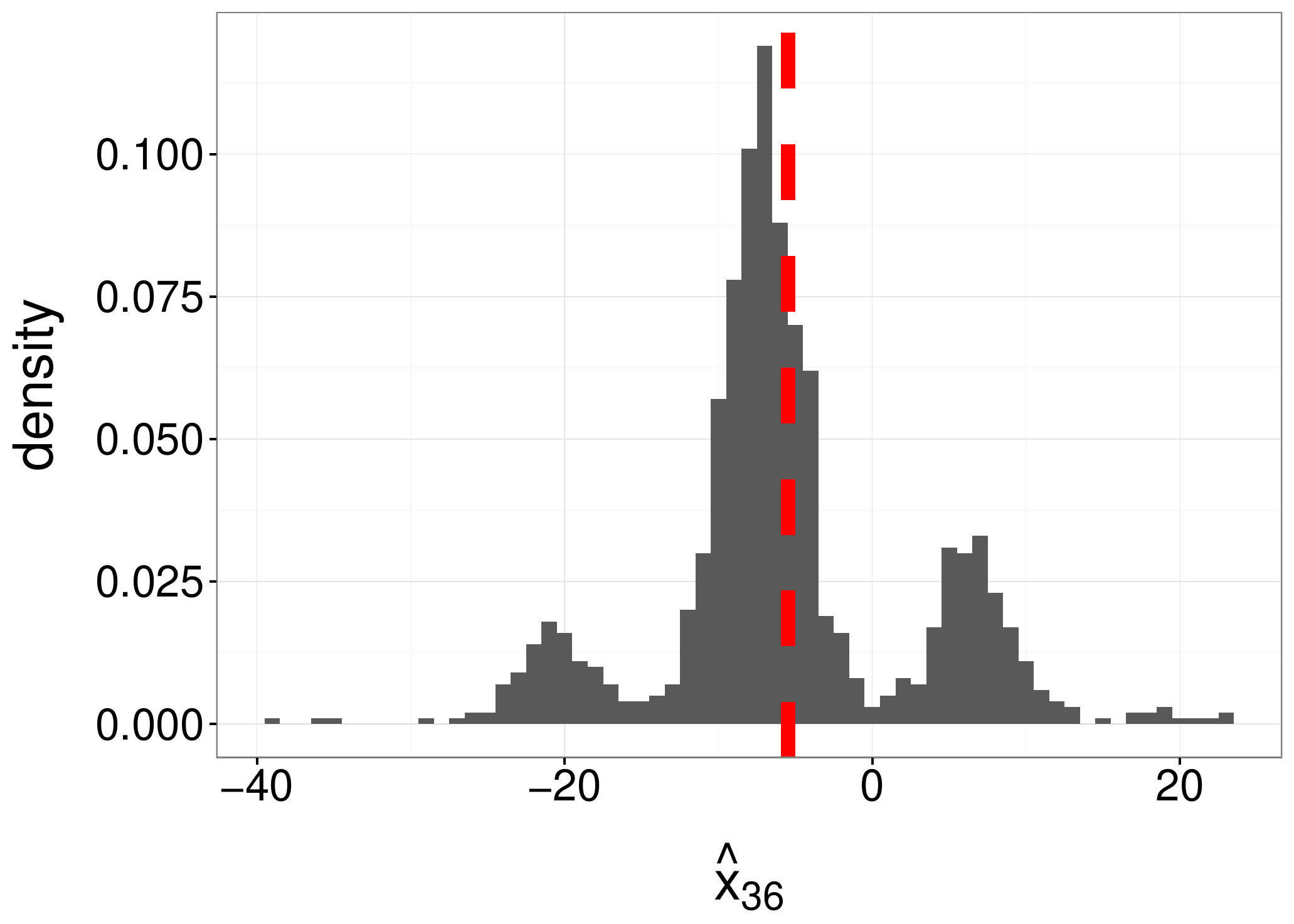}
  \label{fig:gss:rheeglynn}}
  \caption{ Smoothing distribution approximated by conditional particle filters (left), and $R=1,000$ independent Rhee--Glynn estimators
      of the smoothing mean (right), at time  $t = 36$ for the nonlinear growth model with $T = 50$.
      \label{fig:gss:x36} }
\end{figure}

We switch to another model to investigate the behaviour of the Rhee--Glynn estimator
when the smoothing distribution is multimodal.
We consider the nonlinear growth model used by \cite{gordon:salmon:smith:1993}.
We set $x_0 \sim \mathcal{N}(0,2)$, and, for $t\geq 1$,
    \[x_t = 0.5x_{t-1} + 25 x_{t-1} / (1 + x_{t-1}^2) + 8 \cos (1.2(t-1)) + W_{t},\quad \text{and} \quad
    y_t = x_{t-1}^2 / 20 + V_{t},\]
where $W_t$ and $V_t$ are independent normal variables, with variances $1$ and $10$
respectively.  We generate $T = 50$ observations using $x_0 = 0.1$, following
\cite{gordon:salmon:smith:1993}.  Because the measurement distribution
$g(y_t|x_t,\theta)$ depends on $x_t$ through $x_t^2$, the sign of $x_t$ is hard
to identify, and as a result the smoothing distribution has multiple modes.  We
run a conditional particle filter with ancestor sampling, with $N =
1,024$ particles for $M = 50,000$ iterations, and discard the first $25,000$
iterations.  We plot the histogram of the obtained sample for $p(dx_t|y_{1:T},\theta)$
at time $t = 36$ in Figure \ref{fig:gss:pgas}. We notice at least two modes,
located around $-7$ and $+7$, with possibly an extra mode near zero.

We run the Rhee--Glynn smoother with $N = 1,024$ and ancestor sampling.  Each estimator took
less than $10$ iterations of the coupled
conditional particle filter to meet, with a median meeting time of $3$
iterations. The total number of calls to the coupled conditional particle
filter to obtain $R = 1,000$ estimators adds up to $2,984$.  We plot the
histogram of the estimators $H_t^{(r)}$, for $r\in1:R$, of the smoothing mean $\mathbb{E}[x_{t}|y_{1:T}]$ at time $t = 36$
in Figure \ref{fig:gss:rheeglynn}.  We see that the distribution of the
estimator is itself multimodal. Indeed, the two initial reference trajectories might
belong to the mode around $-7$, or to the mode around $+7$, or each trajectory
might belong to a different mode.  Each of these cases leads to a mode in 
the distribution of the Rhee--Glynn estimator.

\begin{figure}
    \centering
    \includegraphics[width=0.8\textwidth]{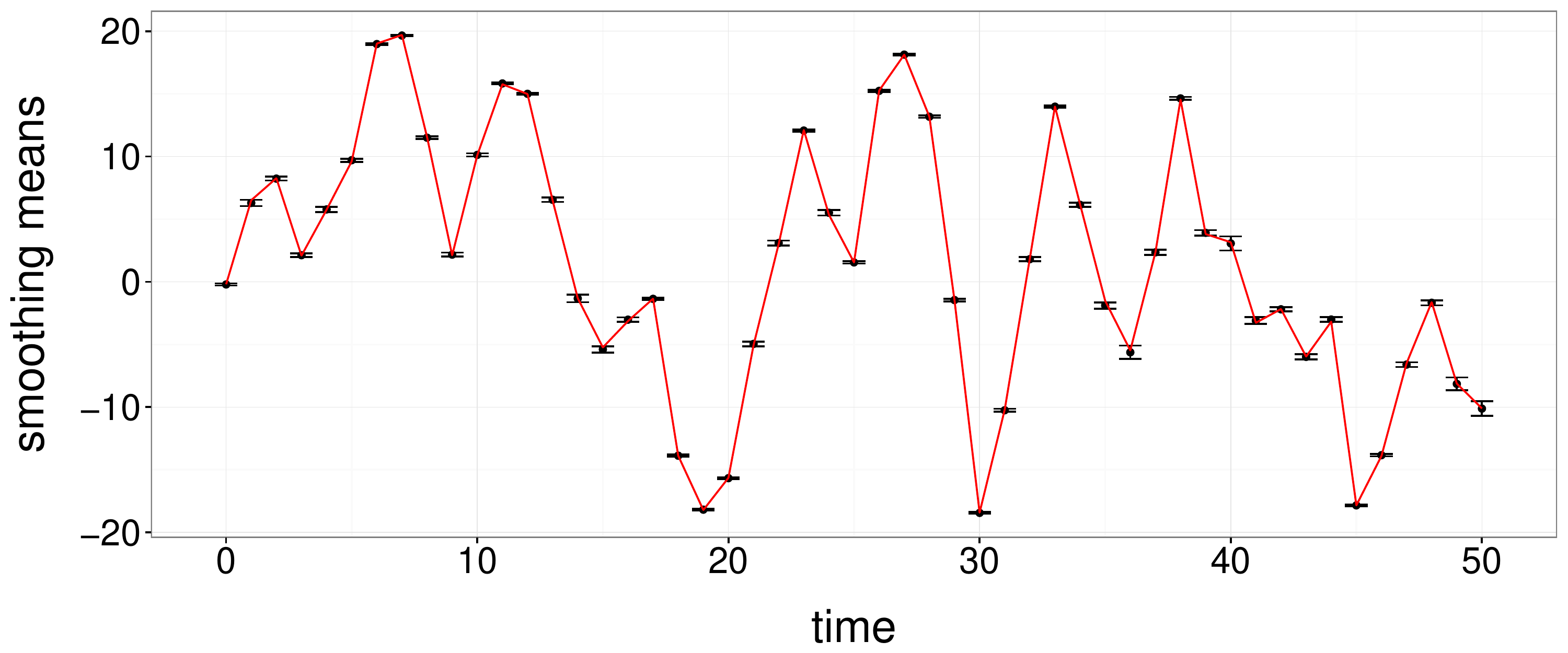}
    \caption{Confidence intervals around the exact smoothing means.
        The intervals are computed as two standard deviations around the mean of
        $R = 1,000$ proposed smoothing estimators. The line represents the exact smoothing means, retrieved by a long run of conditional particle filter,
    for the nonlinear growth model with $T = 50$ observations.}
    \label{fig:gss:smoothingestimators}
\end{figure}

The resulting estimator $\hat{x}_{t}$ of each smoothing mean is obtained by
averaging the $R = 1,000$ independent estimators $H_t^{(r)}$.  We compute the
Monte Carlo standard deviation $\hat\sigma_t$ at each time $t$, and represent
the confidence intervals $[\hat{x}_t - 2\hat{\sigma}_t/\sqrt{R}, \hat{x}_t +
2\hat{\sigma}_t/\sqrt{R}]$ as error bars in Figure \ref{fig:gss:smoothingestimators}.
The line represents the smoothing means obtained by conditional particle filter
with ancestor sampling, taken as ground truth. The agreement shows that the proposed
method is robust to multimodality in the smoothing distribution. 

\section{Pseudo-code for particle filters \label{appendix:pseudocode:pf}}

We provide pseudo-code for the bootstrap particle filter (Algorithm \ref{alg:Particle-Filter}),
the conditional particle filter (Algorithm \ref{alg:conditional-particle-filter}),
the coupled bootstrap particle filter (Algorithm \ref{alg:coupled-particle-filter}),
and the coupled conditional particle filter (Algorithm \ref{alg:coupled-conditional-particle-filter}).

\noindent \begin{center}
\textsf{}
\begin{algorithm}
\textsf{At step $t=0$.}
\begin{enumerate}
\item \textsf{Draw $x_{0}^{k}\sim m_0(dx_0|\theta)$, for all $k\in 1:N$.}\\
\textsf{This can also be written $x_{0}^{k} = M(U_0^k,\theta)$, for all $k\in 1:N$.}
\item \textsf{Set $w_{0}^{k}=N^{-1}$, for all $k\in 1:N$.}
\end{enumerate}
\textsf{At step $t\geq1$.}
\begin{enumerate}
\item \textsf{Draw ancestors $a_{t-1}^{1:N}\sim r(da^{1:N}| w_{t-1}^{1:N})$.}
\item \textsf{Draw $x_{t}^{k}\sim f(dx_{t}| x_{t-1}^{a_{t-1}^{k}},\theta)$, for all $k\in 1:N$.}\\
    \textsf{This can also be written $x_{t}^{k} = F(x_{t-1}^{a_{t-1}^k}, U_t^k,\theta)$, for all $k\in 1:N$.}
\item \textsf{Compute $w_{t}^{k}\propto g(y_{t}| x_{t}^{k},\theta)$, for all $k\in 1:N$,
and normalize the weights.}
\end{enumerate}
\textsf{Return the likelihood estimator $\hat p^N\left(y_{1:T}|\theta\right) = \prod_{t=1}^{T}N^{-1}\sum_{k=1}^{N}g(y_t|x_t^k,\theta)$}.
\protect\caption{Bootstrap particle filter, given a parameter $\theta$. \label{alg:Particle-Filter}}
\end{algorithm}
\par\end{center}

\noindent \begin{center}
    \textsf{}
    \begin{algorithm}
        \textsf{At step $t=0$.}
        \begin{enumerate}
            \item \textsf{Draw $x_{0}^{k} \sim m_0(dx_0 | \theta)$, for $k\in 1:N-1$, and set  $x_0^{N} = x_0$.}\\
                \textsf{This can also be written $x_{0}^{k} = M(U_0^k,\theta)$, for all $k\in 1:N-1$, and $x_0^{N} = x_0$.}
            \item \textsf{Set $w_{0}^{k}=N^{-1}$, for $k\in 1:N$.}
        \end{enumerate}
        \textsf{At step $t\geq 1$.}
        \begin{enumerate}
            \item \textsf{Draw ancestors $a_{t-1}^{1:N-1}\sim r(da^{1:N-1}| w_{t-1}^{1:N})$, and set $a_{t-1}^N = N$.}
            \item \textsf{Draw $x_{t}^{k}\sim f(dx_{t}| x_{t-1}^{a_{t-1}^{k}},\theta)$, for all $k\in 1:N-1$, and set  $x_t^{N} = x_t$.}\\
                \textsf{This can also be written $x_{t}^{k} = F(x_{t-1}^{a_{t-1}^k}, U_t^k,\theta)$, for all $k\in 1:N-1$, and $x_t^{N} = x_t$.}
            \item \textsf{Compute $w_{t}^{k}\propto g(y_{t}| x_{t}^{k},\theta)$, for all $k\in 1:N$,
            and normalize the weights.}
    \end{enumerate}
    \textsf{Draw a trajectory.}
    \begin{enumerate}
        \item  \textsf{Draw $b_T$ from a discrete distribution on $1:N$, with probabilities $w^{1:N}_{T}$.}
        \item  \textsf{For $t= T-1,\ldots,0$, set $b_t = a_{t}^{b_{t+1}}$.}
    \end{enumerate}
    \textsf{Return $x_{0:T}^\prime = (x_0^{b_0}, \ldots, x_T^{b_T})$. }
    \protect\caption{Conditional particle filter, given a reference trajectory $x_{0:T}$ and $\theta$. \label{alg:conditional-particle-filter}}
\end{algorithm}

\par\end{center}

\noindent \begin{center}
\textsf{}
\begin{algorithm}
\textsf{At step $t=0$.}
\begin{enumerate}
    \item \textsf{Draw $U_{0}^{k}$, and compute $x_{0}^{k} = M(U_0^k, \theta)$ and $\tx_{0}^{k} = M(U_0^k, \ttheta)$, for all $k\in 1:N$.}
    \item \textsf{Set $w_{0}^{k}=N^{-1}$ and $\tw_{0}^{k}=N^{-1}$, for all $k\in 1:N$.}
\end{enumerate}
\textsf{At step $t\geq1$.}
\begin{enumerate}
    \item \textsf{Compute a probability matrix $P_{t-1}$, 
        with marginals $w_{t-1}^{1:N}$ and $\tw_{t-1}^{1:N}$. Sample $(a_{t-1}^k, \tilde{a}_{t-1}^k)$ from $P_{t-1}$, for all $k\in 1:N$.}
    \item \textsf{Draw $U_{t}^k$, and compute $x_{t}^{k} = F(x_{t-1}^{a_{t-1}^{k}}, U_{t}^k, \theta)$ and $\tx_{t}^{k} = F(\tx_{t-1}^{\tilde{a}_{t-1}^{k}}, U_{t}^k, \ttheta)$, for all $k\in 1:N$.}
    \item \textsf{Compute $w_{t}^{k}\propto g(y_{t}| x_{t}^{k},\theta)$ and $\tw_{t}^{k}\propto g(y_{t}| \tx_{t}^{k}, \ttheta)$, for all $k\in 1:N$, and normalize the weights.}
\end{enumerate}
\textsf{Return  $\hat p^N\left(y_{1:T}|\theta\right) =
  \prod_{t=1}^{T}N^{-1}\sum_{k=1}^{N}g(y_t|x_t^k,\theta)$ and 
  $\hat p^N(y_{1:T}|\ttheta) = \prod_{t=1}^{T}N^{-1}\sum_{k=1}^{N}g(y_t|\tx_t^k,\ttheta)$}.
\protect\caption{Coupled bootstrap particle filter, given parameters $\theta$ and $\ttheta$. \label{alg:coupled-particle-filter}}
\end{algorithm}
\par\end{center}

\noindent \begin{center}
    \textsf{}
    \begin{algorithm}
        \textsf{At step $t=0$.}
        \begin{enumerate}
            \item \textsf{Draw $U_{0}^{k}$, compute $x_{0}^{k} = M(U_0^k,\theta)$ and $\tx_{0}^{k} = M(U_0^k,\theta)$ for $k\in 1:N-1$.}
            \item \textsf{Set  $x_0^{N} = x_0$ and $\tx_0^N=\tx_0$.}
            \item \textsf{Set $w_{0}^{k}=N^{-1}$ and $\tw_{0}^{k}=N^{-1}$, for $k\in 1:N$.}
        \end{enumerate}
        \textsf{At step $t\geq 1$.}
        \begin{enumerate}
            \item \textsf{Compute a probability matrix $P_{t-1}$, with marginals $w_{t-1}^{1:N}$ and $\tw_{t-1}^{1:N}$. Sample $(a_{t-1}^k, \tilde{a}_{t-1}^k)$ from $P_{t-1}$, for all $k\in 1:N-1$. Set $a_{t-1}^N = N$ and $\ta_{t-1}^N = N$.}
            \item \textsf{Draw $U_{t}^k$, and compute $x_{t}^{k} = F(x_{t-1}^{a_{t-1}^{k}}, U_{t}^k, \theta)$ and $\tx_{t}^{k} = F(\tx_{t-1}^{\tilde{a}_{t-1}^{k}}, U_{t}^k, \theta)$, for all $k\in 1:N-1$. Set  $x_t^{N} = x_t$ and $\tx_t^{N} = \tx_t$.}
            \item \textsf{Compute $w_{t}^{k}\propto g(y_{t}| x_{t}^{k},\theta)$ and
                $\tw_{t}^{k}\propto g(y_{t}| \tx_{t}^{k},\ttheta)$, for all $k\in 1:N$,
            and normalize the weights.}
    \end{enumerate}
    \textsf{Draw a pair of trajectories.}
    \begin{enumerate}
        \item \textsf{Compute a probability matrix $P_{T}$, with marginals $w_{T}^{1:N}$ and $\tw_{T}^{1:N}$. Draw  $(b_{T}, \tilde{b}_{T})$ from $P_{T}$.}
            \item  \textsf{For $t= T-1,\ldots,0$, set $b_t = a_{t}^{b_{t+1}}$ and $\tilde{b}_t = \ta_{t}^{\tilde{b}_{t+1}}$.}
    \end{enumerate}
    \textsf{Return $x_{0:T}^\prime = (x_0^{b_0}, \ldots, x_T^{b_T})$ and $\tx_{0:T}^\prime = (\tx_0^{\tilde{b}_0}, \ldots, \tx_T^{\tilde{b}_T})$. }
    \protect\caption{Coupled conditional particle filter, given reference trajectories $x_{0:T}$ and $\tx_{0:T}$. \label{alg:coupled-conditional-particle-filter}}
\end{algorithm}

\par\end{center}

\end{document}

%% file: ar1effectresampling.tex
\begin{table}[ht]
\centering
\begin{tabular}{rll}
  \hline
 & systematic & index-coupled \\ 
  \hline
N = 50 & 482.87 (472.96) & 7.95 (7.41) \\ 
  N = 100 & 462.88 (448.37) & 4.88 (3.45) \\ 
  N = 150 & 531.69 (546.32) & 4.19 (2.68) \\ 
  N = 200 & 569.49 (575.09) & 4.01 (2.34) \\ 
   \hline
\end{tabular}
\caption{Average meeting time as a function of the number of particles $N$ and of the resampling scheme.
Standard deviations are between brackets. Results obtained in the hidden auto-regressive model with $T = 20$. \label{table:effectresampling}} 
\end{table}

%% file: ar1effectnparticles.tex
\begin{table}[ht]
\centering
\begin{tabular}{rll}
  \hline
 & cost & meeting time \\ 
  \hline
N =   256 & 220567 (241823) & 861.59 (944.62) \\ 
  N =   512 & 17074 (17406) & 33.35 (34) \\ 
  N =  1024 & 7458 (5251) & 7.28 (5.13) \\ 
  N =  2048 & 8739 (4888) & 4.27 (2.39) \\ 
  N =  4096 & 14348 (6631) & 3.5 (1.62) \\ 
   \hline
\end{tabular}
\caption{Average cost and meeting time, as a function of the number of particles $N$.
Standard deviations are between brackets. Results obtained in the hidden auto-regressive model with $T = 500$.
\label{table:effectnparticles}} 
\end{table}

%% file: ar1effecttruncation.tex
\begin{table}[ht]
\centering
\begin{tabular}{ll}
  \hline
 & meeting time \\ 
  \hline
p =  0 & 31.94 (31.26) \\ 
  p =  0.025 & 17.66 (17.35) \\ 
  p =  0.05 & 11.82 (11.68) \\ 
   \hline
\end{tabular}
\caption{Average meeting time, as a function of the Geometric parameter $p$.
Standard deviations are between brackets. Results obtained in the hidden auto-regressive model with $T = 500$.
\label{table:effecttruncation}} 
\end{table}

%% file: ar1effectancestor.tex
\begin{table}[ht]
\centering
\begin{tabular}{rll}
  \hline
 & without ancestor sampling & with ancestor sampling \\ 
  \hline
N =  256 & 861.59 (944.62) & 8.79 (3.33) \\ 
  N =  512 & 33.35 (34) & 5.99 (2.27) \\ 
  N = 1024 & 7.28 (5.13) & 4.51 (1.88) \\ 
  N = 2048 & 4.27 (2.39) & 3.76 (1.63) \\ 
  N = 4096 & 3.5 (1.62) & 3.34 (1.51) \\ 
   \hline
\end{tabular}
\caption{Average meeting time, as a function of the number of particles $N$, with and without ancestor sampling.
Standard deviations are between brackets. Results obtained in the hidden auto-regressive model with $T = 500$.
\label{table:effectancestor}} 
\end{table}

%% file: tablear1horizon.tex
\begin{table}[ht]
\centering
\begin{tabular}{lll}
  \hline
 & without ancestor sampling & with ancestor sampling \\ 
  \hline
N = 128, T = 64 & 11.73 (10.87) & 6.54 (3.91) \\ 
  N = 256, T = 128 & 9.51 (7.61) & 5.77 (2.8) \\ 
  N = 512, T = 256 & 11.25 (9.33) & 5.66 (2.67) \\ 
  N = 1024, T = 512 & 7.8 (6.05) & 4.51 (1.81) \\ 
  N = 2048, T = 1024 & 9.07 (6.82) & 4.58 (1.9) \\ 
   \hline
\end{tabular}
\caption{Average meeting time, as a function of the number of particles $N$ and the time
horizon $T$, with and without ancestor sampling.
Standard deviations are between brackets. Results obtained in the hidden auto-regressive model.
\label{table:effecthorizon}} 
\end{table}

%% file: couplingpf.bbl
\begin{thebibliography}{62}
\expandafter\ifx\csname natexlab\endcsname\relax\def\natexlab#1{#1}\fi
\expandafter\ifx\csname url\endcsname\relax
  \def\url#1{\texttt{#1}}\fi
\expandafter\ifx\csname urlprefix\endcsname\relax\def\urlprefix{URL: }\fi

\bibitem[{Ahnert and Mulansky(2011)}]{ahnert2011odeint}
Ahnert, K. and Mulansky, M. (2011) Odeint-solving ordinary differential
  equations in {C}++.
\newblock \textit{arXiv preprint arXiv:1110.3397}.

\bibitem[{Andrieu et~al.(2010)Andrieu, Doucet and
  Holenstein}]{andrieu:doucet:holenstein:2010}
Andrieu, C., Doucet, A. and Holenstein, R. (2010) Particle {M}arkov chain
  {M}onte {C}arlo (with discussion).
\newblock \textit{Journal of the Royal Statistical Society: Series B
  (Statistical Methodology)}, \textbf{72}, 357--385.

\bibitem[{Andrieu et~al.(2013)Andrieu, Lee and
  Vihola}]{andrieuvihola2013uniform}
Andrieu, C., Lee, A. and Vihola, M. (2013) Uniform ergodicity of the iterated
  conditional {SMC} and geometric ergodicity of particle {G}ibbs samplers.
\newblock \textit{arXiv preprint arXiv:1312.6432}.

\bibitem[{Andrieu and Vihola(2015)}]{andrieu2015convergence}
Andrieu, C. and Vihola, M. (2015) Convergence properties of pseudo-marginal
  {M}arkov chain {M}onte {C}arlo algorithms.
\newblock \textit{The Annals of Applied Probability}, \textbf{25}, 1030--1077.

\bibitem[{Asmussen and Glynn(2007)}]{asmussen2007stochastic}
Asmussen, S. and Glynn, P.~W. (2007) \textit{Stochastic simulation: Algorithms
  and analysis}, vol.~57.
\newblock Springer.

\bibitem[{Aude et~al.(2016)Aude, Cuturi, Peyr{\'e} and
  Bach}]{aude2016stochastic}
Aude, G., Cuturi, M., Peyr{\'e}, G. and Bach, F. (2016) Stochastic optimization
  for large-scale optimal transport.
\newblock \textit{arXiv preprint arXiv:1605.08527}.

\bibitem[{Benamou et~al.(2015)Benamou, Carlier, Cuturi, Nenna and
  Peyr\'e}]{benamou2014iterative}
Benamou, J.-D., Carlier, G., Cuturi, M., Nenna, L. and Peyr\'e, G. (2015)
  Iterative {B}regman projections for regularized transportation problems.
\newblock \textit{SIAM Journal on Scientific Computing}, \textbf{37},
  A1111--A1138.

\bibitem[{B{\'e}rard et~al.(2014)B{\'e}rard, Del~Moral and
  Doucet}]{berard2014lognormal}
B{\'e}rard, J., Del~Moral, P. and Doucet, A. (2014) A lognormal central limit
  theorem for particle approximations of normalizing constants.
\newblock \textit{Electron. J. Probab}, \textbf{19}, 1--28.

\bibitem[{Bouchard-C\^ot\'e et~al.(2012)Bouchard-C\^ot\'e, Sankararaman and
  Jordan}]{BouchardCSJ:2012}
Bouchard-C\^ot\'e, A., Sankararaman, S. and Jordan, M.~I. (2012) Phylogenetic
  inference via sequential {M}onte {C}arlo.
\newblock \textit{Systematic Biology}, \textbf{61}, 579--593.

\bibitem[{Bret{\'o} et~al.(2009)Bret{\'o}, He, Ionides and
  King}]{breto2009time}
Bret{\'o}, C., He, D., Ionides, E.~L. and King, A.~A. (2009) Time series
  analysis via mechanistic models.
\newblock \textit{The Annals of Applied Statistics}, 319--348.

\bibitem[{Capp{{\'e}} et~al.(2005)Capp{{\'e}}, Moulines and
  Ryd{{\'e}}n}]{cappe:ryden:2004}
Capp{{\'e}}, O., Moulines, E. and Ryd{{\'e}}n, T. (2005) \textit{Inference in
  Hidden {M}arkov Models}.
\newblock Springer-Verlag, New York.

\bibitem[{C{\'e}rou et~al.(2011)C{\'e}rou, Del~Moral and
  Guyader}]{CerDelGuy2011nonasymptotic}
C{\'e}rou, F., Del~Moral, P. and Guyader, A. (2011) A nonasymptotic theorem for
  unnormalized {F}eynman--{K}ac particle models.
\newblock \textit{Ann. Inst. Henri Poincarr\'e}, \textbf{47}, 629--649.

\bibitem[{Chopin and Singh(2015)}]{ChopinS:2015}
Chopin, N. and Singh, S.~S. (2015) On particle {G}ibbs sampling.
\newblock \textit{Bernoulli}, \textbf{21}, 1855--1883.

\bibitem[{Cuturi(2013)}]{cuturi2013sinkhorn}
Cuturi, M. (2013) Sinkhorn distances: Lightspeed computation of optimal
  transport.
\newblock In \textit{Advances in Neural Information Processing Systems (NIPS)},
  2292--2300.

\bibitem[{Cuturi and Doucet(2014)}]{CuturiDoucet}
Cuturi, M. and Doucet, A. (2014) Fast computation of {W}asserstein barycenters.
\newblock In \textit{Proceedings of the 31st International Conference on
  Machine Learning (ICML)}, 685--693.

\bibitem[{Dahlin et~al.(2015)Dahlin, Lindsten, Kronander and
  Sch{\"o}n}]{dahlin2015accelerating}
Dahlin, J., Lindsten, F., Kronander, J. and Sch{\"o}n, T.~B. (2015)
  Accelerating pseudo-marginal {M}etropolis--{H}astings by correlating
  auxiliary variables.
\newblock \textit{arXiv preprint arXiv:1511.05483}.

\bibitem[{Del~Moral(2004)}]{del2004feynman}
Del~Moral, P. (2004) \textit{Feynman-Kac Formulae, Genealogical and Interacting
  Particle Systems with Applications}.
\newblock New York: Springer-Verlag.

\bibitem[{Del~Moral et~al.(2006)Del~Moral, Doucet and Jasra}]{DelDouJas:2006}
Del~Moral, P., Doucet, A. and Jasra, A. (2006) {Sequential Monte Carlo
  samplers}.
\newblock \textit{Journal of the Royal Statistical Society: Series B
  (Statistical Methodology)}, \textbf{68}, 411--436.

\bibitem[{Del~Moral and Murray(2015)}]{del2015sequential}
Del~Moral, P. and Murray, L.~M. (2015) Sequential monte carlo with highly
  informative observations.
\newblock \textit{SIAM/ASA Journal on Uncertainty Quantification}, \textbf{3},
  969--997.

\bibitem[{Deligiannidis et~al.(2015)Deligiannidis, Doucet and
  Pitt}]{deligiannidis2015correlated}
Deligiannidis, G., Doucet, A. and Pitt, M.~K. (2015) The correlated
  pseudo-marginal method.
\newblock \textit{arXiv preprint arXiv:1511.04992}.

\bibitem[{Diaconis and Freedman(1999)}]{diaconis1999iterated}
Diaconis, P. and Freedman, D. (1999) Iterated random functions.
\newblock \textit{SIAM review}, \textbf{41}, 45--76.

\bibitem[{Douc and Capp{\'e}(2005)}]{douc2005comparison}
Douc, R. and Capp{\'e}, O. (2005) Comparison of resampling schemes for particle
  filtering.
\newblock In \textit{Proceedings of the 4th International Symposium on Image
  and Signal Processing and Analysis (ISPA)}, 64--69.

\bibitem[{Doucet et~al.(2001)Doucet, {{de Freitas}} and
  Gordon}]{doucet:defreitas:gordon:2001}
Doucet, A., {{de Freitas}}, N. and Gordon, N. (2001) \textit{Sequential {M}onte
  {C}arlo methods in practice}.
\newblock Springer-Verlag, New York.

\bibitem[{Doucet and Johansen(2011)}]{doucet2011tutorial}
Doucet, A. and Johansen, A. (2011) A tutorial on particle filtering and
  smoothing: Fifteen years later.
\newblock In \textit{Handbook of Nonlinear Filtering}. Oxford, UK: Oxford
  University Press.

\bibitem[{Doucet et~al.(2015)Doucet, Pitt, Deligiannidis and
  Kohn}]{doucet2015efficient}
Doucet, A., Pitt, M., Deligiannidis, G. and Kohn, R. (2015) Efficient
  implementation of {M}arkov chain {M}onte {C}arlo when using an unbiased
  likelihood estimator.
\newblock \textit{Biometrika}, \textbf{102}, 295--313.

\bibitem[{Gerber and Chopin(2015)}]{gerber2015sequential}
Gerber, M. and Chopin, N. (2015) Sequential quasi {M}onte {C}arlo.
\newblock \textit{Journal of the Royal Statistical Society: Series B
  (Statistical Methodology)}, \textbf{77}, 509--579.

\bibitem[{Glasserman and Yao(1992)}]{glasserman1992some}
Glasserman, P. and Yao, D.~D. (1992) Some guidelines and guarantees for common
  random numbers.
\newblock \textit{Management Science}, \textbf{38}, 884--908.

\bibitem[{Glynn and Rhee(2014)}]{glynn2014exact}
Glynn, P.~W. and Rhee, C.-H. (2014) Exact estimation for {M}arkov chain
  equilibrium expectations.
\newblock \textit{J. Appl. Probab.}, \textbf{51A}, 377--389.

\bibitem[{Glynn and Whitt(1992)}]{glynn1992asymptotic}
Glynn, P.~W. and Whitt, W. (1992) The asymptotic efficiency of simulation
  estimators.
\newblock \textit{Operations Research}, \textbf{40}, 505--520.

\bibitem[{Gordon et~al.(1993)Gordon, Salmond and
  Smith}]{gordon:salmon:smith:1993}
Gordon, N., Salmond, J. and Smith, A. (1993) A novel approach to
  non-linear/non-{G}aussian {B}ayesian state estimation.
\newblock \textit{IEE Proceedings on Radar and Signal Processing},
  \textbf{140}, 107--113.

\bibitem[{Guarniero et~al.(2015)Guarniero, Johansen and
  Lee}]{guarniero2015iterated}
Guarniero, P., Johansen, A.~M. and Lee, A. (2015) The iterated auxiliary
  particle filter.
\newblock \textit{arXiv preprint arXiv:1511.06286}.

\bibitem[{Jacob(2015)}]{jacob2015sequential}
Jacob, P.~E. (2015) Sequential {B}ayesian inference for implicit hidden
  {M}arkov models and current limitations.
\newblock \textit{ESAIM: Proceedings and Surveys}, \textbf{51}, 24--48.

\bibitem[{Jacob et~al.(2015)Jacob, Murray and Rubenthaler}]{jacob2015path}
Jacob, P.~E., Murray, L.~M. and Rubenthaler, S. (2015) Path storage in the
  particle filter.
\newblock \textit{Statistics and Computing}, \textbf{25}, 487--496.

\bibitem[{Jasra et~al.(2015)Jasra, Kamatani, Law and
  Zhou}]{jasra2015multilevel}
Jasra, A., Kamatani, K., Law, K.~J. and Zhou, Y. (2015) Multilevel particle
  filter.
\newblock \textit{arXiv preprint arXiv:1510.04977}.

\bibitem[{Jones et~al.(2010)Jones, Parslow and Murray}]{jones2010bayesian}
Jones, E., Parslow, J. and Murray, L. (2010) A {B}ayesian approach to state and
  parameter estimation in a phytoplankton-zooplankton model.
\newblock \textit{Australian Meteorological and Oceanographic Journal},
  \textbf{59}, 7--16.

\bibitem[{Jun et~al.(2012)Jun, Wang and
  Bouchard-C{\^o}t{\'e}}]{jun2012entangled}
Jun, S.-H., Wang, L. and Bouchard-C{\^o}t{\'e}, A. (2012) Entangled {M}onte
  {C}arlo.
\newblock In \textit{Advances in Neural Information Processing Systems
  ({NIPS})}, 2726--2734.

\bibitem[{Kahn and Marshall(1953)}]{kahn1953methods}
Kahn, H. and Marshall, A.~W. (1953) Methods of reducing sample size in {M}onte
  {C}arlo computations.
\newblock \textit{Journal of the Operations Research Society of America},
  \textbf{1}, 263--278.

\bibitem[{Kantas et~al.(2015)Kantas, Doucet, Singh, Maciejowski and
  Chopin}]{kantas2015particle}
Kantas, N., Doucet, A., Singh, S.~S., Maciejowski, J. and Chopin, N. (2015) On
  particle methods for parameter estimation in state-space models.
\newblock \textit{Statistical science}, \textbf{30}, 328--351.

\bibitem[{Lee(2008)}]{lee2008towards}
Lee, A. (2008) \textit{Towards smooth particle filters for likelihood
  estimation with multivariate latent variables}.
\newblock Master's thesis, University of British Columbia.

\bibitem[{{Lee} et~al.(2014){Lee}, {Doucet} and
  {{\L}atuszy{\'n}ski}}]{leedoucetperfectsimulation}
{Lee}, A., {Doucet}, A. and {{\L}atuszy{\'n}ski}, K. (2014) {Perfect simulation
  using atomic regeneration with application to sequential Monte Carlo}.
\newblock \textit{ArXiv e-prints}.

\bibitem[{Lee and Holmes(2010)}]{leecommentonpmcmc}
Lee, A. and Holmes, C. (2010) Comment on {P}article {M}arkov chain {M}onte
  {C}arlo by {A}ndrieu, {D}oucet and {H}olenstein.
\newblock \textit{Journal of the Royal Statistical Society: Series B
  (Statistical Methodology)}, \textbf{72}, 357--385.

\bibitem[{Lee and Whiteley(2015{\natexlab{a}})}]{lee2015forest}
Lee, A. and Whiteley, N. (2015{\natexlab{a}}) Forest resampling for distributed
  sequential {M}onte {C}arlo.
\newblock \textit{Statistical Analysis and Data Mining: The ASA Data Science
  Journal}.

\bibitem[{Lee and Whiteley(2015{\natexlab{b}})}]{lee2015variance}
--- (2015{\natexlab{b}}) Variance estimation and allocation in the particle
  filter.
\newblock \textit{arXiv preprint arXiv:1509.00394}.

\bibitem[{Lindsten et~al.(2015)Lindsten, Douc and Moulines}]{LindstenDM:2015}
Lindsten, F., Douc, R. and Moulines, E. (2015) Uniform ergodicity of the
  particle {G}ibbs sampler.
\newblock \textit{Scandinavian Journal of Statistics}, \textbf{42}, 775--797.

\bibitem[{Lindsten et~al.(2014)Lindsten, Jordan and
  Sch\"{o}n}]{LindstenJS:2014}
Lindsten, F., Jordan, M.~I. and Sch\"{o}n, T.~B. (2014) Particle {G}ibbs with
  ancestor sampling.
\newblock \textit{Journal of Machine Learning Research (JMLR)}, \textbf{15},
  2145--2184.

\bibitem[{Lindsten and Sch\"{o}n(2013)}]{LindstenS:2013}
Lindsten, F. and Sch\"{o}n, T.~B. (2013) Backward simulation methods for
  {M}onte {C}arlo statistical inference.
\newblock \textit{Foundations and {T}rends in {M}achine {L}earning},
  \textbf{6}, 1--143.

\bibitem[{Lindvall(2002)}]{lindvall2002lectures}
Lindvall, T. (2002) \textit{Lectures on the coupling method}.
\newblock Courier Corporation.

\bibitem[{Malik and Pitt(2011)}]{malik2011particle}
Malik, S. and Pitt, M.~K. (2011) Particle filters for continuous likelihood
  evaluation and maximisation.
\newblock \textit{Journal of Econometrics}, \textbf{165}, 190--209.

\bibitem[{McLeish(2012)}]{McLeish:2012}
McLeish, D. (2012) A general method for debiasing a {M}onte {C}arlo estimator.
\newblock \textit{Monte Carlo methods and applications}, \textbf{17}, 301--315.

\bibitem[{Murray et~al.(2015)Murray, Lee and Jacob}]{murray2015parallel}
Murray, L.~M., Lee, A. and Jacob, P.~E. (2015) Parallel resampling in the
  particle filter.
\newblock \textit{Journal of Computational and Graphical Statistics}.

\bibitem[{Naesseth et~al.(2014)Naesseth, Lindsten and
  Sch\"{o}n}]{NaessethLS:2014}
Naesseth, C.~A., Lindsten, F. and Sch\"{o}n, T.~B. (2014) Sequential {M}onte
  {C}arlo for graphical models.
\newblock In \textit{Advances in Neural Information Processing Systems (NIPS)
  27}. Montreal, Quebec, Canada.

\bibitem[{Pitt(2002)}]{pitt2002smooth}
Pitt, M.~K. (2002) Smooth particle filters for likelihood evaluation and
  maximisation.
\newblock \textit{Technical report, University of Warwick, Department of
  Economics}.

\bibitem[{Poyiadjis et~al.(2011)Poyiadjis, Doucet and
  Singh}]{poyiadjis2011particle}
Poyiadjis, G., Doucet, A. and Singh, S.~S. (2011) Particle approximations of
  the score and observed information matrix in state space models with
  application to parameter estimation.
\newblock \textit{Biometrika}, \textbf{98}, 65--80.

\bibitem[{Rhee(2013)}]{rhee:phd}
Rhee, C. (2013) \textit{Unbiased Estimation with Biased Samplers}.
\newblock Ph.D. thesis, Stanford University.
\newblock \urlprefix\url{http://purl.stanford.edu/nf154yt1415}.

\bibitem[{Rhee and Glynn(2012)}]{Rhee:Glynn:2012}
Rhee, C. and Glynn, P.~W. (2012) A new approach to unbiased estimation for
  {SDE}'s.
\newblock In \textit{Proceedings of the Winter Simulation Conference},
  17:1--17:7.

\bibitem[{Ruiz and Kappen(2016)}]{ruiz2016particle}
Ruiz, H.-C. and Kappen, H. (2016) Particle smoothing for hidden diffusion
  processes: Adaptive path integral smoother.
\newblock \textit{arXiv preprint arXiv:1605.00278}.

\bibitem[{Sen et~al.(2016)Sen, Thiery and Jasra}]{sen2016coupling}
Sen, D., Thiery, A. and Jasra, A. (2016) On coupling particle filter
  trajectories.
\newblock \textit{arXiv preprint arXiv:1606.01016}.

\bibitem[{{The CGAL Project}(2016)}]{cgal:eb-16a}
{The CGAL Project} (2016) \textit{{CGAL} User and Reference Manual}.
\newblock {CGAL Editorial Board}, {4.8} edn.
\newblock \urlprefix\url{http://doc.cgal.org/4.8/Manual/packages.html}.

\bibitem[{Vihola(2015)}]{vihola2015unbiased}
Vihola, M. (2015) Unbiased estimators and multilevel {M}onte {C}arlo.
\newblock \textit{arXiv preprint arXiv:1512.01022}.

\bibitem[{Whiteley(2010)}]{whiteleycommentonpmcmc}
Whiteley, N. (2010) Comment on {P}article {M}arkov chain {M}onte {C}arlo by
  {A}ndrieu, {D}oucet and {H}olenstein.
\newblock \textit{Journal of the Royal Statistical Society: Series B
  (Statistical Methodology)}, \textbf{72}, 357--385.

\bibitem[{Whiteley(2013)}]{whiteley2011stability}
--- (2013) Stability properties of some particle filters.
\newblock \textit{Ann. Appl. Probab.}, \textbf{23}, 2500--2537.

\bibitem[{Williams(1991)}]{williams1991probability}
Williams, D. (1991) \textit{Probability with martingales}.
\newblock Cambridge university press.

\end{thebibliography}
